\documentclass[aps,prb,reprint,superscriptaddress,amsmath,amssymb,showpacs]{revtex4-1}

\usepackage{appendix}
\usepackage{graphicx}
\usepackage[usenames]{color}
\usepackage{multirow}

\newcommand{\specialcell}[2][c]{\begin{tabular}[#1]{@{}c@{}}#2\end{tabular}}

\begin{document}

\title{Complex time, shredded propagator method for large-scale GW calculations}

\author{Minjung Kim}
\affiliation{
Department of Applied Physics, Yale University,
New Haven, Connecticut 06520, USA
}
\author{Glenn J. Martyna}
\affiliation{
IBM TJ Watson Laboratory, Yorktown Heights, 10598, New York, USA 
}
\affiliation{Pimpernel Science, Software and Information Technoglogy, Westchester, NY 10598, USA}

\author{Sohrab Ismail-Beigi}
\email{sohrab.ismail-beigi@yale.edu}
\affiliation{
Department of Applied Physics, Yale University,
New Haven, Connecticut 06520, USA 
}

\date{\today}
\begin{abstract}
The GW method is a many-body electronic structure technique capable of generating accurate quasiparticle properties for realistic systems spanning physics, chemistry, and materials science.  Despite its power, GW is not routinely applied to large complex assemblies due to its large computational overhead and quartic scaling with particle number. Here, the GW equations are recast, exactly, as Fourier-Laplace time integrals over complex time propagators. The propagators are then ``shredded" via energy partitioning and the time integrals approximated in a controlled manner using generalized Gaussian quadrature(s) while discrete variable methods are employed to represent the required propagators in real-space. The resulting cubic scaling GW method has a sufficiently small prefactor to outperform standard quartic scaling methods on small systems ($\gtrapprox$ 10 atoms) and also represents a substantial improvement over other cubic methods tested for all system sizes studied. The approach can be applied to any theoretical framework containing large sums of terms with  energy differences in the denominator. 
\end{abstract}
\maketitle

\section{Introduction}
Density Functional Theory (DFT)~\cite{hohenberg_inhomogeneous_1964,kohn_self-consistent_1965} within the local density (LDA) or generalized gradient (GGA)~\cite{perdew_self-interaction_1981,perdew_atoms_1992} approximation provides a solid workhorse capable of realistically modeling an ever increasing number and variety of physical systems spanning condensed matter, chemistry, and biology.  Generally, this approach provides a highly satisfactory description of the total energy, electron density,  atomic geometries, vibrational modes, etc. However, DFT is a ground-state theory for electrons and DFT band energies do not have direct physical meaning because DFT is not formally a quasiparticle theory. Therefore, there can be significant failures when DFT band structure is used to predict electronic excitations \cite{perdew_density-functional_1982, lundqvist_theory_2013, 
anisimov_first-principles_1997}.

The GW approximation to the electron self-energy \cite{hedin_new_1965,hybertsen_electron_1986, aryasetiawan_gw_1998,onida_electronic_2002} is one of the most accurate fully {\it ab initio} methods for the prediction of electronic excitations. Despite its power, GW is not routinely applied to complex materials systems due to its unfavorable computational scaling: the cost of a standard GW calculation scales as ${\cal O}(N^4)$ where $N$ is the number of atoms in the simulation cell whereas the standard input to a GW study, a Kohn-Sham DFT calculation, scales as ${\cal O}(N ^3)$.  

Reducing the computational overhead of GW calculations has been the subject of much prior research.  First, GW methods scaling as ${\cal O}(N ^4)$ but with smaller prefactors either avoid the use of unoccupied states via iterative matrix inversion~\cite{wilson_efficient_2008, wilson_iterative_2009, rocca_ab_2010,lu_dielectric_2008, giustino_gw_2010,umari_gw_2010,govoni_large_2015} or use sum rules or energy integration to greatly reduce the number of unoccupied states required for convergence~\cite{bruneval_accurate_2008,berger_ab_2010,gao_speeding_2016}. Second, cubic-scaling ${\cal O}(N ^3)$ methods including both a spectral representation approach~\cite{foerster_on3_2011} and a space/imaginary time method~\cite{liu_cubic_2016} utilizing analytical continuation from imaginary to real frequencies, have been proposed.  Third, a linear scaling GW  technique~\cite{neuhauser_breaking_2014} has recently been developed that employs stochastic approaches for the total density of electronic states with the caveat that the non-deterministic stochastic noise must be added to the list of usual convergence parameters.

Here, we present a deterministic, small prefactor, ${\cal O}(N ^3)$ scaling GW approach that does not require analytic continuation. The GW equations are first recast exactly using Fourier-Laplace identities into the complex time domain where products of propagators expressed in real-space using discrete variable techniques~\cite{dvrbook} are integrated over time to generate an ${\cal O}(N ^3)$ GW formalism. However, the time integrals are ill-posed due to the multiple time scales inherent in the propagators. Second, the time scale challenge is met by shredding the propagators in energy space, again exactly, to allow windows of limited dynamical bandwidth to be treated via generalized Gaussian quadrature numerical integration with low overhead and high accuracy. The unique combination of a (complex) time domain formalism, bandwidth taming propagator partitioning, and discrete variable real-space forms of the propagators, permits a fast ${\cal O}(N ^3)$ to emerge.  Last, our approach is easy to implement in standard GW applications~\cite{deslippe_berkeleygw:_2012,our_own_GW} because the formulae follow naturally from those of the standard approach(es) and much of the existing software can be refactored to utilize our reduced order technique.

The resulting GW formalism is tested to ensure both its accuracy and high performance in comparison to the standard ${\cal O}(N^4)$ approach for crystalline silicon, magnesium oxide, and aluminium as a function of supercell size. The new method's accuracy and performance were also compared to that of reduced overhead quartic scaling methods as well as existing ${\cal O}(N^3)$ scaling techniques. Importantly we provide estimates of the speed-up over conventional GW computations and the memory requirement in the application of the new method to study technologically and scientifically interesting systems consisting of $\lessapprox 200-300$ atoms - the sweet spot for the approach.

\section{Theory}
\subsection{Summary of GW}

The theoretical object of interest for understanding one-electron properties such as quasiparticle bands and wave functions is the one-electron Green's function $G(x,t,x',t')$, which describes the propagation amplitude of an electron starting at $x'$ at time $t'$ and ending at $x$ at time $t$:\cite{negele_quantum_1998}
\[
iG(x,t,x',t') = \langle T\left\{\, \hat \psi (x,t) \, \hat \psi(x',t')^\dag \, \right\} \rangle\,,
\]
where the electron coordinate $x=(r,\sigma)$ specifies electron position ($r$) and spin ($\sigma$).
Here, $\hat \psi(x,t)$ is the electron annihilation field operator at $(x,t)$, $T$ is the time-ordering operator, and the average is over the statistical  ensemble of interest.  We focus primarily on the zero-temperature case (i.e., ground-state averaging); however, to treat systems with small gaps, the grand 
canonical ensemble is invoked. As is standard, henceforth atomic units are employed, $\hbar=1$ and the quantum of charge $e=1$.

The Green's function in the frequency domain obeys Dyson's equation
\[
G^{-1}(\omega) = \omega I  - \left[ T + V_{ion} + V_H + \Sigma(\omega) \right]
\]   
where the $x,x'$ indices have been suppressed; a more compact but complete notation shall be employed henceforth
\[
G(\omega)_{x,x'} = G(x,x',\omega)\,.
\]
Above, $I$ is the identity operator, $T$ is the electron kinetic operator, $V_{ion}$ is the electron-ion interaction potential operator (or pseudopotential for valence electron only calculations), $V_H$ is the Hartree potential operator, and $\Sigma(\omega)$ is the self-energy operator encoding all the many-body interaction effects on the electron Green's function. 

The GW approximation to the self-energy is
\[
\Sigma(t)_{x,x'} = iG(t)_{x,x'}W(t^+)_{r,r'}
\] 
where $t^+$ is infinitesimally larger than $t$ and $W(t)_{r,r'}$ is the dynamical screened Coulomb interaction between an external test charge at $(r',0)$ and $(r,t)$:
\[
W(\omega)_{r,r'} = \int dr''\ \epsilon^{-1}(\omega)_{r,r''}V_{r'',r'}\,.
\]
Here, $\epsilon$ is the linear response, dynamic and nonlocal microscopic dielectric screening matrix and $V_{r,r'}=1/|r-r'|$ is the bare Coulomb interaction.  The GW self-energy includes the effects due to dynamical and nonlocal screening on the propagation of electrons in a many-body environment.  The notation introduced above to be continued below is that parametric functional dependencies are placed in parentheses and explicit dependencies are given as subscripts; the alternative notation wherein all variables are in parentheses with explicit dependencies given first followed by parametric dependencies separated by a semicolon is also employed where convenient (e.g. $W(r,r';\omega)\equiv W(\omega)_{r,r'}$).

To provide a closed and complete set of equations, one must approximate $\epsilon$. The most common approach is the random-phase approximation (RPA):  one first writes $\epsilon$ in terms of the dynamic irreducible polarizability $P$ via
\begin{equation}
\epsilon(\omega)_{r,r'} = \delta(r-r') - \int dr''\  V_{r,r''} P(\omega)_{r'',r}
\label{eq:epsfromP}
\end{equation}
and $P$ is related to $G$ by the RPA
\[
P(t)_{r,r'} = -i\sum_{\sigma,\sigma'} G(t)_{x,x'}G(-t)_{x',x}\,.
\]

In the vast majority of GW calculations, including the formalism given here, the Green's function is approximated by an independent electron form (band theory) specified by a complete set of one-particle eigenstates $\psi_n(x)$ (compacted to $\psi_{x,n})$ and eigenvalues $E_n$
\begin{equation}
G(\omega)_{x,x'} = \sum_n \frac{\psi_{x,n}\psi_{x',n}^*}{\omega - E_n}\,.
\label{eq:Gindep}
\end{equation}
The $\psi_n$ and $E_n$ are typically obtained as eigenstates of a non-interacting one-particle Hamiltonian from a first principles method such as Density Functional Theory \cite{hohenberg_inhomogeneous_1964,kohn_self-consistent_1965} although it is not limited to this choice.  Although not central to the analysis given here, formally $E_n$ has a small imaginary part that is positive for occupied states (i.e., energies below the chemical potential) and negative for unoccupied states.
We have suppressed the non-essential crystal momentum index $k$ in Eq.~(\ref{eq:Gindep}) for simplicity - including it simply amounts to adding the $k$ index to the eigenstates $\psi_{x,n}\rightarrow\psi_{x,nk}$ and energies $E_n\rightarrow E_{nk}$ and averaging over the $k$ sampled in the first Brillouin zone.

For our purposes, the frequency domain representations of all quantities are useful.  The Green's function $G$ in frequency space is given in Eq.~(\ref{eq:Gindep}) while the frequency dependent polarizability, $P$, is
\begin{eqnarray}
P(\omega)_{r,r'} &=& \sum_{c,v,\sigma,\sigma'}\psi_{x,c}\psi_{x,v}^*\psi_{x',c}^*\psi_{x',v}[f(E_v)-f(E_c)]\nonumber\\
& & \times \frac{2(E_c-E_v)}{\omega^2-(E_c-E_v)^2} 
\label{eq:Pomegadef}
\\
&=& \sum_{c,v,\sigma,\sigma'} 
     \psi_{x,c}\psi_{x,v}^*\psi_{x',c}^*\psi_{x',v}[f(E_v)-f(E_c)] \nonumber \\
     & & \ \ \ \times \left [\frac{1}{(\omega-(E_c-E_v))} - \frac{1}{(\omega+(E_c-E_v))}\right ] \nonumber
\end{eqnarray}
Here, $v$ labels occupied (valence) eigenstates while $c$ labels unoccupied (conduction) eigenstates.  The occupancy function $f(E)$ required to handle finite temperatures for zero/small gap systems is explicitly included (see Sec.~\ref{sec:metals} for more details); for gapped systems at zero temperature $f(E_v)=1$ and $f(E_c)=0$. Note for the non-expert, the occupancy formally depends parametrically on the thermodynamic variables, the inverse temperature $\beta=1/k_BT$ and the chemical potential $\mu$, $f(E;\beta,\mu)$.

Of particular practical importance is the zero-frequency or static polarizability $P(\omega=0)$ (which we also simply denote as $P$ below) 
\begin{equation}
P_{r,r'} = \sum_{c,v,\sigma,\sigma'} \frac{-2\psi_{x,c}\psi_{x,v}^*\psi_{x',c}^*\psi_{x',v}[f(E_v)-f(E_c)]}{E_c-E_v}
\label{eq:Pstatic}
\end{equation}
which is employed both as part of plasmon-pole models of the frequency dependent screening\cite{hedin_new_1965,hedin_effects_1970,hybertsen_electron_1986,aryasetiawan_gw_1998,onida_electronic_2002} as well as within the COHSEX approximation\cite{hedin_new_1965} (see below).  
Again, the crystal momentum index has been suppressed for simplicity; including it requires the replacements $P\rightarrow P^q$ where $q$ is the momentum transfer, $\psi_{x,v}\rightarrow \psi_{x,vk}$ and $E_v\rightarrow E_{vk}$, $\psi_{x,c}\rightarrow\psi_{x,ck+q}$ and  $E_c\rightarrow E_{ck+q}$, and averaging Eq.~(\ref{eq:Pomegadef}) and Eq.~(\ref{eq:Pstatic}) over the $k$ (Brillouin zone sampling).  We note that current numerical methods for computing $P$ based on the sum-over-states formulae, e.g., that  of Eq.~(\ref{eq:Pstatic}), have an ${\cal O}(N^4)$ scaling (e.g.,  see Ref.~[\onlinecite{deslippe_berkeleygw:_2012}]). 

Formally, the screened interaction $W$ can always be represented as a sum of ``plasmon'' screening modes indexed by $p$,
\begin{equation}
W(\omega)_{r,r'} = V_{r,r'} + \sum_p  \frac{ 2\omega_p\, B^{(p)}_{r,r'} } { \omega^2 - \omega_p^2}
\label{eq:Wplasmonrep}
\end{equation}
Here $B_p$ is the mode strength for screening mode $p$ and $\omega_p>0$ is its frequency.  This form is directly relevant when making computationally efficient plasmon-pole models for the screened interaction~\cite{hedin_effects_1970}.  The self-energy is then given by
\begin{multline}
\Sigma(\omega)_{x,x'} = -\sum_v \psi_{x,v}\psi_{x',v}^* W(\omega-E_v)_{r,r'} \\+ \sum_{n,p} \frac{\psi_{x,n}B^{(p)}_{r,r'}\psi_{x',n}^*}{\omega-E_n-\omega_p}\,.
\label{eq:Sigmafreq}
\end{multline}
Inclusion of crystal momentum in Eq.~(\ref{eq:Sigmafreq}) means $\Sigma(\omega)$ carries a $k$ index, $\psi_{x,v}\rightarrow\psi_{x,vk-q}$ and $E_v\rightarrow E_{v,k-q}$,  all screening quantities derived from $P$ such as $W$, $\omega_p$ and $B^{(p)}$ now also carry a $q$ index, and Eq.~(\ref{eq:Sigmafreq}) is then averaged over the $q$ sampling.

Within the COHSEX approximation, when the applicable screening frequencies $\omega_p$ are much larger than the interband energies of interest,
the frequency dependence of $\Sigma$ can be neglected
\begin{multline}
\Sigma^{COHSEX}_{x,x'} = -\sum_v \psi_{x,v}\psi_{x',v}^* W(0)_{r,r'} \\+ \frac{1}{2}\delta(x-x')\left[W(0)_{r,r'}-V_{r,r'}\right] \,.
\label{eq:SigmaCOHSEX}
\end{multline}
The numerically intensive part of the COHSEX approximation is the computation of the static polarizability,  Eq.~(\ref{eq:Pstatic}) --- once $P$ is on hand, the static $W(0)$ is completely determined by $P$ via matrix multiplication and inversion,
\[
W(0) = \epsilon^{-1}(0)V = (I-VP)^{-1}V\,.
\]

Eqs.~(\ref{eq:Pomegadef},\ref{eq:Pstatic},\ref{eq:Sigmafreq},\ref{eq:SigmaCOHSEX}) are of primary interest here as evaluating them scales as ${\cal O}(N^4)$ as written. Terms with manifestly cubic scaling terms will not be discussed further.

\subsection{Complex time shredded propagator formalism}\label{sec:formalism}

We now describe the main ideas and merits of our new approach to cubic scaling GW calculations. The resulting formalism is general and can be applied to a broad array of theoretical frameworks whose evaluation involves sums over states with energy differences in denominators.  

The analytic structure of the equations central to GW calculations, outlined in the prior section, necessitates the evaluation of terms of the form 
\begin{equation}
\chi(\omega)_{r,r'} = \sum_{i=1}^{N_a}\sum_{j=1}^{N_b} \frac{A^i_{r,r'}B^j_{r,r'}}
{\omega+a_i-b_j} 
\label{eq:Xdoublesum}
\end{equation}
as can be discerned from Eqs.~(\ref{eq:Pomegadef},\ref{eq:Pstatic},\ref{eq:Sigmafreq},\ref{eq:SigmaCOHSEX}). The input energies $a_i$ and $b_j$ and the matrices $A^i$ and $B^j$ are either direct outputs of the ${\cal O}(N^3)$ ground state calculation (i.e., single particle energies and products of wave functions when $\chi=P$), or are obtained from ${\cal O}(N^3)$ matrix operations on the frequency dependent polarizability $P(\omega)$, or other such derived quantities.

The analytic form of $\chi$  in Eq. (\ref{eq:Xdoublesum}) arises because we have chosen to work in the frequency or energy representation.  However, one can equally well represent such an equation in real, imaginary or complex time by changing the structure of the theory to involve time integrals over propagators. Here, we will effect the change of representation from time to frequency directly through the introduction of Fourier-Laplace identities which allows us to reduce the computational complexity of the GW calculation as shown below.  This imaginary time formalism has strong connection to the work presented in  Refs.~[\onlinecite{rieger_gw_1999,kaltak_low_2014,liu_cubic_2016}]. 

While the frequency representation has advantages, the evaluation of Eq.~(\ref{eq:Xdoublesum}) scales as ${\cal O}(N_aN_bN_r^2)$ because the numerator is separable but the energy denominator is not. This basic structure of the frequency representation leads to the familiar ${\cal O}(N^4)$ computational complexity of GW as the number of states or modes ($N_a,N_b$) and the number real-space points ($N_r$) required to represent them, here by discrete variable methods, scale as the number of electrons, $N$.  For the widely used plane wave / Fourier basis, adopted herein, a uniform grid in $r$-space that is dual to the finite $g$-space representation is indicated - fast Fourier transforms (FFTs) switch between the dual spaces, $g$- and $r$-space, both efficiently and {\it exactly} (without information loss);
for other basis sets, appropriate real-space discrete variable representations (DVRs) with similar dual properties can be adopted~\cite{baye_heenen_1986,friesner_solution_1986,dvrbook}.

In the following, a time domain formalism that reduces the computational complexity of Eq.~(\ref{eq:Xdoublesum}) by $N$ to achieve ${\cal O}((N_a+N_b)N_r^2)\sim {\cal O}(N^3)$ scaling in a controlled and rapidly convergent manner, is developed. Again, this will be accomplished through the introduction of time integrals and associated propagators which we shall then shred (i.e., partition) to tame the multiple time scales inherent to the theory. The resulting formulation is general - it can be directly applied to any theory of the structure of Eq.~(\ref{eq:Xdoublesum}).

Reduced scaling is enabled by replacing the energy denominator of Eq.~(\ref{eq:Xdoublesum}) by a separable form through the introduction of the generalized Fourier-Laplace transform
\begin{equation}
F(E;\zeta) =   \int_0^\infty d\tau\, h(\tau;\zeta)\exp[-\zeta E\tau ]
\label{eq:Fdef}
\end{equation}
That is, inserting the transform,  Eq.~(\ref{eq:Xdoublesum}) becomes
\begin{equation}
\chi(\omega)_{r,r'} = \sum_{i=1}^{N_a}\sum_{j=1}^{N_b}
F(\omega+a_i-b_j;\zeta) A^i_{r,r'}B^j_{r,r'}
\label{eq:XdoublesumwithF}
\end{equation}
Here, $\zeta$ is a complex constant with $|\zeta|$ akin to an inverse Planck's constant that sets the energy scale, and $h(\tau;\zeta)$ a  weight function. The desired separability arises from the exponential function in the integrand of $F(E;\zeta)$, and allows us to reduce the computational complexity of GW as follows.

To motivate the utility of Eq.~(\ref{eq:XdoublesumwithF}), consider the case where $\forall\ i,j$ either $\omega+a_i-b_j>0$ or $\omega+a_i-b_j<0$: here, $\zeta$ is chosen to be real (positive for the first case and negative for the second), and we set $h(\tau;\zeta)=\zeta$. This corresponds  to a  textbook Laplace transform~\cite{noteonLaplace} and  yields an {\em exact} expression for the energy denominator:
\begin{equation}
\lim_{h(\tau;\zeta)\rightarrow\zeta} F(\omega+a_i-b_j;\zeta) = \frac{1}{\omega+a_i-b_j} \,.
\label{eq:genFourLap}
\end{equation}
For this case, the introduction of the transform involves no approximation, and $h(\tau;\zeta)=\zeta$ will be employed to establish and describe our formalism.  It is directly applicable to the static limit of $\chi(\omega)$ where $\omega\rightarrow0$ and $b_j-a_i<0$ $\forall\ i,j$ (i.e., a gapped systems when computing the static polarizability matrix of Eq.~(\ref{eq:Pstatic})). The importance of non-unit $\zeta$ will become apparent below. A yet more general treatment, applicable to gapless systems and finite frequencies $\omega\ne0$, requiring non-trivial $h(\tau;\zeta)$ will then be given wherein $F$ becomes an approximation to the inverse of the energy denominator within the class of regularization procedures commonly employed in standard GW computations.

Inserting the generalized Fourier-Laplace identity into Eq.~(\ref{eq:Xdoublesum}) yields
\begin{eqnarray}
\label{eq:Xintegral}
\chi(0)_{r,r'} &=& \int_0^\infty d\tau \, h(\tau;\zeta)\, \left[\sum_{i=1}^{N_a}A^i_{r,r'}e^{-\zeta (E_{\mathrm{off}}-a_i)\tau}\right] \nonumber \\
& & \ \ \ \ \ \ \ \ \ \ \ \ \ \times \left[\sum_{j=1}^{N_b}B^j_{r,r'}e^{-\zeta (b_j-E_{\mathrm{off}})\tau}\right]\, \\
&=& \int_0^\infty d\tau\ h(\tau;\zeta) \rho^A_{r,r'}(\zeta\tau)\rho^B_{r,r'}(\zeta\tau) \nonumber \\
&=& \int_0^\infty d\tau \, h(\tau;\zeta)\, \tilde{\chi}(\zeta\tau;0)_{r,r'} \nonumber\ .
\end{eqnarray}
Here, $E_{\mathrm{off}}$ is an energy offset for numerical convenience such that all the exponential functions are decaying and 
\begin{eqnarray}
\rho^A(\zeta\tau)_{r,r'}&=& \sum_{i=1}^{N_a}A^i_{r,r'}e^{-\zeta (E_{\mathrm{off}}-a_i)\tau} \nonumber \\
\rho^B(\zeta\tau)_{r,r'} &=& \sum_{j=1}^{N_b}B^j_{r,r'}e^{-\zeta (b_j-E_{\mathrm{off}})\tau} \nonumber \\
\tilde{\chi}(\zeta\tau;0)_{r,r'} &=& \rho^A(\zeta\tau)_{r,r'}\rho^B(\zeta\tau)_{r,r'}
\end{eqnarray}
where the $\rho^{A,B}(\zeta\tau)$ are imaginary time propagators. The result is a separable form for $\tilde{\chi}(\tau\zeta;0)_{r,r'}$ (a product of $A$ and $B$ propagators), whose zero frequency transform over $h(\tau;\zeta)$ yields the desired $\chi(0)_{r,r'}$.  This exact reformulation can be evaluated in ${\cal O}(N^3)$ given that an ${\cal O}(N^0)$ scaling discretization (i.e., quadrature) of the time integral can be defined.  

Consider that the largest energy difference in the argument of the exponential terms defining $\tilde{\chi}(\zeta\tau;0)_{r,r'}$, is the bandwidth $E_{\mathrm{BW}}=\max(a_i)-\min(b_j)$ while the smallest energy difference is the gap $E_{\mathrm{g}}=\min(a_i)-\max(b_j)$ which are both known from input.  Both  energy differences are independent of system size $N$ (i.e., they are intensive and not extensive); hence the longest and shortest time scales ($1/E_{\mathrm{BW}}$ and $1/E_{\mathrm{g}}$)  in $\tilde{\chi}(\tau\zeta;0)_{r,r'}$ are independent of $N$. Therefore, barring non-analytic behavior in the density of states or modes, a system size independent discretization scheme can be devised to generate  
$\chi(0)_{r,r'}$ from $\tilde{\chi}(\zeta\tau;0)_{r,r'}$.
Of course, the formulation is most useful when the discrete form converges rapidly to the integral with increasing number of discretizations (i.e., quadrature points).

The development of a rapidly convergent discretization scheme is, however, challenged by the large dynamic range present in the electronic structure of complex systems, $E_{\mathrm{BW}}/E_{\mathrm{g}}\gtrapprox 100$. Simply selecting the free parameter $|\zeta|\approx 1/E_{\mathrm{BW}}$ to treat such large bandwidths is insufficient to allow a small number of discretizations (i.e., number of quadrature points) to represent the time integrals, accurately.  Hence, an efficient approach capable of taming the multiple time scale challenge  presented by the large dynamic range in the integrand, $\tilde{\chi}(\zeta\tau;0)_{r,r'}$, of $\chi(0)_{r,r'}$ will be given.  Once such an approach has been developed for gapped systems, the solution will be generalized to treat gapless systems and response functions at finite frequencies through use of imaginary $\zeta$ and non-trivial $h(\tau;\zeta)$.

\begin{figure}
    \centering
    \includegraphics[width=2.8in]{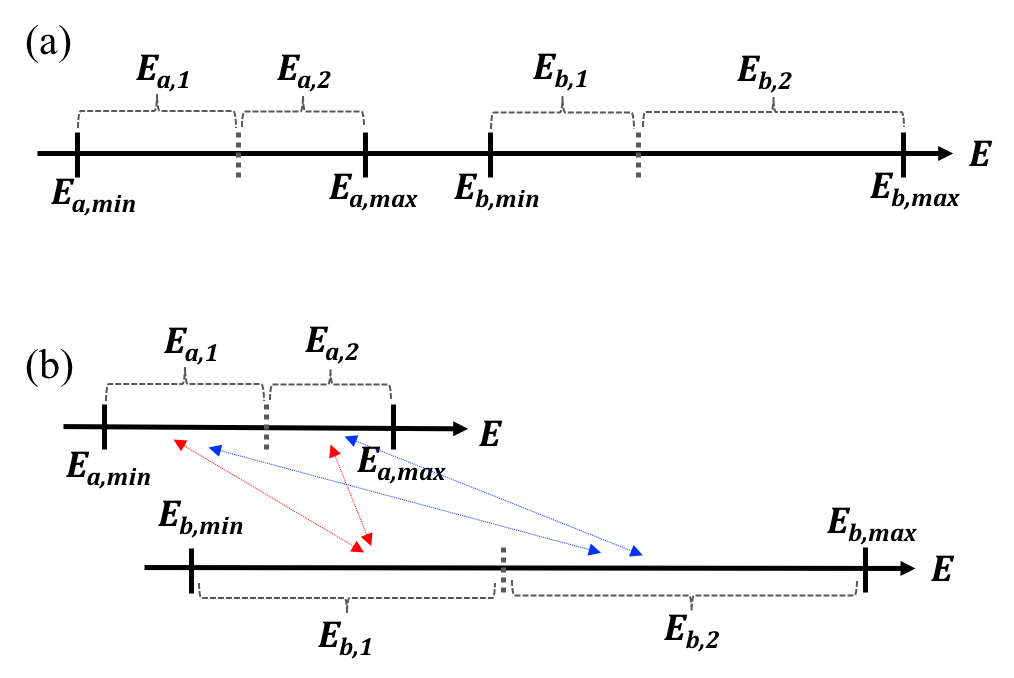}
    \caption{An example of windowing with $N_{aw}=N_{bw}=2$ (a) for a gapped system: the range of energies in $a_i$ and $b_j$ do not overlap and (b) for a system with an overlapping energy range. The energy window pairs with energy crossing are indicated with red arrows, and window pairs without crossing are shown with blue arrows.}
    \label{fig:2x2w}
\end{figure}

In order to tame the multiple time scales inherent in the present time domain approach to $\chi(0)_{r,r'}$, the propagators $\rho^{A,B}$ must be modified. Borrowing ideas from Feynman's path integral approach, the propagators are ``shredded''  (sliced into pieces) here in energy space.  That is, the energy range spanned by $a_i$ is partitioned into $N_{aw}$ contiguous energy windows indexed by $l=1,\ldots,N_{aw}$ and $b_j$ is similarly partitioned into $N_{bw}$ windows indexed by $m=1,\ldots,N_{bw}$;  to illustrate this shredding, 
a 2$\times$2 energy window decomposition for a gapped system is shown in Fig.~\ref{fig:2x2w}(a).
Shredding the propagators allows $\tilde{\chi}(\tau\zeta;0)_{r,r'}$ to be recast {\it exactly} as a sum over window pairs $(l,m)$,
\begin{equation}
\chi(0)_{r,r'} = \sum_{l=1}^{N_{aw}}\sum_{m=1}^{N_{bw}}
\int_0^\infty \!\!\! d\tau \, h(\tau;\zeta_{lm}) \tilde{\chi}^{lm}(\zeta_{lm}\tau;0)_{r,r'} ,
\label{eq:chishredlm}
\end{equation}
where for each window pair $(l,m)$,
\begin{eqnarray}
\tilde{\chi}^{lm}(\zeta_{lm}\tau;0)_{r,r'} &=& \rho^{Alm}(\zeta_{lm}\tau)_{r,r'}\rho^{Blm}(\zeta_{lm}\tau)_{r,r'} \nonumber \\
\rho^{Alm}(\zeta_{lm}\tau)_{r,r'} &=& \sum_{i\in l} A^i_{r,r'} e^{-\zeta_{lm} (a_i-E^{\mathrm{off}})\tau} \nonumber \\
\rho^{Blm}(\zeta_{lm}\tau)_{r,r'} &=& \sum_{j\in m} B^j_{r,r'} e^{-\zeta_{lm} (E^{\mathrm{off}}-b_j)\tau}  \ .
\label{eq:chirhoArhoBlm}
\end{eqnarray}
Here, the shorthand $i\in l$ or $j\in m$ represents a sum over only the states or modes inside each energy window. The energy windows need not be equally spaced in energy; in fact, the optimal choice of windows is not equally spaced even for a uniform density of states or modes as shown in Sec.~\ref{sec:n3staticP}. 

The shredded form of $\chi(0)_{r,r'}$ given in Eq.~(\ref{eq:chishredlm}) has computational complexity of ${\cal O}(N^3)$ because the operation count to evaluate it, is simply
\begin{equation}
\label{eq:opcnt}
 N_r^2\sum_{lm} (L_{la}+L_{mb})N_{lm}^{(\tau,h)} \sim {\cal O}(N^3) \ .
\end{equation}
Here the $L_{la},L_{mb}\sim {\cal O}(N)$ are the number of states or modes in the $l$th and $m$th energy windows, respectively, and  $N_{lm}^{(\tau,h)}\sim {\cal O}(N^0)$ is the number of quadrature points, which depends on $h(\tau;\zeta_{lm})$, required for accurate integration in a specific window pair $(l,m)$  (see   Sec.~\ref{sec:n3staticP} for details).

The shredded propagator formulation of $\chi(0)_{r,r'}$ has four important advantages. First, every term in the double sum over window pairs $(l,m)$ has its own intrinsic bandwidth which is controlled by its own $\zeta_{lm}$ while yet preserving the desired separability. Second each window pair can be assigned its own quadrature optimized to treat its limited dynamic range. Third, the windows can be selected to minimize the dynamic range in the window pairs which allows for small $N_{lm}^{(\tau,h)}$ (i.e., efficient quadrature) in all pairs. These first three advantages are sufficient to tame the multiple time scale challenge. Fourth, finite frequency expressions for gapped systems as well as gapless systems at finite temperature can be addressed utilizing simple extensions of Eq.~(\ref{eq:chishredlm}) as demonstrated below.


The next theoretical issue to tackle is to show that the optimal windows can be found in ${\cal O}(N^3)$ or less computational effort given the input energies $a_i$ and $b_j$. Since the computational intensive part of $\chi(0)_{r,r'}$ involves its $r,r'$ spatial dependence, it is best to choose an optimal windowing scheme in the limit $A^i_{r,r'},B^j_{r,r'}\rightarrow 1$ as, within a limited energy range of a window pair, the spatial dependence of the $A^i$ or $B^j$ are to good approximation roughly similar. If the density of states for $a_i$ and $b_j$ is taken to be locally flat, then the optimal number and placement of windows can be determined in ${\cal O}(N^0)$; if the actual density of states is taken into account, the scaling remains ${\cal O}(N^0)$ as the density of states is an input from the electronic structure computation (typically, KS-DFT). Here, optimal indicates the windows are selected to minimize the operation count, Eq.~(\ref{eq:opcnt}), required to 
compute Eq.~(\ref{eq:chishredlm}) over the number and placement (in energy space) of the windows. In practice, as given in Sec.~\ref{sec:n3staticP}, we take $N_{lm}^{(\tau,h)}$ to be the number of quadrature points required to guarantee a prespecified error tolerance on time integral of each window pair when $A^i_{r,r'},B^j_{r,r'}\rightarrow 1$; again, each window pair, $lm$, has its own tuned quadrature and energy scale, $\zeta_{lm}$.

The control given by the energy windowed formulation of $\chi(0)_{r,r'}$ in Eq.~(\ref{eq:chishredlm}) is the key to extending our efficient ${\cal O}(N^3)$ method to gapless systems and to finite frequencies. For gapless systems at zero frequency, there will be some few energy windows pairs (mostly likely one) where $a_i=b_j$ happens at least once. This is not problematic because, e.g., for the case of computing the polarizability matrix of Eqs.~(\ref{eq:Pomegadef},\ref{eq:Pstatic}), the occupancy difference $f(E_v)-f(E_c)$ regularizes the singularity of the denominator via L'H\^opital's rule applied to $[f(E_v)-f(E_c)]/(E_c-E_v)$. Adding the occupancy factors presents no difficulties: all that is required is to take the difference between two terms of the same form as Eq.~(\ref{eq:Xdoublesum}) in the problematic overlapping window pair(s) --- a small added expense (see Sec.~\ref{sec:metals} for details).  However, a more general approach that can handle finite frequencies, described next, can also be adopted to handle gapless systems.

For the case of finite frequency $\omega\ne0$, in some window pair(s) $e_{ij}=\omega+a_i-b_j$ can change sign. In standard GW implementations, these potential singularities are tamed by either dropping their contributions to the sum when $|e_{ij}|$ is small\cite{HL} or by regularizing the fraction $1/e_{ij}$ through the introduction of a Lorentzian factor that approaches $1/e_{ij}$ asymptotically for large $e_{ij}$~\cite{bgw}, i.e., replacing $1/e_{ij}$ by $e_{ij}/(e_{ij}^2+|\zeta|^{-2})$ which implicitly invokes the broadening caused by complex scattering mechanisms present in real materials. Lorentzian regularization can easily be accommodated within our time domain formalism simply by selecting $h(\tau;\zeta)=|\zeta|\exp(-\tau)$ for the weight function in Eq.~(\ref{eq:genFourLap}) and choosing $\zeta$ to be a pure imaginary number for the small number of window pairs where $e_{ij}$ changes sign.  In more detail, for  the Lorentzian regularization choice in the literature, 
\begin{eqnarray}
\frac{e_{ij}}{e_{ij}^2+|\zeta|^{-2}} & = & Im \left[\int_0^\infty d\tau\, |\zeta|e^{-\tau}\,e^{i|\zeta|e_{ij}\tau}\right]\nonumber\\
& = & |\zeta|\int_0^\infty d\tau\,e^{-\tau}\Big[
\sin(|\zeta|(\omega-b_j))\cos(|\zeta|a_i) \nonumber\\
& &  - \cos(|\zeta|(\omega-b_j))\sin(|\zeta|a_i) \Big]
\end{eqnarray}
where we have chosen to separate the energy $e_{ij}=\omega+a_i-b_j$ into the sum of $\omega-b_j$ and $a_i$ prior to factorizing the exponential; splitting into $\omega+a_i$ and $-b_j$ is also possible if desired. However, a large number of quadrature points must be taken to accurately discretize the integral.

As discussed in more detail in Sec.~\ref{sec:n3sigma}, the alternative weight function
\begin{eqnarray}
h(\tau;\zeta)&=&|\zeta|\exp(-\tau-\tau^2/2) \nonumber
\end{eqnarray}
and its transform
\begin{eqnarray}
F(e_{ij};\zeta) &=& |\zeta| Im\left \{\sqrt{\frac{\pi}{2}}\exp\left(-\frac{(e_{ij}|\zeta|+i)^2}{2}\right)\times\right. \nonumber \\
& & \left. \left [ 1 + i\,\mathrm{erfi}\left (\frac{e_{ij}|\zeta|+i}{\sqrt{2}}\right)\right ]\right \} \,,
\end{eqnarray}
is preferable since this transform $F$ still approaches $1/e_{ij}$ at large $e_{ij}$ but generates much higher quadrature accuracy compared to the exponential weight $|\zeta|\exp(-\tau)$ (for the same number of quadrature points) and is well behaved for all $e_{ij}$.  The benefits of the alternative weight function, an asymptotic analysis, and the associated rapidly convergent quadrature are detailed in Sec.~\ref{sec:newquad} and associated appendices.

Last we note that the new formalism can handle problematic regions of the density of states, such as van Hove singularities, by simply assigning them their own window in a Lebesgue type approach. As long as the {\it  number} of problematic regions is independent of systems size, the scaling of the method remains ${\cal O}(N^3)$. 

\begin{figure}
\includegraphics[width=2.5in]{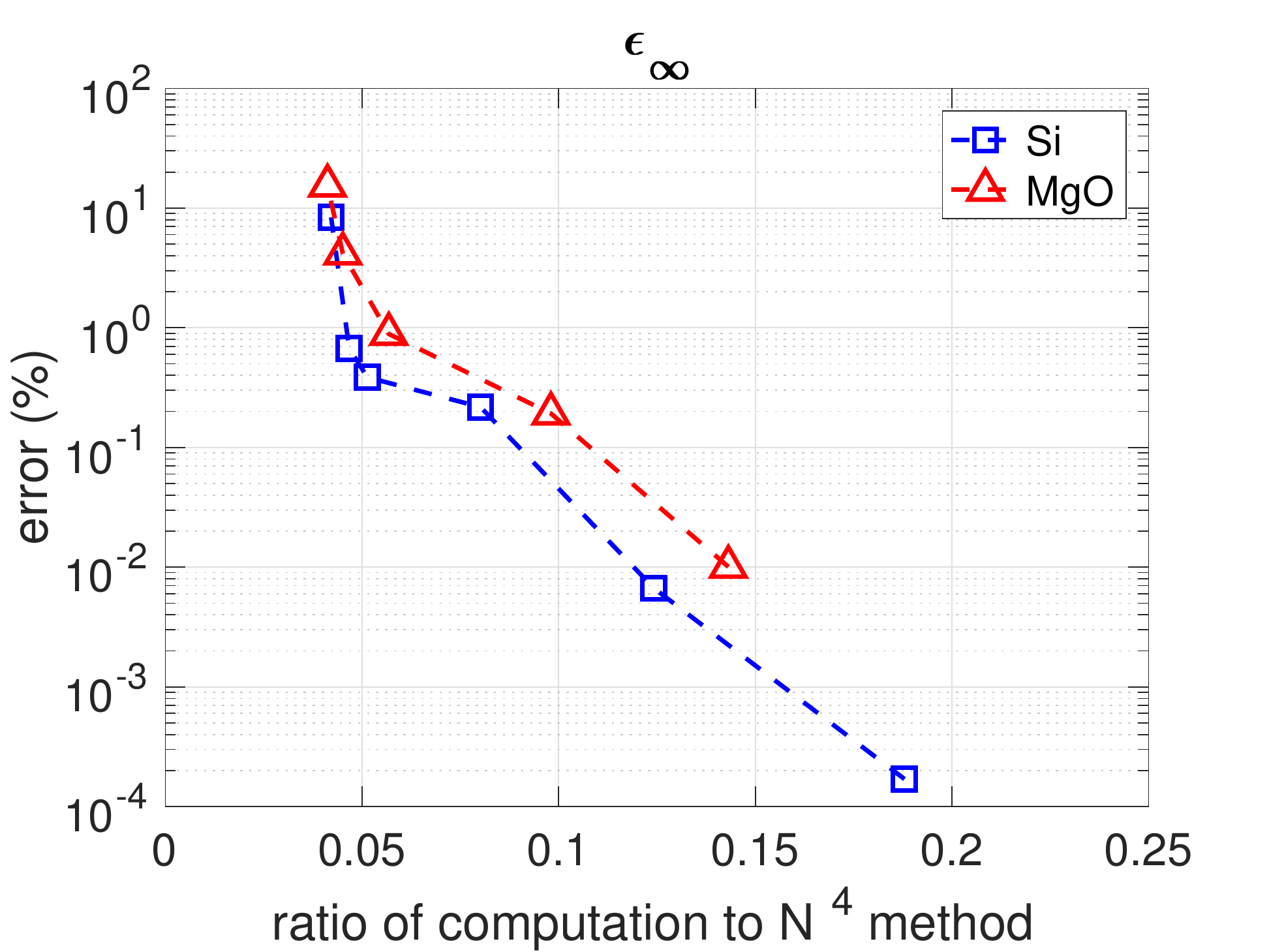}
\includegraphics[width=2.5in]{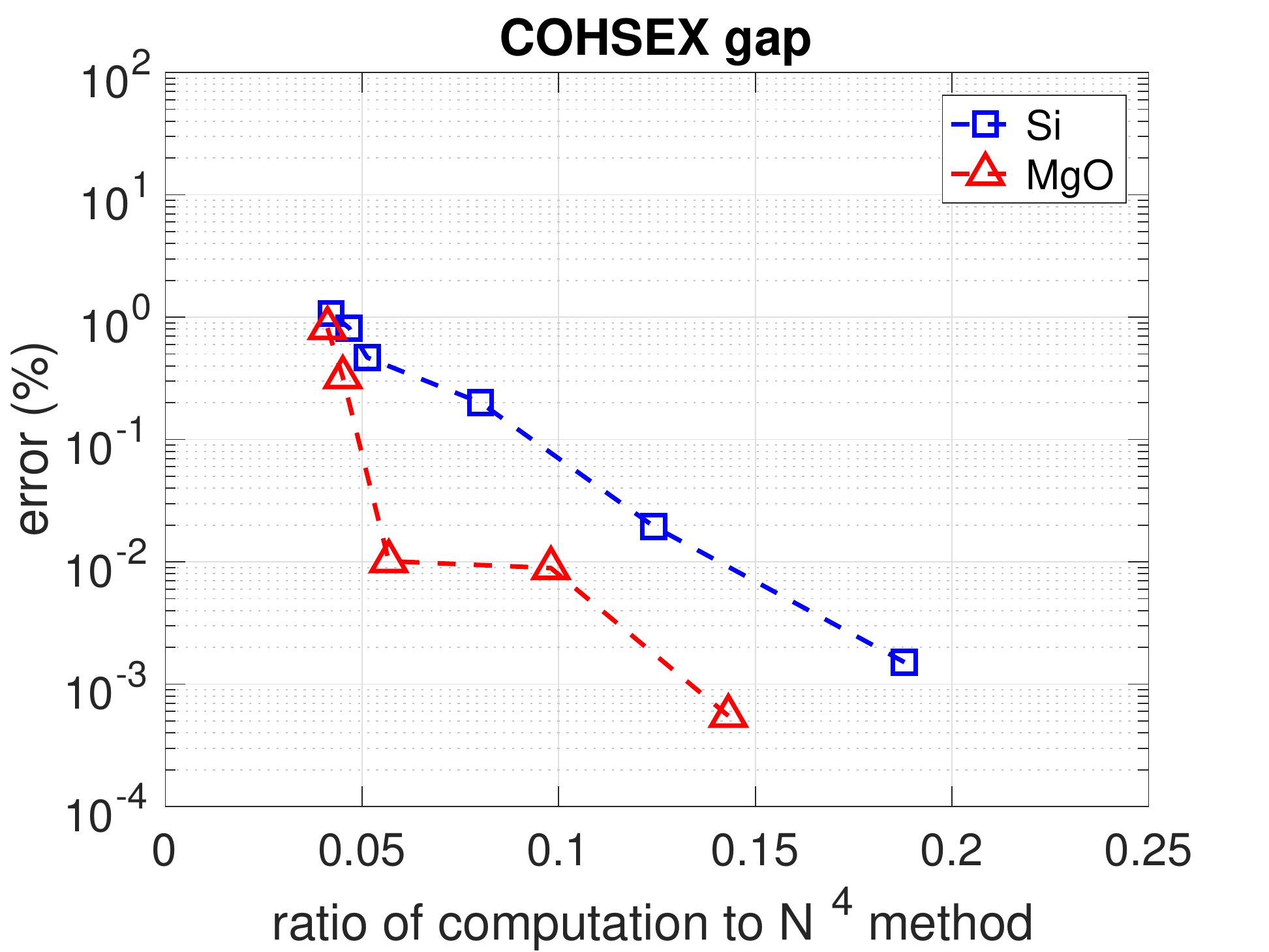}
\caption{Numerical error versus computational savings for our cubic scaling formalism, CTSP-W, compared to the standard quartic GW formulation for bulk Si and MgO modeled by 16 atom unit cells.
Top: error in the macroscopic dielectric constant 
$\epsilon_\infty^{\rm (MgO)}=6.35$, $\epsilon_\infty^{\rm (Si)}=64.85$.  
Bottom: error in the COHSEX band gap.
 $E_g^{\rm (COHSEX,MgO)}=7.56eV$, $E_g^{\rm(COHSEX,Si)}=1.92eV$. }
\label{fig:teaserfigure}
\end{figure}

In order to convince the reader that the new formalism represents an important improvement, we provide a comparison of our ${\cal O}(N^3)$ time domain results to those of the corresponding ${\cal O}(N^4)$ frequency domain computation in Fig.~\ref{fig:teaserfigure} for two standard test systems, crystalline silicon and magnesium oxide. In the figure, the new method is referred to via the sobriquet, complex time shredded propagator (CTSP) method where CTSP-W indicates the use of optimal windowing and in the discussion to follow, CTSP-1 the use of 1 window. Even for small unit cells, the ${\cal O}(N^3)$ computation outlined above delivers a significant reduction in computational effort compared to the standard approach. Numerical details are now presented in Secs.~\ref{sec:n3staticP}-\ref{sec:dynP} and associated appendices where we show that the new method's parameters can be reduced to one, the fractional quadrature error $\epsilon^{(q)}$, which allows for the easily tunable convergence demonstrated in the figures.

\subsection{Static polarization matrix in ${\cal O}(N^3)$ for gapped systems}
\label{sec:n3staticP}
The static polarizability matrix defined in  Eq.~(\ref{eq:Pstatic}) reduces, for systems with large
energy gaps compared to $k_BT$, to
\[
P_{r,r'} = -2\sum_{v=1}^{N_v} \sum_{c=1}^{N_c} \frac{\psi_{r,v}^*\psi_{r,c}\psi_{r',c}^*\psi_{r',v}}{E_c-E_v} 
\]
as the occupation number functions for this special case are $0/1$; these will be reintroduced to
treat zero gap systems in Sec.~\ref{sec:metals}. 
Here, $N_v$ and $N_c$ are the number of valence and conduction states respectively.  Non-essential indices / quantum numbers such as spin $\sigma$ and Bloch $k$-vectors have been suppressed.

\subsubsection{Laplace identity and shredded propagators}
Employing the energy windowing approach of Eqs.~(\ref{eq:chishredlm},\ref{eq:chirhoArhoBlm}),  the energy range of the valence / conduction band is divided into $N_{vw}/N_{cw}$ partitions with the valence / conduction partition $l/m$ ranging from $E_{vl/cm}^{min}$ to $E_{vl/cm}^{max}$. Thus, the static polarizability can be written as 
\begin{equation}
P_{r,r'} = \sum_l^{N_{vw}} \sum_m^{N_{cw}} P^{lm}_{r,r'}
\label{eq:Pwindowed}
\end{equation}
where each window pair $(l,m)$ contributes 
\begin{equation*}
P^{lm}_{r,r'} = -2\zeta_{lm} \int_{0}^\infty \!\!\!d\tau\, e^{-\zeta_{lm}E_{g}^{lm}\tau } \bar\rho^{m}(\zeta_{lm}\tau)_{r,r'}\rho^{l}(\zeta_{lm}\tau)_{r',r}.
\end{equation*}
via the Laplace identity where the choice $h=\zeta$ generates exactly the desired energy denominator $1/(E_c-E_v)$ (i.e., $F(x;\zeta)=1/x$ in Eq.~(\ref{eq:XdoublesumwithF})). Each window pair $(l,m)$ has its own energy gap $E_g^{lm}=E_{cm}^{min}-E_{vl}^{max}$, energy scale $\zeta_{lm}$ and bandwidth 
$E_{BW}^{lm}=E_{cm}^{max}-E_{vl}^{min}$.
The imaginary time density matrices for the windows are given by
\begin{eqnarray}
\bar\rho^m(\tau)_{r,r'} & = & \sum_{c\in m} e^{-\tau\Delta E_{cm}}\psi_{r,c}\psi_{r',c}^*\,,
\label{eq:rhorhobarlm1}\\
\rho^l(\tau)_{r,r'} & = & \sum_{v
\in l} e^{-\tau\Delta E_{vl}}\psi_{r,v}^*\psi_{r',v}\,.
\label{eq:rhorhobarlm2}
\end{eqnarray}
Here, $\Delta E_{vl}=E_{vl}^{max}-E_v$ and $\Delta E_{cm}=E_c-E_{cm}^{min}$ are defined with respect to the edges of each energy window.  A good choice of windows can significantly reduce the dynamic range, the band width to band gap ratio $E^{lm}_{BW}/E^{lm}_{g}$, for all window pairs such that a coarse quadrature grid can be employed in all window pairs allowing for an efficient numerical method as described next.

\subsubsection{Discrete approximation to the time integral}

The continuous imaginary time integral of Eq.~(\ref{eq:Pwindowed}) must be discretized in an efficient and error-controlled manner to form an effective numerical method.  The natural choice of is Gauss-Laguerre (GL) quadrature
\begin{equation}
\int_0^\infty d\tau\ e^{-\tau}s(\tau) \approx \sum_{k=1}^{N^{(\tau,GL)}} w_k\,s(\tau_k)
\label{eq:GLquad}
\end{equation}
where $N^{(\tau,GL)}$ is the number of quadrature points, $s(\tau)$ is the funciton to be integrated 
over the exponential function, $\exp(-\tau)$, and
$\{w_k\}$ and $\{\tau_k\}$ are the weights and nodes~\cite{abramowitz_handbook_1972} whose $N^{(\tau,GL)}$ dependence has been suppressed for clarity. Inserting the discete approximation,
the contribution from each window pair $(l,m)$ to Eq.~(\ref{eq:Pwindowed}) is
\begin{multline}
P^{lm}_{r,r'} = -2\zeta_{lm}\sum_{k=1}^{N^{(\tau,GL)}_{lm}} w_k e^{-\tau_k(\zeta_{lm}E_{g}^{lm}-1)} \\ \times\ \ 
\bar\rho^{m}(\zeta_{lm}\tau_k)_{r,r'}\rho^{l}(\zeta_{lm}\tau_k)_{r',r}\,.
\label{eq:GLquadPlm}
\end{multline}

\paragraph{Optimal energy scale, $\zeta_{lm}$:} 
\label{choosinga}

The energy scale $\zeta_{lm}$ is selected to minimize the error of all integrals in a window pair. The
geometric mean, $\zeta_{lm}^{-1}\approx\sqrt{E_{BW}^{lm}E_{g}^{lm}}$, is  close to the optimal choice as  described in Appendix~\ref{app:optima}.

\paragraph{Estimating the number of quadrature points:}
For any set of interband transition energies \{$E_{cm}-E_{vl}$\}, the largest quadrature errors occur at the largest interband transition energy ($E_{BW}^{lm}$) and the smallest interband transition energy ($E_g^{lm}$). Taking $\zeta_{lm}^{-1}=\sqrt{E_{BW}^{lm}E_g^{lm}}$, Appendix~\ref{app:ngl} shows that the number of quadrature points required to generate fractional error level, $\epsilon^{(q)}$, scales as $N^{(\tau,GL)}_{lm}\sim\sqrt{E_{BW}^{lm}/E_g^{lm}}=\alpha_{lm}$ with proportionality constant of order unity, such that
\begin{equation}
\label{eq:Ngtext}
N^{(\tau,GL)}(\alpha;\epsilon^{(q)})=\alpha(0.4-0.3\ln{\epsilon^{(q)}})
\end{equation}
is a good approximation.  Since $\alpha \geq 1$ and $N^{(\tau,GL)}(\alpha;\epsilon^{(q)})\geq 1$ the valid error range is $0<\epsilon^{(q)}<0.125$; in order to treat larger errors, $0<\epsilon^{(q)}<0.7$, $N^{(\tau,GL)}(\alpha;\epsilon^{(q)})\rightarrow N^{(\tau,GL)}(\alpha;\epsilon^{(q)})+0.5$ which has little effect on the predictions of the formula but does ensure physical results.

\subsubsection{Optimal windowing}
Given the number of points required to generate fractional quadrature error, $\epsilon^{(q)}$, in a given window pair can be neatly determined, next, consider the construction of the optimal set of windows. This can be accomplished via minimization of the cost to compute the static polarizability over the window positions, 
the sets $\{E_{v,min},E_{v,max}\}$ and $\{E_{c,min},E_{v,max}\}$, and the number of windows, $N_{vw},N_{cw}$,
\begin{multline}
\label{eq:Ccost}
C^{(GL)}(\epsilon^{(q)})=
\sum_l^{N_{vw}}\sum_m^{N_{cw}} N^{(\tau,GL)}(\alpha_{lm};\epsilon^{(q)})\ \times \\\left(\int_{E_{v,min}^l}^{E_{v,max}^l}D(E)dE+\int_{E_{c,min}^m}^{E_{c,max}^m}D(E)dE\right)\ .
\end{multline}
which for clarity are omitted from the dependencies of $C^{(GL)}(\epsilon^{(q)})$.
Here $\alpha_{lm}$ is square root of the window dynamic range,
$N^{(\tau,GL)}_{lm}(\alpha;\epsilon^{(q)})$ is given in Eq.~(\ref{eq:Ngtext}) and $D(E)$ is the density of states which can be assigned band indices, if desired, when considering k-point sampling.  The integrals over the density of states, $D(E)$, are simply the number/fraction of states in the appropriate energy window. For a density of states with problematic points, we simple assign windows to those regions {\it a priori} (fixed position in energy space) allowing for fast minimization over the smooth parts of $D(E)$. For example, if there is a singularity in $D(E)$ in the range $E_{singular}\pm \Delta E/2$, a window is simply pinned or fixed at this position in energy space, allowing the minimization to proceed over slowly varying regions of the DOS integral, in a Lebesque inspired approach.

The cost estimator, Eq.~(\ref{eq:Ccost}), can be minimized straightforwardly, as detailed in Appendix~\ref{app:winmin}, once at the start of a GW calculation.  The computational complexity of the minimization procedure is negligible ${\cal O}(N^0)$ compared to both the ${\cal O}(N^3)$ computational complexity of both $P$ and the input band structure. We note that for the form of $N^{(\tau,GL)}(\alpha;\epsilon^{(q)})$ in Eq.~(\ref{eq:Ngtext}), the optimal windowing, both the number of windows and their positions in energy, is independent of error level as $N^{(\tau,GL)}(\alpha;\epsilon^{(q)})=s(\alpha)u(\epsilon^{(q)})$ is separable. Importantly, all parameters of the method are completely determined by the usual set, input from band structure and a choice of energy cutoff in the conduction band, and \textit{one} new parameter, $\epsilon^{(q)}$, the fractional quadrature error required to accurately transform from the time domain to the frequency domain. The quadrature error will be connected to the error in physical quantities in Sec.~\ref{sec:results}.

\subsection{Static $P$ for gapless systems}
\label{sec:metals}

The standard approach employed to treat gapless systems is to introduce a smoothed  step function $f(E;\mu,\beta)$ for the electron occupation numbers as a function of energy, $E$, centered on the chemical potential, $\mu$ (Fermi level) with ``smoothing" parameter or inverse temperature $\beta$~\cite{fu_first-principles_1983,needs_total-energy_1986,gillan_calculation_1989}.
Examples include the Fermi-Dirac distribution of the grand canonical ensemble
\[
f(E)=\frac{1}{1+\exp[\beta(E-\mu)]}
\]
where formally, $\beta=1/k_BT$, or the more rapidly (numerically) convergent and hence convenient 
\[
f(E) = \frac{\beta\int_E^\infty dx \,\exp(-[\beta(x-\mu)]^2/2)}{\sqrt{2\pi}} = \frac{\mathrm{erfc}(\beta(E-\mu))}{2} \ .
\]
Typical literature values of  $\beta$ correspond to temperature above ambient conditions (e.g., $\beta^{-1}=0.1$ eV $\approx 1000$ K).  The static RPA irreducible polarizability matrix including the occupation functions is given in Eq.~(\ref{eq:Pstatic}).

To proceed, note that the energy-dependent part of the sum,
\begin{equation}
J_{cv} = \frac{f(E_v)-f(E_c)}{E_c-E_v}\,,
\label{eq:Jcvgapless}
\end{equation}
is smooth for all energies and has the finite value $-f'(\mu)$ as $E_v,E_c\rightarrow \mu$ (note, 
$E_c \geq E_v\ \forall\ c,v$).  Hence, for a calculation with a small but finite gap, the terms in the sum for $P$ are finite and well behaved such that windowing plus quadrature approach will work well.  As before, we split $P$ into a sum over window pairs

with the contributions from each window pair now given by 
\begin{multline*}
P^{lm}_{r,r'} = -2\zeta_{lm} \sum_{k=1}^{N_{GL}^{lm}} w_k\, e^{-\tau_k(\zeta_{lm}E_{g}^{lm}-1)}\times \\
\Big\{ D^{lm}_{r,r'}E^{lm}_{r,r'}-F^{lm}_{r,r'}G^{lm}_{r,r'}\Big\}
\end{multline*}
where
\[
D^{lm}_{r,r'} = \sum_{v\in l} f(E_v)
e^{-\tau_k \zeta_{lm}\Delta E_{vl}}\psi_{r,v}^*\psi_{r',v}
\]
\[
E^{lm}_{r,r'} = \sum_{c\in m} e^{-\tau_k \zeta_{lm}\Delta E_{cm}}\psi_{r,c}\psi_{r',c}^*
\]
\[
F^{lm}_{r,r'} = \sum_{v\in l} 
e^{-\tau_k \zeta_{lm}\Delta E_{vl}}\psi_{r,v}^*\psi_{r',v}
\]
\[
G^{lm}_{r,r'} = \sum_{c\in m} f(E_c)e^{-\tau_k \zeta_{lm}\Delta E_{cm}}\psi_{r,c}\psi_{r',c}^*\,.
\]
The matrices $D,E,F,G$ can be computed with $O(N_vN_r^2)$ or $O(N_cN_r^2)$ operations (i.e., cubic scaling) where $N_r$ is the number of $r$ grid points.
Since $f(E_c)$ becomes small as a function of increasing $E_c$, the $FG$ term need only be computed for the few window pairs where $\beta(E_c-\mu)$ is sufficiently small.  Hence, the additional work required to treat gapless systems is, in fact, modest.

Direct application of the cost-optimal energy windowing method for gapped systems in Sec.~\ref{sec:n3staticP} to a finite gapless material is feasible since, generically, a finite system without special symmetry will have a small but finite energy gap $E_{gap}={\rm min}(E_c-E_v)>0$.  However, direct application of the procedure generates infinite quadrature grids in situations where the gap is exactly zero due to degeneracy at the Fermi energy.  The solution is straightforward: the key quantity that is to be represented by quadrature is $J_{cv}$ of Eq.~\ref{eq:Jcvgapless}.  For $E_c-E_v\rightarrow0$, $J_{cv}\rightarrow-f'(\mu)$ where $-f'(\mu)=\beta/4$ for the Fermi-Dirac distribution and $\beta/\sqrt{2\pi}$ for the erfc form above.  Thus, the system has an effective gap of $\sim\beta^{-1}$.  For energy window pairs $(l,m)$ that contain degenerate states at the Fermi energy, we manually set their gap to $E_{g}^{lm}=1/\beta$ via "scissors" (shifting the conduction band up by $1/(2\beta)$ and valence bound down by $1/(2\beta)$ in the offending window pair and then applying the method of Sec.~\ref{sec:n3staticP}. Alternatively, the method of the next subsection can be generalized.

\subsection{$\Sigma(\omega)$ in cubic computational complexity}
Given the poles of the screened interaction $W(\omega)_{r,r'}$ are at $\omega_p$ with residues $B_{r,r'}^p$, the dynamic (frequency-dependent) part of the GW self-energy can be expressed as
\begin{multline}
\Sigma(\omega)_{r,r'}
= \sum_{p,v}\frac{B^p_{r,r'}\psi_{rv}\psi_{r'v}^*}{\omega-E_v+\omega_p} + \sum_{p,c}\frac{B^p_{r,r'}\psi_{rc}\psi_{r'c}^*}{\omega-E_c-\omega_p}.
\label{eq:sigmatocalc}
\end{multline}
The task is to develop an energy window-plus-quadrature technique to generate a cubic scaling method that delivers  $\Sigma(\omega)$ directly 
\footnote{To avoid excessive memory use, one can compute the large matrix $\Sigma(\omega)_{r,r'}$ for a fixed $\omega$ and then compute and only store the much smaller number of desired matrix elements $<n| \Sigma(\omega) |n'>$  before moving
to the next $\omega$ value.} 
for real frequencies $\omega$, such that analytical continuation is not required. 

\label{sec:n3sigma}
\subsubsection{Windowing}
To proceed, two sets of windows are created for the energies in the denominator, $\{e_n=\omega-E_n\}$ and $\{\omega_p\}$. Next, $\Sigma(\omega)$ is expressed as a sum over window pairs, where each window pair has its own quadrature as above
\begin{equation}
\Sigma(\omega)_{r,r'} = \sum_{l}^{N_{pw}}\sum_m^{N_{e w}} \Sigma^{lm}(\omega)_{r,r'}
\label{eq:sigmawin}
\end{equation}
Here energy window $l$ contains the excitation energies $\Omega_{l}^{min}\le \pm\omega_p < \Omega_{l}^{max}$ and  energy window $m$ contains the band energies satisfying $e_m^{min}\le e_n=\omega-E_n<e_m^{max}$. The contribution from window pair $(l,m)$ is
\[
\Sigma^{lm}(\omega)_{r,r'} = \sum_{p\in l}\sum_{n\in m}\frac{B^p_{r,r'}\psi_{rn}\psi_{r'n}^*}{\omega-E_n\pm\omega_p}
\]
The notation $\pm\omega_p$ permits us to treat of both signs of $\omega_p$ in Eq.~(\ref{eq:sigmatocalc}) in a unified manner. 
 
Analyzing the result, we note almost all the window pairs $(l,m)$ in Eq.~(\ref{eq:sigmawin}) can use the Laplace with GL quadrature scheme of Sec.~\ref{sec:n3staticP} because the denominator $x=\omega-E_n\pm\omega_p$ is finite and does not change sign.  The difficulty is that, for some windows, the denominator $x$ changes sign inside the energy windows such that the Laplace integral (Eq.~(\ref{eq:genFourLap})) does not apply. Thus, a scheme to treat window pairs with energy crossings is required.
 
\subsubsection{Specialized quadrature for energy crossings}
\label{sec:newquad}

The quantity of interest is
\begin{equation}
\Sigma^{lm}(\omega)_{r,r'} = 
\sum_{p\in l}\sum_{n\in m}B^p_{r,r'}\psi_{rn}\psi_{r'n}^*\, F(\omega-E_n\pm\omega_p;\zeta)
\label{eq:sigmaOmega}
\end{equation}
where $x=\omega-E_n\pm\omega_p$ changes sign as the sum over $p$ and $n$ is performed.  As discussed in Sec. \ref{sec:formalism}, standard choices in the GW literature are to either these zero contributions for small $x$ (i.e., set $F(x;\zeta)=0$ for small $x$) or to use a Lorentzian smoothing function 
\[
\frac{x}{x^2+\gamma^{-2}} = Im \int_0^\infty d\tau\, \gamma\, e^{-\tau}e^{i\tau \gamma x}
\]
Note, $\gamma>0$ corresponds to the weight function $h(\tau;\zeta)=|\zeta|e^{-\tau}$ in Eq.~(\ref{eq:Fdef}) with $\zeta=-i\gamma$. Below we shall eschew $\zeta$ and work in terms of $\gamma$ which is more natural.

As detailed in Appendix~\ref{app:weight} and above, a better choice of the weight function and resulting transform are
\begin{eqnarray}
h(\tau;\gamma) &=& \gamma \exp(- \tau -\tau^2/2) \\
F(x;\gamma) &=& \gamma Im\left \{\sqrt{\frac{\pi}{2}}e^{-\frac{(x\gamma+i)^2}{2}}
\left [ 1 + i\mathrm{erfi}\left (\frac{x\gamma+i}{\sqrt{2}}\right)\right ]\right \} \nonumber \,.
\end{eqnarray}
The new weight generates smaller quadrature errors than the exponential choice for the same sized quadrature grid, has a transform that both approaches $1/x$ faster than a Lorentzian in the large $x$ limit, and is regular for all $x$.  A Gaussian-type quadrature for 
the new weight function can be generated following the standard procedure\cite{orthpolys} to create a set of nodes $\{\tau_j\}$ and weights $\{w_j\}$ for a given quadrature grid size $N^{(\tau,HGL)}$ (see Appendix \ref{app:matlabcode}). The superscript $HGL$ denotes Hermite-Gauss-Laguerre quadrature since the weight function has both linear and quadratic terms in the exponent.  
Inserting the result, the discrete approximation becomes
\begin{equation}
F(x) \approx \gamma Im \!\!\!\sum_{j=1}^{N^{(\tau,HGL)}} \!\!\! w_j\,e^{i\tau_jx\gamma}
= \gamma \!\!\! \sum_{j=1}^{N^{(\tau,HGL)}} \!\!\! w_j\,\sin(\tau_jx\gamma)\,.
\label{eq:quadFx}
\end{equation}

Finally, the use of the quadrature of Eq.~(\ref{eq:quadFx}) leads to the desired separable form for the window pairs $(l,m)$ with an energy crossing
\begin{multline}
\Sigma^{lm}(\omega)_{r,r'} = \gamma\sum_{j=1}^{N^{(\tau,HGL)}} w_j \Big\{\\
\left[\sum_{p\in l} B^p_{r,r'}
\sin(\pm \tau_j\omega_p\gamma)\right]\times\\
\left[\sum_{n\in m} \psi_{rn}\psi_{r'n}^* \cos(\tau_j (\omega-\epsilon_n)\gamma)\right]\\
+ \left[\sum_{p\in l} B^p_{r,r'}
\cos(\pm \tau_j\omega_p\gamma)\right]\times\\
\left[\sum_{n\in m} \psi_{rn}\psi_{r'n}^* \sin(\tau_j(\omega-\epsilon_n)\gamma)\right]
\Big\}\,.
\label{eq:finalsigmalmexpr}
\end{multline}
One value of broadening parameter, $\gamma \sim 1/k_BT$, is selected for all windows with energy crossings. The number of grid points will vary depending on the bandwidth in the window pair scaled by $\gamma$ and the desired fractional error.

\subsubsection{Quadrature points for specified error level}
For window pairs without an energy crossing, $\omega-E_n\pm \omega_p$ does not change sign, and the GL quadrature previously analyzed is utilized.  For window pairs with energy crossings $HGL$ quadrature is required. Appendix~\ref{app:HGLgrid} details the construction of $N^{(\tau,HGL)}(x;\epsilon^{(q)})$ at a fixed fractional quadrature error level,
\begin{equation}
N^{(\tau,HGL)}(x;\epsilon^{(q)}) =
c_2(\epsilon^{(q)})x^2+c_1(\epsilon^{(q)})x+c_0(\epsilon^{(q)})
\end{equation}
where $x=E_{\rm max}-E_{\rm min}$ is the bandwidth of the window pair with energy crossings, and $c_2$, $c_1$, and $c_0$ are low order polynomial functions of $\ln\epsilon^{(q)}$. The values of the coefficients are given in Appendix~\ref{app:HGLgrid}. 

\subsubsection{Optimal window choice}

Next consider the computational cost to compute $\Sigma(\omega)$ for window pairs with an energy crossing,
\begin{multline*}
C^{lm,HGL}(\epsilon^{(q)}) = 2 N^{(\tau,HGL)}_{lm}(x_{lm};\epsilon^{(q)}) \times \\
\left( \int_{\omega_{p,min}^{m}}^{\omega_{p,max}^{m}} D_p(\omega)d\omega + \int_{E_{n,min}^{l}}^{E_{n,max}^{l}} D(E)dE \right)
\end{multline*}
where the $m^{th}$ plasmon energy window spans the energy range $[\omega_{p,min}^{m},\omega_{p,max}^{m}]$, the $l^{th}$ band energy window spans the energy range $[E_{n,min}^l,E_{n,max}^l]$, the density of plasmon modes is $D_p(\omega)$ and the density of band states is $D(E)$ (the explicit dependence on the window edges is suppressed). This formula is simply employed whenever a window pair generates a crossing in the minimization procedure to determine the optimal windowing given above. We have observed that this insertion, although potentially discontinuous as the window ranges change, does not prevent rapid numerical minimization of the total cost. Further details can be found in  Appendix~\ref{app:cost4overlap}.

\subsection{Cubic-scaling dynamic $P(\omega)$}
\label{sec:dynP}

The windowed quadrature methods developed to compute the static $P$ and the dynamic $\Sigma(\omega)$ can be applied directly and without modification to the computation of the frequency-dependent polarizability $P(\omega)$ of Eq.~(\ref{eq:Pomegadef}) with ${\cal O}(N^3)$ computational effort.  The key observation is that $P(\omega)$ can be rewritten as the sum of two simple energy denominator poles:
\begin{multline}
P(\omega)_{r,r'} = \sum_{c,v,\sigma,\sigma'}\psi_{x,c}\psi_{x,v}^*\psi_{x',c}^*\psi_{x',v} \times \\ 
\left(  \frac{1}{\omega - (E_c-E_v)} - \frac{1}{\omega + (E_c-E_v)} \right)\,.
\label{eq:Pomegasimple}
\end{multline}
Since $P(\omega)=P(-\omega)$, we need only focus on $P(\omega)$ for $\omega>0$.  The second energy denominator $\omega+E_c-E_v$ is always positive definite since $E_c-E_v>0$ and can be evaluated in ${\cal O}(N^3)$ with the same GL quadrature methodology developed for evaluating static $P$ in Sec.~\ref{sec:n3staticP}; the presence of $\omega>0$ in the second denominator enlarges the effective energy gap and enhances convergence of our method. The first energy denominator $\omega-(E_c-E_v)$ can change sign once $\omega$ is larger than the energy gap.  However, this term  can be evaluated with ${\cal O}(N^3)$ effort using the energy crossing quadrature method developed for $\Sigma(\omega)$ in Sec.~\ref{sec:n3sigma}.

\section{Results: Standard Benchmarks}
\label{sec:results}
Here, the application of the new CTSP method to standard benchmark systems is presented. Results for the dielectric constant and the energy band gap within the COHSEX approximation are given for crystalline Si and MgO. Next, results for the static polarization of crytsalline Al, a gapless systems, are presented. Last, a 
G$_0$W$_0$ computation of the band gap of crystalline Si is performed. 

\subsection{Dielectric constant \& COHSEX energy gap}
In order to evaluate the performance of the new reduced order method, CTSP, we study two standard benchmark  materials: Si and MgO.  We first performed plane wave pseudopotential DFT calculations for both materials to generate the DFT band structure and then employed the results in the reported GW computations. Appendix~\ref{app:details} contains the details of the DFT and GW calculations.

Si is a prototypical covalent crystal (diamond structure) with a moderate band gap (0.5 eV in DFT-LDA) while rocksalt MgO is an ionic crystal with a relatively large  gap (4.4 eV with LDA). To judge the performance of new method, the convergence of two basic observables was studied: the macroscopic optical dielectric constant $\epsilon_{\infty}$ and the band gap with the COHSEX approximation.  

\begin{figure}
\includegraphics[width=2.4in]{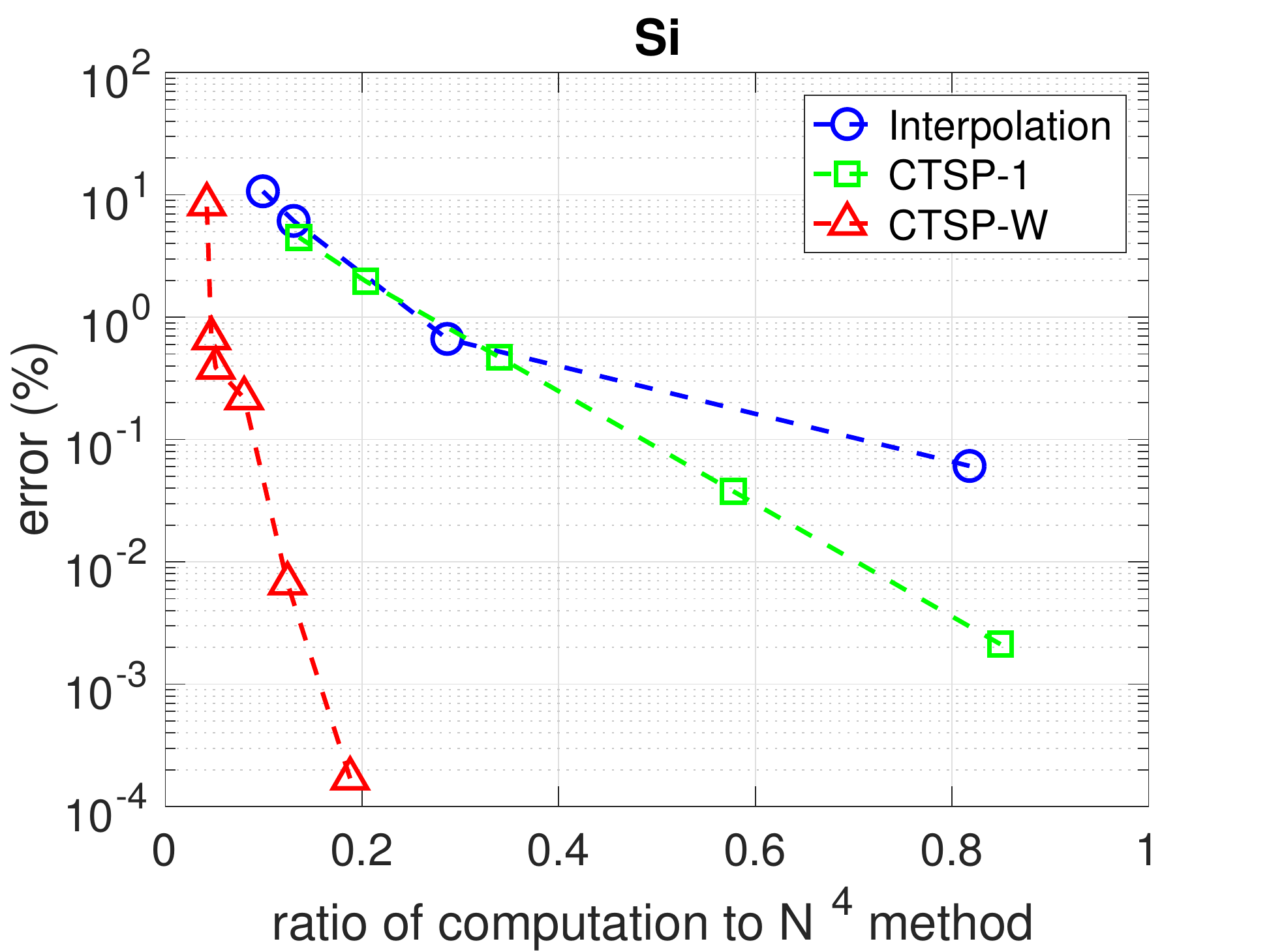}
\includegraphics[width=2.4in]{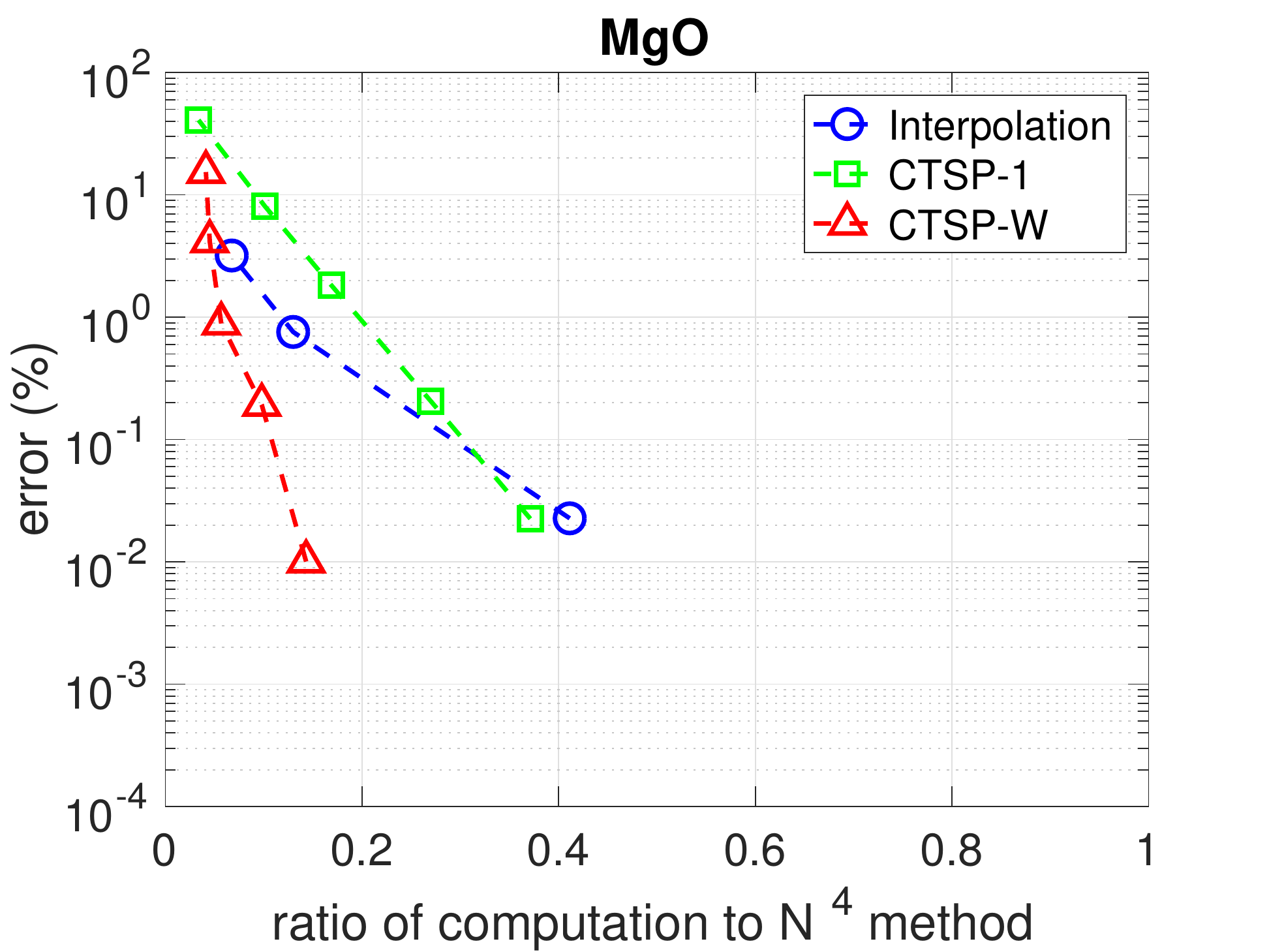}
\caption{Error in the macroscopic RPA optical dielectric constant $\epsilon_{\infty}$ for the interpolation, the CTSP-W and CTSP-1 methods with respect to the quartic ${\cal O}(N^4)$ method.  The horizontal axis is the ratio of computational load of the cubic to ${\cal O}(N^4)$ methods for a system  of 16 Si atoms.  Upper: Bulk Si data generated by using input fractional quadrature errors $\epsilon^{(q)}$ in percents of of 0.1, 1, 10 and 20 for interpolation; 0.1, 1, 10, 30, and 50 for CTSP-1; and 0.1, 1, 10, 30, 50, and 80 for CTSP-W.   Middle: same for bulk MgO; percent input errors are set to be 0.1, 1, and 10 for interpolation; 0.1, 1, 10, 30, and 70 for CTSP-1 and 0.1, 1, 10, 20, and 40 for CTSP-W.}
\label{fig:inveps}
\end{figure}

\begin{figure}
\includegraphics[width=2.4in]{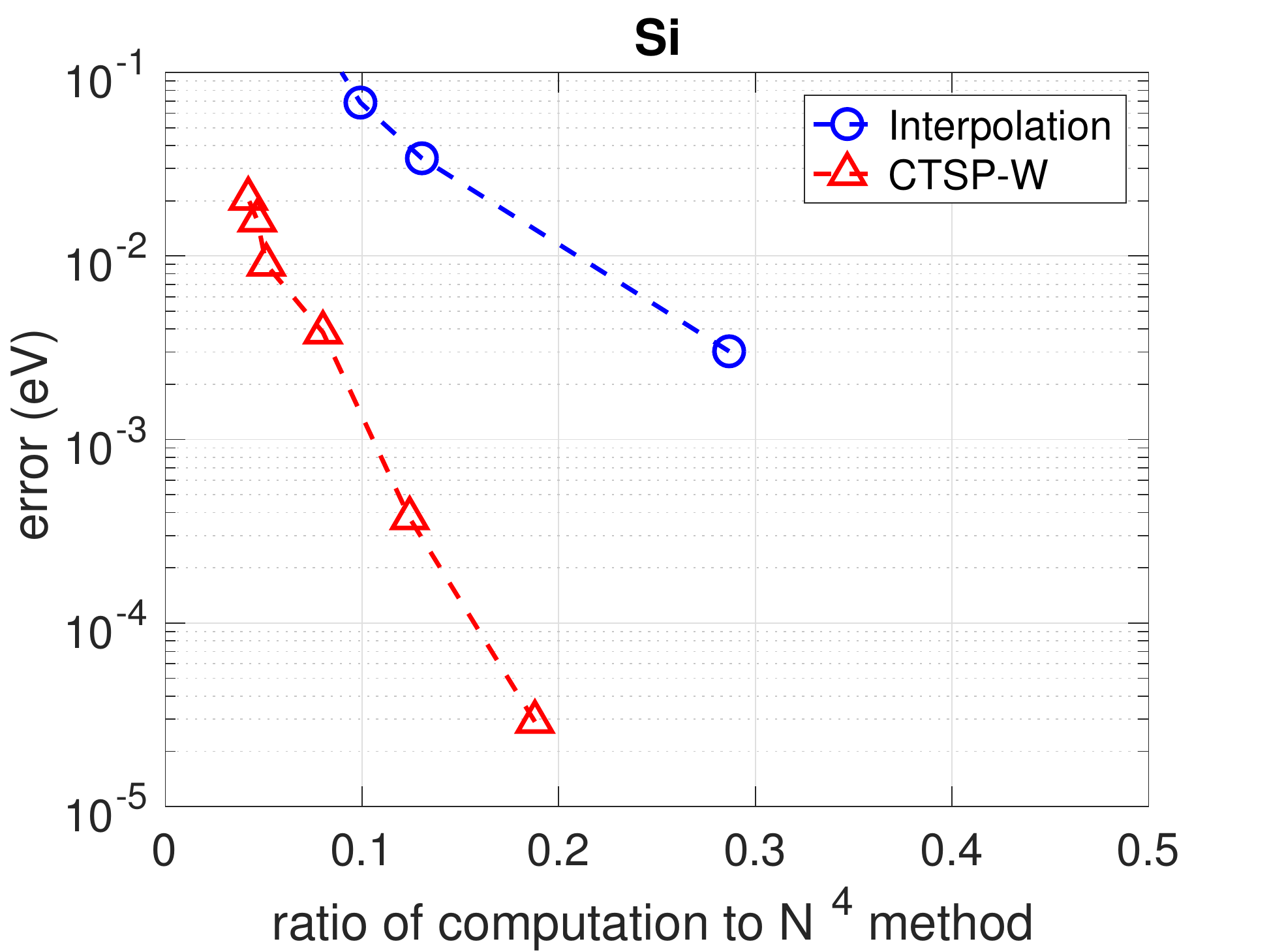}
\includegraphics[width=2.4in]{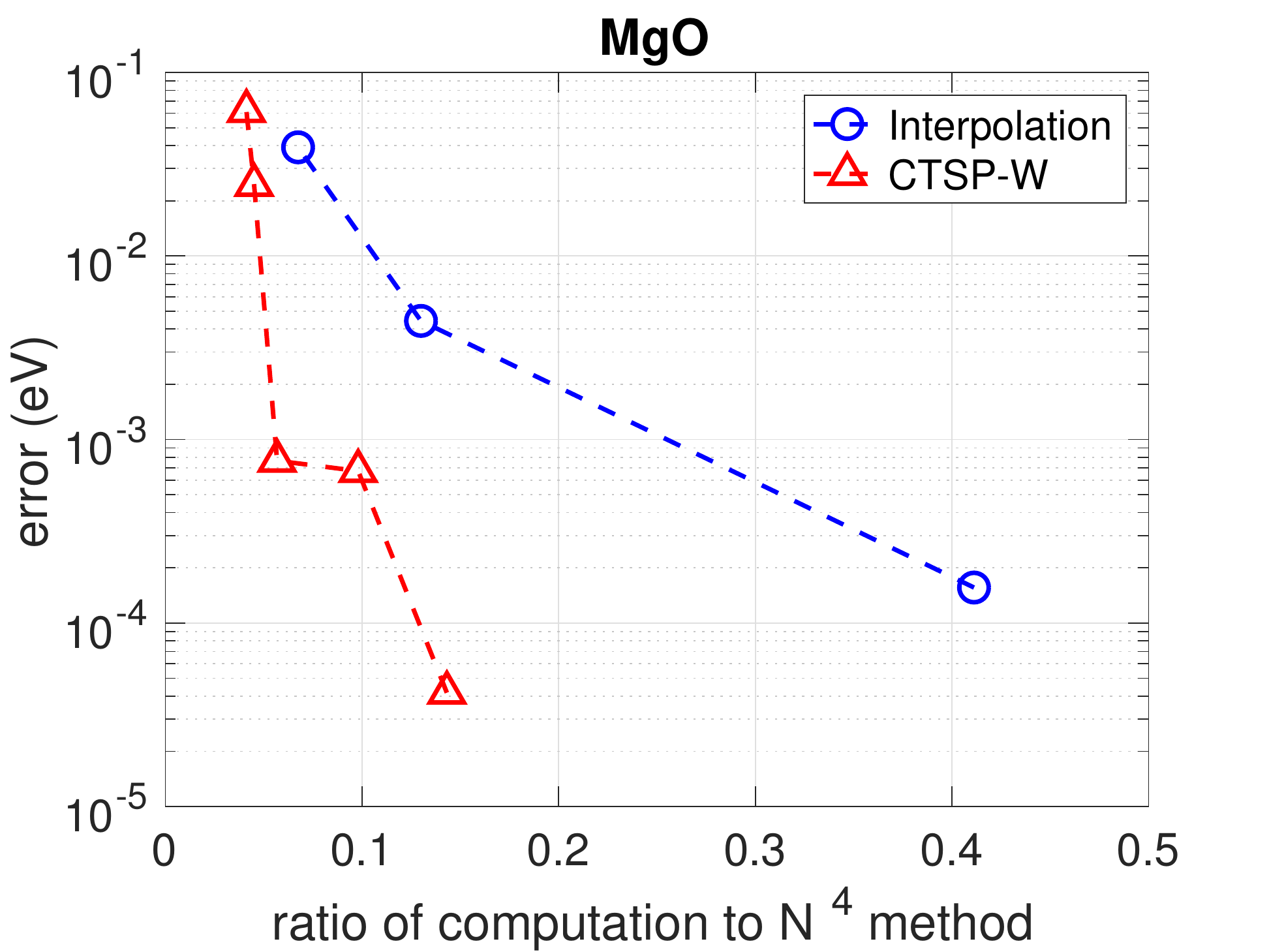}
\caption{Error in the bulk band gap from the COHSEX approximation ($\Gamma-X$ gap for Si and at $\Gamma$ for MgO) for different methods as a function of computational savings over the quartic method.  All data are for a fixed supercell size of 16 atoms.  The nomenclature and numerical tolerances are those of Fig.~\ref{fig:inveps}. }
\label{fig:bandgap}
\end{figure}

Figure~\ref{fig:inveps} shows the error in $\epsilon_{\infty}$ as a function of the computational savings achieved by the present ${\cal O} (N^3)$ method, both CTSP-W and CTSP-1 forms, and the ${\cal O} (N^3)$ interpolation method described in Appendix~\ref{app:interp}, relative to the standard ${\cal O}(N^4)$ method for 16 atom periodic supercells of MgO and Si.   Each data point is generated by fixing a fractional quadrature error, $\epsilon^{(q)}$, followed by  minimization of our cost function for CTSP-W.  Figure~\ref{fig:bandgap} shows the band gap within the COHSEX approximation for the GW self-energy~\cite{hedin_new_1965}, for the CTSP-W and interpolation methods on the same two physical systems. 

The results show the CTSP-W approach exhibits high performance both for accuracy and efficiency; the improvement over CTSP-1 and the interpolation method is clear for Si.  The larger MgO gap can be treated with fewer integration points and also leads to functions that are easier to interpolate.  For both materials, the CTSP-W method achieves better than 0.1 eV accuracy of the band gap with at least an order of magnitude reduction in computation compared to the ${\cal O}(N^4)$ approach.  Note, the savings of the new method will improve linearly as the number of atoms is increased (beyond $N=16$).

\begin{figure}
\includegraphics[width=2.4in]{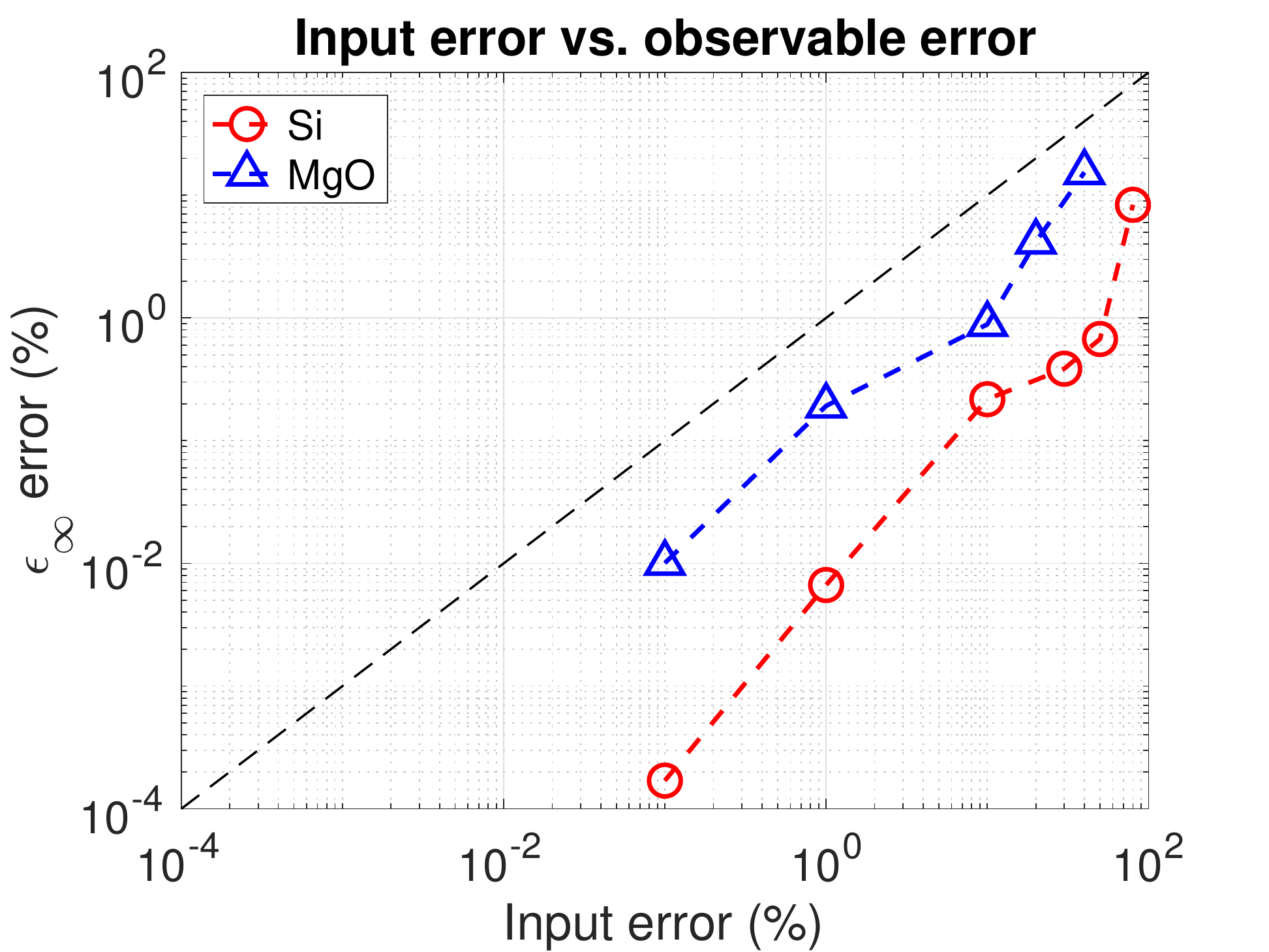}
\caption{The relation between the input fractional quadrature error tolerance $\epsilon^{(q)}$ and the error in the physical observable $\epsilon_{\infty}$ for the CTSP-W method applied to Si and MgO. The dotted line represents quadrature error equal to observable error.}
\label{fig:in_vs_out}
\end{figure}

Figure~\ref{fig:in_vs_out} shows the correlation between the log of the fixed fractional quadrature error, $\epsilon^{(q)}$, and the log of the fractional error of macroscopic dielectric constant for 
the application of CTSP-W to Si and MgO. The data indicate the error in $\epsilon_{\infty}$ is at least one order of magnitude smaller than the input fractional quadrature error and the slopes of the log-log curves are approximately unity. Although these results do not represent a rigorous bound, they nonetheless show that the error of calculated physical quantities from the CTSP method can be controlled by turning a simple ``knob" and the integration accuracy required for converged results is (surprisingly) modest (see also Sec.~III.D).

\subsection{Zero gap materials}
\begin{figure}
\includegraphics[width=2.4in]{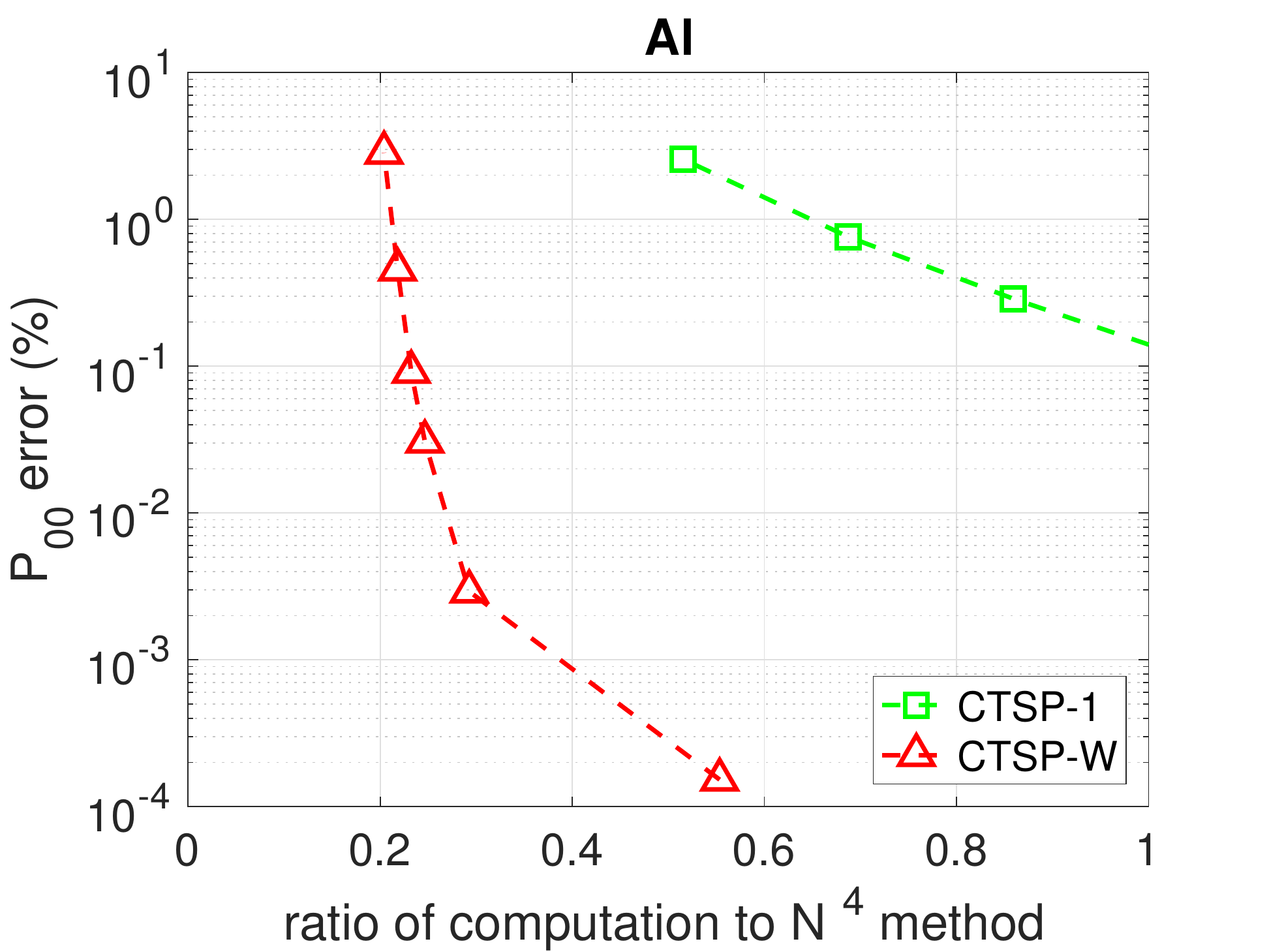}
\caption{Error in the $P^{q=(\frac{1}{2},\frac{1}{2},\frac{1}{2})}_{0,0}$ element of Al using CTSP-W. The horizontal axis is the ratio of computational load of the cubic to ${\cal O}(N^4)$ method for a supercell of 8 Al atoms. A total 400 states were used, and the broadening parameter was set to 0.03~Ry. The fractional input errors in percents are 1, 10, 30, 50, 70, and 80.}
\label{fig:Al}
\end{figure}

In order to test the performance of the reduced order approach for static $P$ in (nearly) zero gap materials, we study crystalline aluminum (Al) (Appendix~\ref{app:details} contains the details of DFT calculations performed to obtain the band structure). Gaussian broadening ($\beta=0.03~eV$) was employed to treat the occupation numbers and for simplicity the occupation numbers were set to 0 or 1 when these differ from the limits by less than $10^{-6}$.  Although there is certainly an energy gap in an extended metallic Al system, calculations in a finite periodic supercell will have discrete eigenvalues and a (small) artificial energy gap. However, the new method is robust to zero energy gap as described above and associated appendices.

Figure~\ref{fig:Al} shows the error of the $P_{0,0}$ polarization element (where these matrix indices correspond to reciprocal space) for $q=(\frac{1}{2},\frac{1}{2},\frac{1}{2})$. Similar to the above results for Si and MgO, the performance of the CTSP-W method is compared to both the quartic scaling method and the single window limit, CTSP-1. Here, the single window limit is not as efficient as it was for Si or MgO.  This is because the very small $E_g$ requires a large quadrature grid for the single window method to obtain the desired accuracy.  However, the CTSP-W method completely removes any trace of difficulties associated with the small $E_g$ and delivers  accurate results with high efficiency for (nearly) zero-gap systems. 

\subsection{G$_0$W$_0$ gap}

\begin{figure}
\includegraphics[width=2.4in]{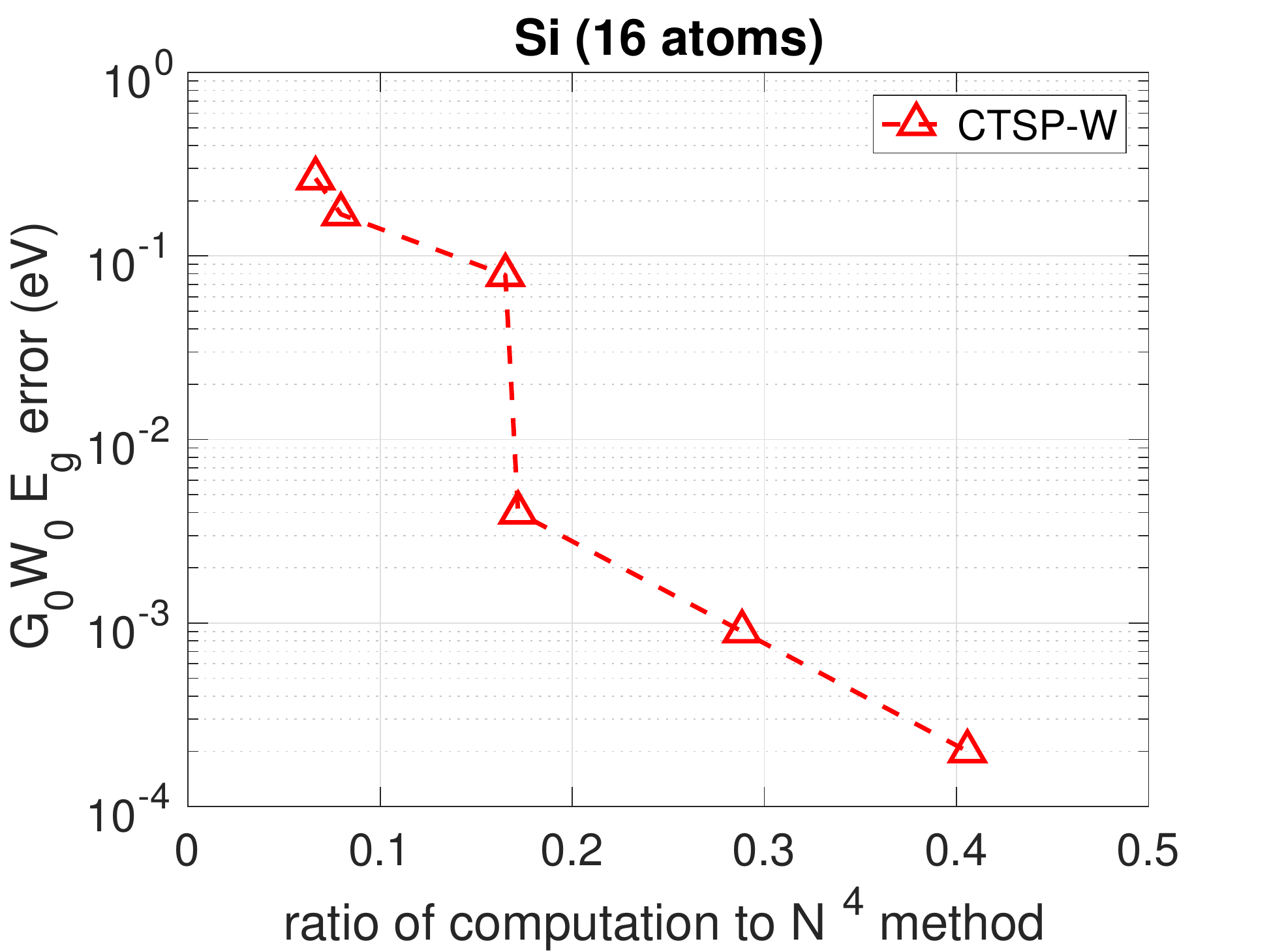}
\caption{Error in the bulk G$_0$W$_0$ band gap ($\Gamma-X$ gap for Si) for CTSP-W as a function of computational savings over the ``exact'' quartic method (horizontal dashed line).  All data are for a fixed system size of 16 atoms. The fractional input errors in percents are 0.1, 1, 5, 10, 50 and 100. $E_g^{(\rm G_0W_0)}=1.37eV$. }
\label{fig:g0w0gap}
\end{figure}

Figure~\ref{fig:g0w0gap} shows the convergence of the band gap of Si for the cubic-scaling CTSP-W method applied to $\Sigma(\omega)$ as described in Sec.~\ref{sec:n3sigma}.  The dynamic behavior of $W$ (i.e., the pole energies $\omega_p$ and pole strengths $B^p$ of Eq.~(\ref{eq:Wplasmonrep})) is determined using the generalized plasmon-pole (GPP) model of Hybertsen and Louie~\cite{HL}.  The figure shows that high accuracy is possible with large computational savings  when compared to the standard ${\cal O}(N^4)$ approach.

\subsection{Single convergence parameter}

Finally, compared to standard ${\cal O}(N^4)$ GW calculations, our cubic scaling CTSP approach has a single added input parameter which is the {\it a priori} desired fractional quadrature error, $\epsilon^{(q)}$.  Due to the CTSP method's construction, the choice of window parameters and quadrature grids are all determined by this single parameter.  As illustrated in Fig.~\ref{fig:in_vs_out}, the output error in computed observables is smaller in magnitude than the input quadrature error.  Hence, one can  estimate quickly the value of the input quadrature error that bounds the desired accuracy in the output observables (although the bound is not rigorous). 

A simple quantity that can be computed in order $N$ complexity with CTSP that provides an estimate of the error in physical observables generated by the CTSP, is the model static polarizability,
\begin{equation}
\label{eq:modelPtest}
P^{(model)} = \sum_{cv} \frac{f(E_v)-f(E_c)}{E_c-E_v} \ .
\end{equation}
That is, Eq.~(\ref{eq:modelPtest}), can be computed for a series of input quadrature errors $\epsilon^{(q)}$ at the startup of the GW calculation: monitoring the output error in $P^{(model)}$ provides a refined estimate of the input error level required to reach a desired error in observables.

If a more quantitative estimate of the error in physical observables is required, we recommend performing a convergence study on a small model system representative of the system of interest (e.g., a small unit cell of bulk material instead of a large unit cell of bulk with defects, an idealized surface with small in-plane lattice parameters instead of a complex surface reconstruction, etc.). This procedure will again necessitate performing a series of cubic scaling CTSP computations at various levels of quadrature error, $\epsilon^{(q)}$, to refine the parameter choice through a direct study of the convergence of the physical observables. Due to the small size of the model system, the process will require small or negligible compute time.  Finally, a full convergence study of the CTSP prediction of the observables of interest as a function of $\epsilon^{(q)}$ on the (large) system can be performed. This latter procedure retains cubic computational complexity but with increased prefactor. In general, performing convergence studies using CTSP on model systems is likely to provide sufficient error control for validation purposes (The model system approach is commonly used to select the cutoff energy in the conduction band, for instance).

\section{Results: Scaling analysis and comparison to other methods}
First, the cubic scaling and small prefactor of the new CTSP-W ${\cal{O}}(N^3)$ method of this paper are verified - in actual computations. Next, the ability of the new technique to treat physical systems of scientific and technological interest that heretofore have been too computationally intensive to study, is evaluated. Last, the performance of the new method is compared to small prefactor quartic scaling methods and other ${\cal{O}}(N^3)$ techniques.
\subsection{Verification of cubic scaling}
\begin{figure}
\centering
\includegraphics[width=2.4in]{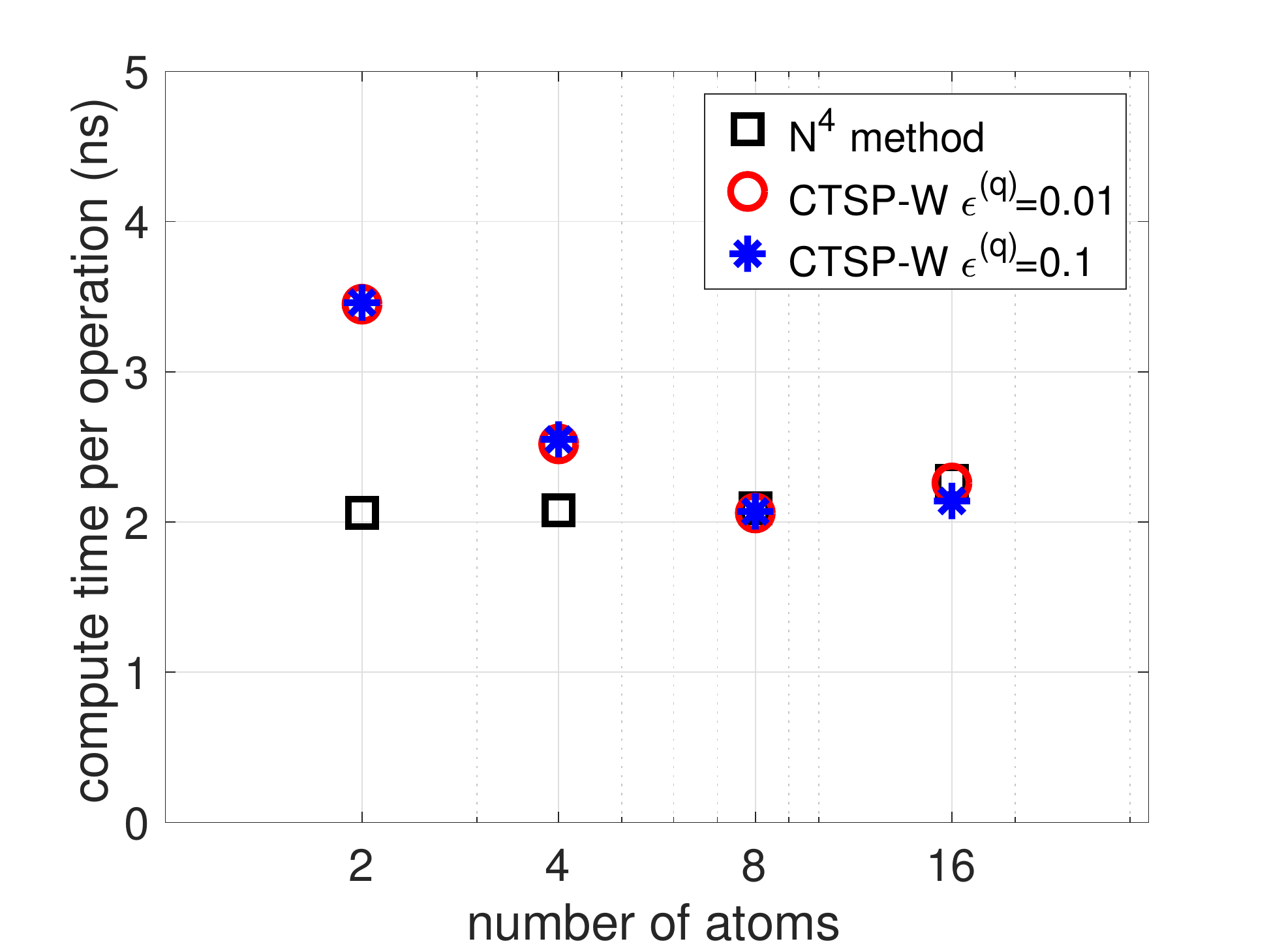}
\caption{Compute time per operation for evaluation of $P$. Black squares indicate the $N^4$ method, and red circles and blue asterisks indicate the $N^3$ CTSP-W method with maximum fractional quadrature error of 1\% and 10\%. A serial linux computer is used. The flattening of the curves for increasing number of atoms for both the ${\cal{O}}(N^4)$ and ${\cal{O}}(N^3)$ methods indicate their computation time increase as $N^4$ and $N^3$: there is no additional hidden cost to using the ${\cal O}(N^3)$ method in an actual calculation.
}
\label{fig:scaling}
\end{figure}

The scaling of the CTSP-W method in realistic calculations at two input fractional quadrature error levels ($\epsilon^{(q)}$) is verified. The computer time required to compute the static $P$ is measured as a function system size (number of atoms in the supercell) and the compute time {\it per operation} presented in Fig.~\ref{fig:scaling}. The number of operations are $N_v N_c N_r^2$ for the $O(N^4)$ method and $\sum_{l,m}N^{(\tau,GL)}_{lm}(N_c^m+N_v^l)N_r^2$ for the CTSP-W ${\cal O}(N^3)$ method. The result is a flat line --  the algorithms scale as they should on a present-day desktop computer
\footnote{The CTSP-W runs slightly slower per operation for few atoms which is due to inefficient caching and pipelining at small problem sizes.}.
It is important to recognize that compute times {\it per operation}  are very close to each other indicating that the ${\cal O}(N^3)$ method has a prefactor that is comparable to the ${\cal O}(N^4)$ method even in small systems $N\gtrapprox10$. Thus, the reduced order method is highly efficient. 

\subsection{Sizing for large systems}

The computational effort required to generate the static $P(0)_{r,r'}$ polarizability matrix with the quartic, CTSP-1 and CTSP-W methods will now be analyzed for two systems: ``medium''  and ``large''. The medium-sized system is a 72-atom GaN unit cell, while the large system is a 100-atom photovoltaic interfacial system. The number of plane-waves in the basis set is 19,200 and 128,000, and the number of FFT grid points is 39,000 and 262,000 for 72-atom and 100-atom systems, respectively (see Table I). 

\begin{table}
\begin{tabular}{ c| c| c}
\hline
& Medium  system & Large  system \\
\hline
$N$   & 72  & 100 \\
$N_v$ & 144 & 800\\
$N_c$ & 2,806 & 16,000\\
$N_k$ & 8 & 2\\
$N_{FFT}$ & 39,000 & 262,000\\
Gap & 2.2~eV & 1~eV\\
Bandwidth & 110~eV & 110~eV\\
\hline
\end{tabular}
\caption{Representative medium and large physical systems that at present are very computationally challenging with existing $O(N^4)$ GW methods.  Here, $N_v$ and $N_c$ are number of valence and conduction bands respectively. $N_k$ is number of k-points to be sampled. $N_{FFT}$ is the number of FFT grids, which is the same as the rank of $P$ matrix. Gap and Bandwidth are the target  energy gap and energy bandwidth of the system. } 
\label{tab:sys}
\end{table}
\begin{table*}
\begin{center}
\begin{tabular}{ c   c  |  c  | c  }
\hline
   &   &  Medium & Large \\
\hline
\multirow{3}{*} {\specialcell{Operation count \\ $P_{r,r'}^q$ }} & Standard $O(N^4)$ &  4.92$\times10^{15}$ & 1.76$\times10^{18}$ \\
 & CTSP-1 & 5.38$\times10^{14}$ &  4.84$\times10^{16}$\\
 & CTSP-W & {1.18$\times10^{14}$} & 8.13$\times10^{15}$\\
\hline
\multicolumn{2}{c|} {Memory for $P_{r,r'}^q$} & 23~GB& 1~TB \\
\hline
\end{tabular}
\caption{Number of operations and memory required to calculate $P_{r,r'}^q$ matrix for medium and large size physical systems of Table~\ref{tab:sys}.  }
\label{tab:sizing}
\end{center}
\end{table*}

Table~\ref{tab:sizing} shows the number of operations~\cite{our_own_GW} for computing $P(0)_{r,r'}$ for the two systems using the standard quartic scaling method, CTSP-1 and CTSP-W. For the CTSP-W method, the parameters are selected by computational cost function as described in Sec.~\ref{sec:n3staticP}. The quadrature grids are chosen to achieve less than 0.1\% error in the calculation of $P^{(model)}$. We emphasize that 0.1\% error in $\epsilon^{(q)}$ will achieve accurate results as presented in Figs.~\ref{fig:inveps}-\ref{fig:in_vs_out}.
Table~\ref{tab:sizing} shows that the CTSP-W method yields an efficient computation of $P(0)_{r,r'}$ without sacrificing accuracy -- for the medium  system, the CTSP-W method delivers about a 40$\times$ reduction in operation count while the reduction is 220$\times$ for the large system (compared to the standard quartic scaling technique). Thus, these technologically interesting problems are now approachable in terms of computer time typically available on supercomputer centers.

It is important to also consider the memory requirements to store a large matrix such as $P_{r,r'}$ and whether this requirement can be satisfied by today's supercomputers (see Table II). The Blue Waters machine at NCSA~\footnote{https://bluewaters.ncsa.illinois.edu/}, a leading HPC platform, has 64~GB of memory per node and the installation has 23K nodes for a total of 1.4 petabytes of memory. The BlueGene/Q installation at Argonne National Laboratory, Mira, has 16 GB of memory per node and 49K nodes for a total of 0.8 petabytes. Thus, using only a fractional allocation of such computers, the $P$ matrix, even for the large system, can easily be accommodated. Of course, the effective utilization of distributed memory supercomputers requires a well-parallelized GW implementation such as that developed by us in Ref.~\onlinecite{our_own_GW} and that by others in Ref.~\onlinecite{deslippe_berkeleygw:_2012}; both applications can be modified to implement the method of this paper straightforwardly.


\subsection{Comparison with small prefactor ${\cal O}(N^4)$ methods}
\label{sec:on4comparison}
\begin{figure}[h]
\includegraphics[width=2.5in]{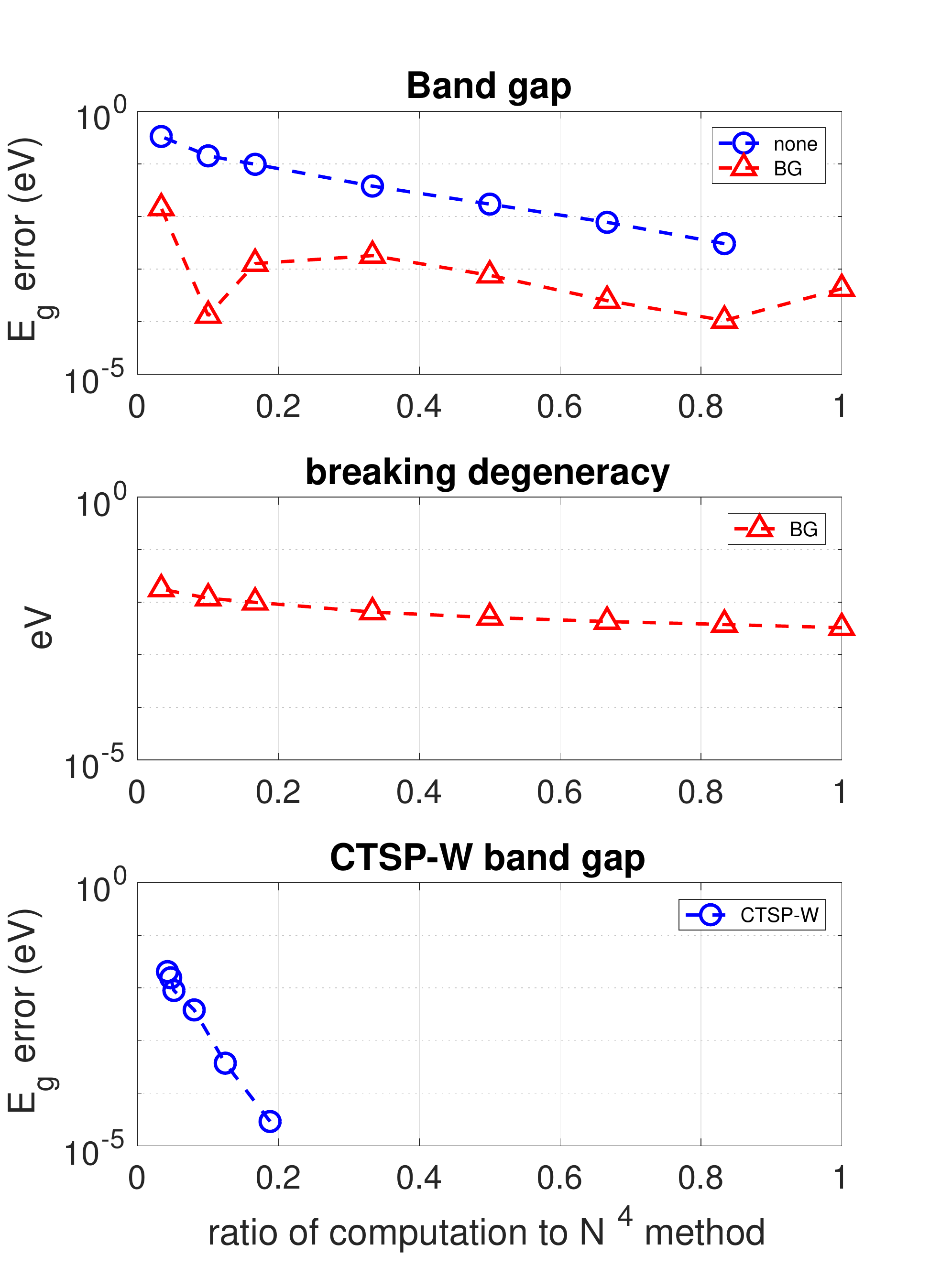}
\caption{Comparison between the ${\cal O}(N^4)$ standard GW method labeled none, the ${\cal O}(N^4)$ GW terminator method of Ref.~\onlinecite{BG} labeled BG as implemented in Yambo code  and the ${\cal O}(N^3)$ CTSP-W method for a 2-atom Si cell with 8 k points. The $N^4$ work is computed for a 300 band computation which also provides the reference results that define the errors.
(Top) Error in the $\Gamma-X$ band gap. The BG band gap is defined as the energy difference of the valence band maximum and the average of the conduction band minimum doublet versus computational work relative to a large $N^4$ computation.  (Middle) The artificial degeneracy-breaking of the conduction band minimum doublet for Si computed using Yambo implementation of the BG method. (Bottom) Error in the $\Gamma-X$ band gap for CTSP-W versus computational work relative to the standard quartic scaling method.}
\label{fig:term}
\end{figure}

Next, we compare the CTSP-W method of this paper to existing, low prefactor, quartic scaling GW methods. We choose to employ the Yambo GW software  (http://www.yambo-code.org/) which implements a quartic scaling sum-over-states technique within the ``terminator'' acceleration approach of Bruneval and Gonze~(BG)~\cite{BG}. A two-atom Si cell with 8 k points is employed as the benchmark.  Figure~\ref{fig:term} shows the error in the band gap versus the computational savings using different number of bands for the self-energy $\Sigma$ band summation calculation. The savings are referenced to a standard ${\cal O}(N^4)$ calculation (no-terminator) with 300 bands which converges the band gap to within 1 meV.  (Comparisons with terminator method of Berger, Reining and Sottile~\cite{BRS} yield similar results, not shown). 
\begin{figure}[t!]
\includegraphics[width=2.5in]{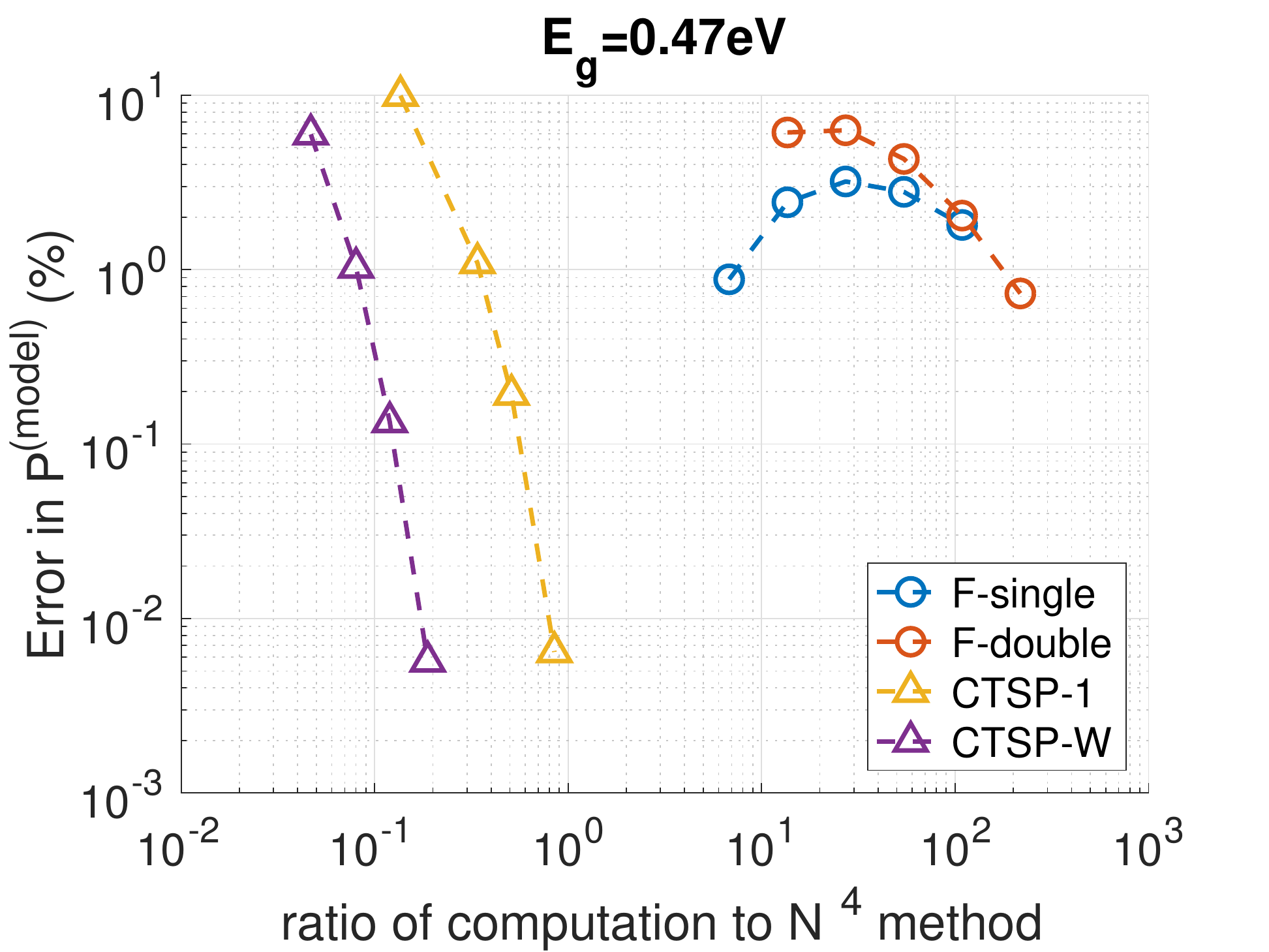}
\includegraphics[width=2.5in]{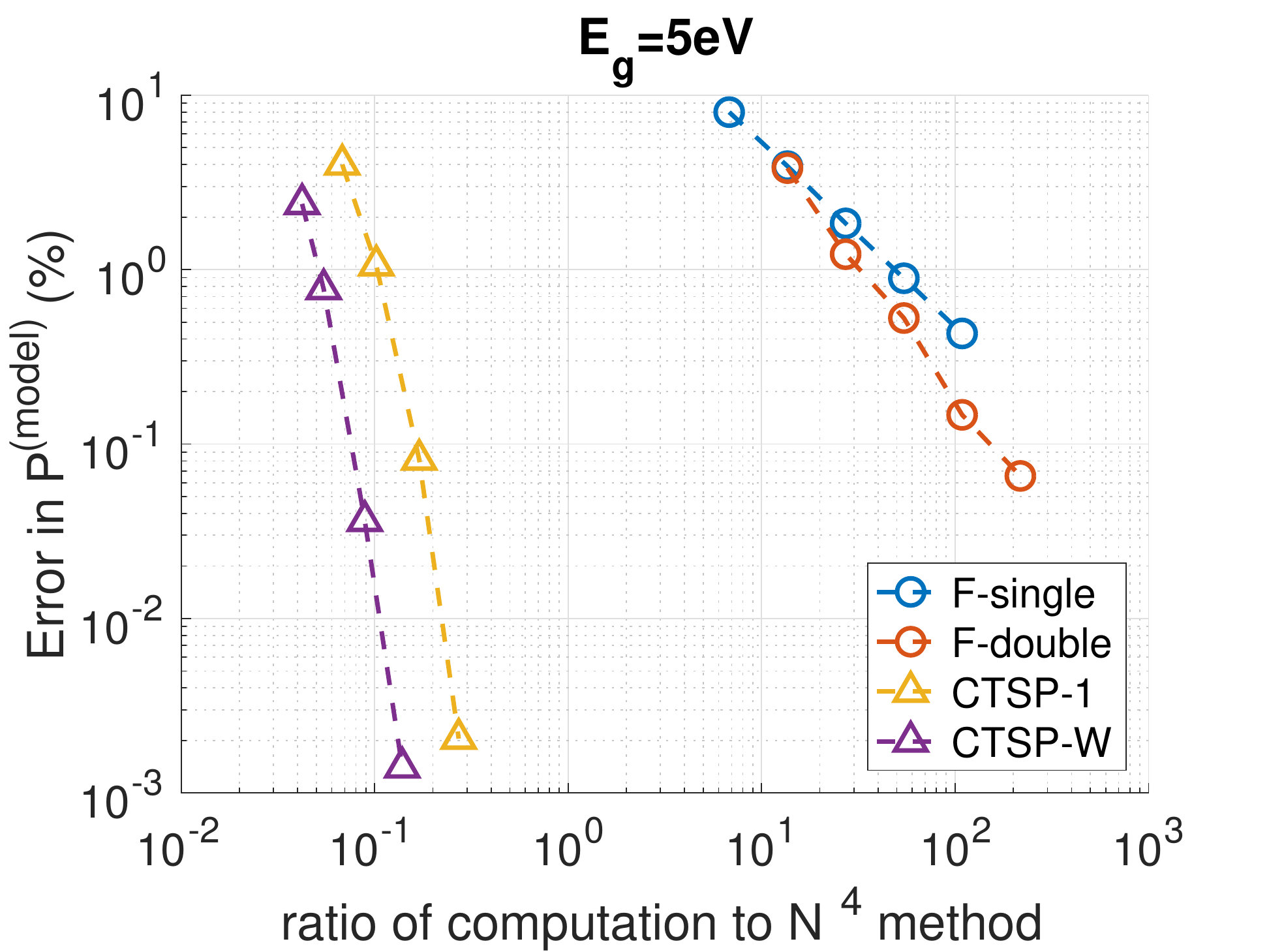}
\caption{Comparison between the cubic scaling Foerster et al.~\cite{Foe} and the CTSP-W method. The y-axis is the error (\%) of  $\hat P$ for a 16-atom Si unit cell with 399 bands; a linear scale is used to enhance the comparison here.  The x-axis is the log of the computational work needed to reach that accuracy: for the Foerster method it is $N_{\omega}(N_c+N_v)$ where $N_{\omega}$ is the number of frequency grid points and for the CTSP-W method the workload is defined in  Eq.~(\ref{eq:opcnt}) where $N_r$ is set to one. The band gap $E_{gap}$ is manually adjusted to investigate the effect of $E_{gap}$. F-single and F-double mean the Foerster method using single and double windows.}
\label{fig:foerster}
\end{figure}

Both the ${\cal O}(N^4)$ terminator method and the ${\cal O}(N^3)$ CTSP-W method deliver more than 90\% savings in the computational workload compared to the standard quartic scaling method for an accuracy of 10 meV in the band gap of the two-atom cell.  
The middle figure shows the artificial removal of the degeneracy of the conduction band minimum due to the use of terminators in Yambo (this may be a feature of the method or the implementation ). We observe that the CTSP-W method is already competitive with the terminator method for this small two-atom system. Since the CTSP-W method scales cubically, it only becomes more efficient as system size is increased.  We thus conclude that the new cubic method has a sufficiently small prefactor to be competitive with existing accelerated quartic scaling GW methods even for unit cells with as few as 10 atoms.
\begin{figure}[t!]
\includegraphics[width=2.5in]{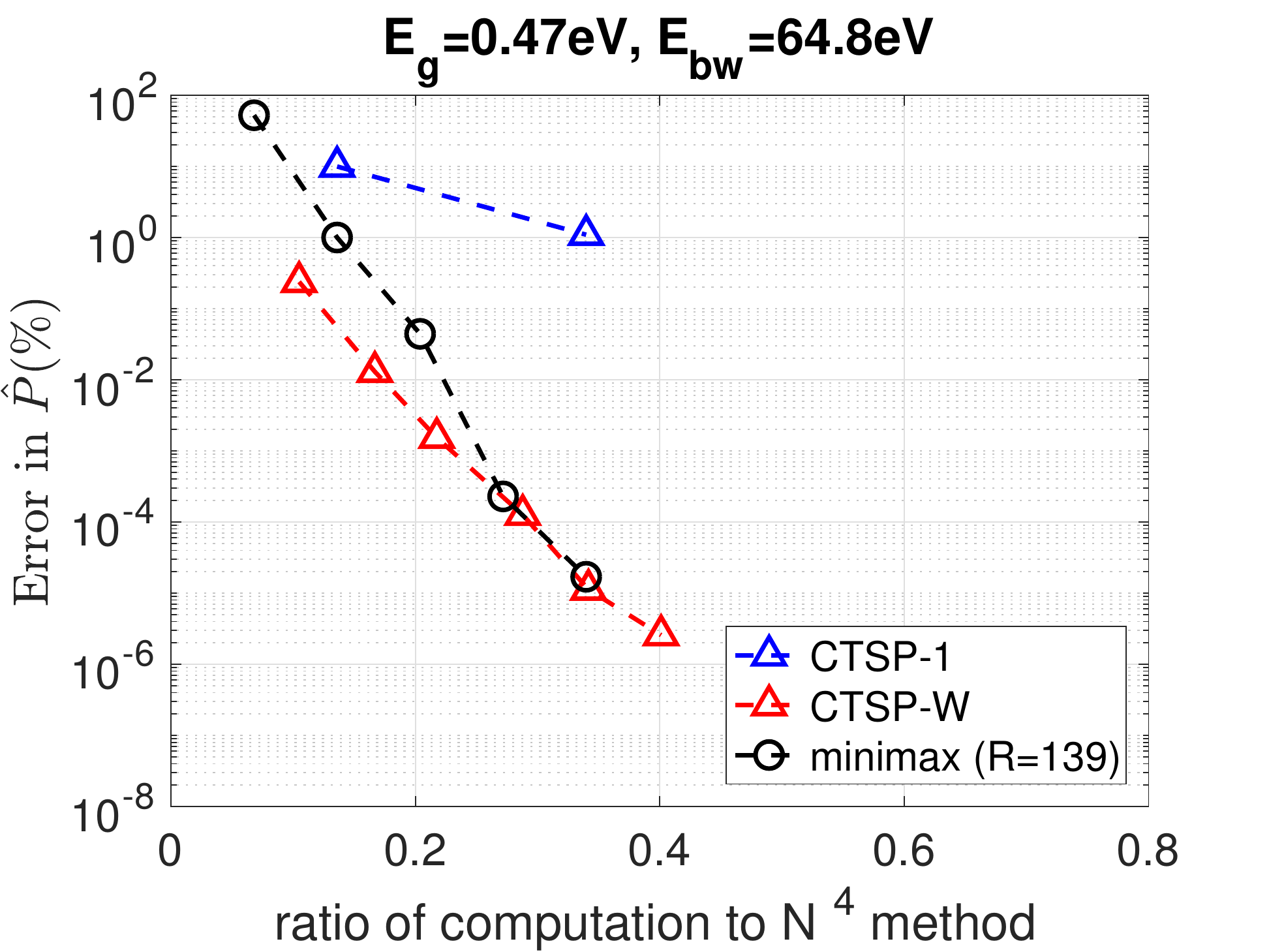}
\includegraphics[width=2.5in]{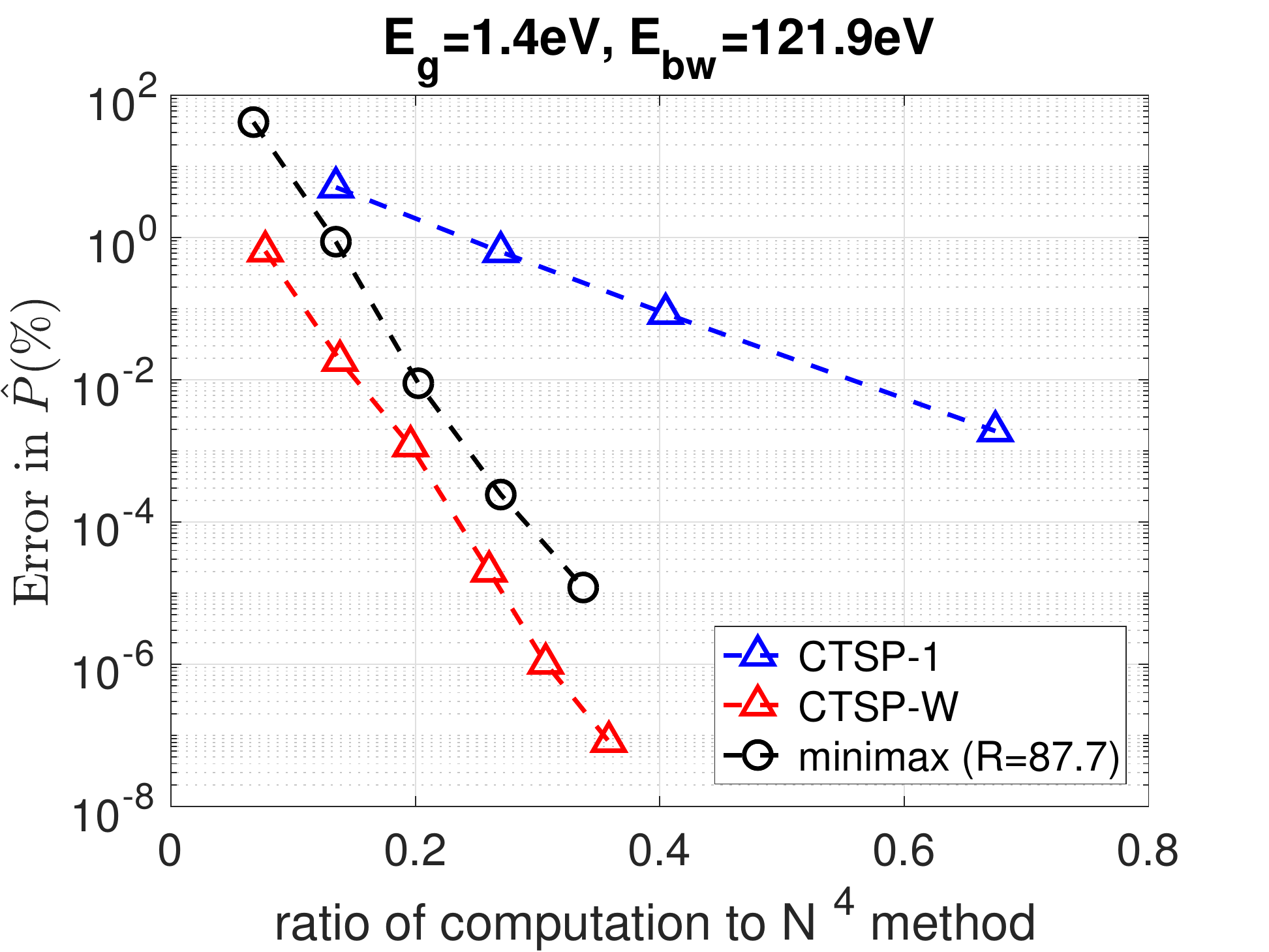}
\caption{Comparison between the cubic scaling Minimax method of Liu et al.~\cite{Liu} and our windowed GL method for computation of $\hat P$. The computational work (x-axis) for the Minimax method is   $N_{grid}(N_c+N_v)$ where $N_{grid}$ is the number of imaginary time grid points.}
\label{fig:minimax}
\end{figure}

\subsection{Comparison with other ${\cal O}(N^3)$ methods}

Last, the performance of the new CTSP-W method is compared to other cubic scaling GW methods -- that of Foerster et al.~\cite{Foe} and Liu et al.~\cite{Liu}, respectively -- in Figs.~\ref{fig:foerster}- \ref{fig:minimax}. The comparison is for a model $\hat{P}$ where all $\psi_n(r)=1$,
\[
\hat{P}=\sum_{c,v}\frac{1}{E_c-E_v},
\]
The input energies are taken from a 16-atom Si unit cell with 399 bands generated at the $\Gamma$ point. The workload is defined as $N_{\omega}(N_c+N_v)$ where $N_{\omega}$ is a number of frequency grid points used in Eq. (32) of Ref.~\onlinecite{Foe}.  We calculate $\hat{P}$ with the single and double window methods of Foerster et al. To investigate the effect of the band gap, we manually adjusted the band gap from 0.47 eV to 5 eV by uniformly shifting the conduction bands up in energy. Our main observation is that the method of Foerester et al. is orders of magnitude more computationally expensive that of the CTSP-W for the same level of accuracy.

Figure~\ref{fig:minimax} shows a comparison between the Minimax grid technique (the approach of Liu et al.) and the CTSP-W method for computing the static polarizability, $P$. We used two sets of data, 399 Si eigenvalues from 16-atom cell and 435 MgO eigenvalues from 16-atom cell. For the Minimax method, the computational work is defined as $N_{grid}(N_c+N_v)$ where $N_{grid}$ is the number of imaginary time grid points used in the Minimax technique. We find that the Minimax method  is competitive with the CTSP-W method but slightly inferior in performance.

To compare two methods more deeply, we note two points in favor the CTSP-W method.  First, the choice of quadrature grids and energy windows is a straightforward and robust exercise, requiring only the minimization of a simple cost function.  By contrast, finding the $N_{grid}$ energy grid points that solve the minmax problem~\cite{Liu} is quite challenging, and we found it required significant (human and computer) effort to do so.  Second, CTSP-W computes frequency-dependent spectral quantities such as the polarizability, $P(\omega)$ and the self-energy $\Sigma(\omega)$ directly on the real $\omega$ frequency axis which is the final desired and useful physical representation of any spectral function.  Namely, using CTSP-W, there is no need to compute quantities along the imaginary energy or time axis and then analytically continue to real frequencies. This is highly desirable as it avoids (i) the use of analytical continuation methods that are based on assumptions on the analytical form of the functions, and / or (ii) the numerical instabilities inherent in analytical continuation when high accuracy is desired~\cite{beach_reliable_2000}.

\section{Conclusion}
In summary, the GW equations have been recast, exactly, as Fourier-Laplace time integrals over complex time propagators. The propagators are then  partitioned in energy space and the time integrals approximated in a controlled manner using generalized Gaussian quadratures. Coupled with discrete variable methods to represent the propagators in real-space, a cubic scaling GW method emerges. Comparisons show that the method has sufficiently small prefactor to outperform standard and accelerated quartic scaling methods on small systems ($N \gtrapprox$ 10 atoms). For large systems (up to $N \lessapprox$ 200-300 atoms), we show the method fits comfortably in today's supercomputers both in terms of memory requirement and  computational load. Its efficiency and easy of use indicate that it has the potential for wide adoption and we are currently developing a fine grained parallel version of the method based on our previous work.\cite{our_own_GW}

Lastly, consider further development of the new CTSP method aimed at reducing its prefactor and / or its computational complexity (scaling).  The key CTSP expressions of Eqs.~(\ref{eq:rhorhobarlm2}) and (\ref{eq:finalsigmalmexpr}) contains sums over many high energy conduction bands which are computationally expensive to generate and manipulate.  Thus, one may develop a modified terminator method~\cite{BG} to reduce the number of needed high energy bands significantly, thereby reducing the ${\cal O}(N^3)$ prefactor.  Reducing the scaling to ${\cal O}(N^2)$ or ${\cal O}(N^2\log{N})$, the latter for a plane-wave basis, is more challenging.  Note, restricting band sums to an energy window (CTSP-W) is equivalent to summing over all bands with occupancies given by the difference between two different Fermi-Dirac distributions whose chemical potentials are set at the start and end of the window, respectively. Thus, we can, in principle, apply the Fermi Operator Expansion~\cite{goedecker_efficient_1994,goedecker_linear_1999} approach to describe the Fermi-Dirac functions using polynomials of the Hamiltonian and matrix-vector multiplications without reference to the bands themselves. Such an approach should lead to an ${\cal O}(N^2)$ (or ${\cal O}(N^2\log{N})$) scaling method, but significant development is required to realize the concept with both low prefactor and effective error control - future work.

\begin{acknowledgments}
We thank Jack Deslippe, Gian-Marco Rignanese and Dennis Newns for helpful discussions. This work was supported by the NSF via grant ACI-1339804. The present address of GJM is Pimpernel Science, Software and Information Technology from whom GJM acknowledges funding; however, the views expressed are those of the authors and do not reflect Pimpernel policy.
\end{acknowledgments}

\vspace{.25in}
\appendix
\section*{Appendix overview}
In order to improve the readability of the paper, we have chosen to place computational details in appendices. In Appendix A, the Gauss-Laguerre quadrature for window pairs without energy crossing is discussed. In Appendix B, the determination of the optimal windowing by minimization of the cost function for quantities without energy crossings is described. Appendices C and D discuss the weight function and quadrature employed to treated window pairs with energy crossings, respectively, while Appendix E describes minimization of the cost function for quantities whose evaluation involves treating energy window pairs with energy crossings. In Appendix~\ref{app:interp} an alternative ${\cal O}(N^3)$ method is given and in Appendix~\ref{app:details} computational details related to the results presented in the main text are present. Last, matlab code to generate the weights and nodes of the Hermite-Gauss-Laguerre quadrature is presented.

\setcounter{section}{0}
\section{Gauss-Laguerre quadrature optimization}
The optimizations required to evaluate energy denominators by discrete approximation to time domain integrals for a set of energies in a window pair are given.

\subsection{Optimal choice for energy scale $\zeta$}
\label{app:optima}
First, we describe the optimal choice of $\zeta_{lm}$ for the Gauss-Laguerre (GL) quadrature of Eq.~(\ref{eq:GLquadPlm}).
We suppress the energy window index $lm$ and  describe why $\zeta^{-1}\approx\sqrt{E_{BW}E_{g}}$ is a good choice for the energy scale $\zeta$ that minimizes the error of the GL quadrature across a given window pair.  

First, we seek to optimally approximate
\[
\frac{1}{\Delta} = \zeta\int_0^\infty e^{-\zeta\Delta\tau}d\tau \approx \zeta\sum_{k=1}^{N^{(\tau,GL)}} w_k e^{-\tau_k(\zeta\Delta-1)}
\]
where $\Delta=E_c-E_v>0$ are the interband transition energies by quadrature for the given range. That is, defining the dimensionless quantity, $u=\zeta\Delta$,  we wish to minimize is the universal error function
\begin{equation}
\frac{\epsilon^{(q)}(u)}{u} = \frac{1}{u} - \sum_{k=1}^{N^{(\tau,GL)}} w_k \exp(-\tau_k(u-1))\,.
\label{eq:erroru}
\end{equation}
for $u$ spanning the range in a window.
\begin{figure}
\centering
\includegraphics[width=2.4in]{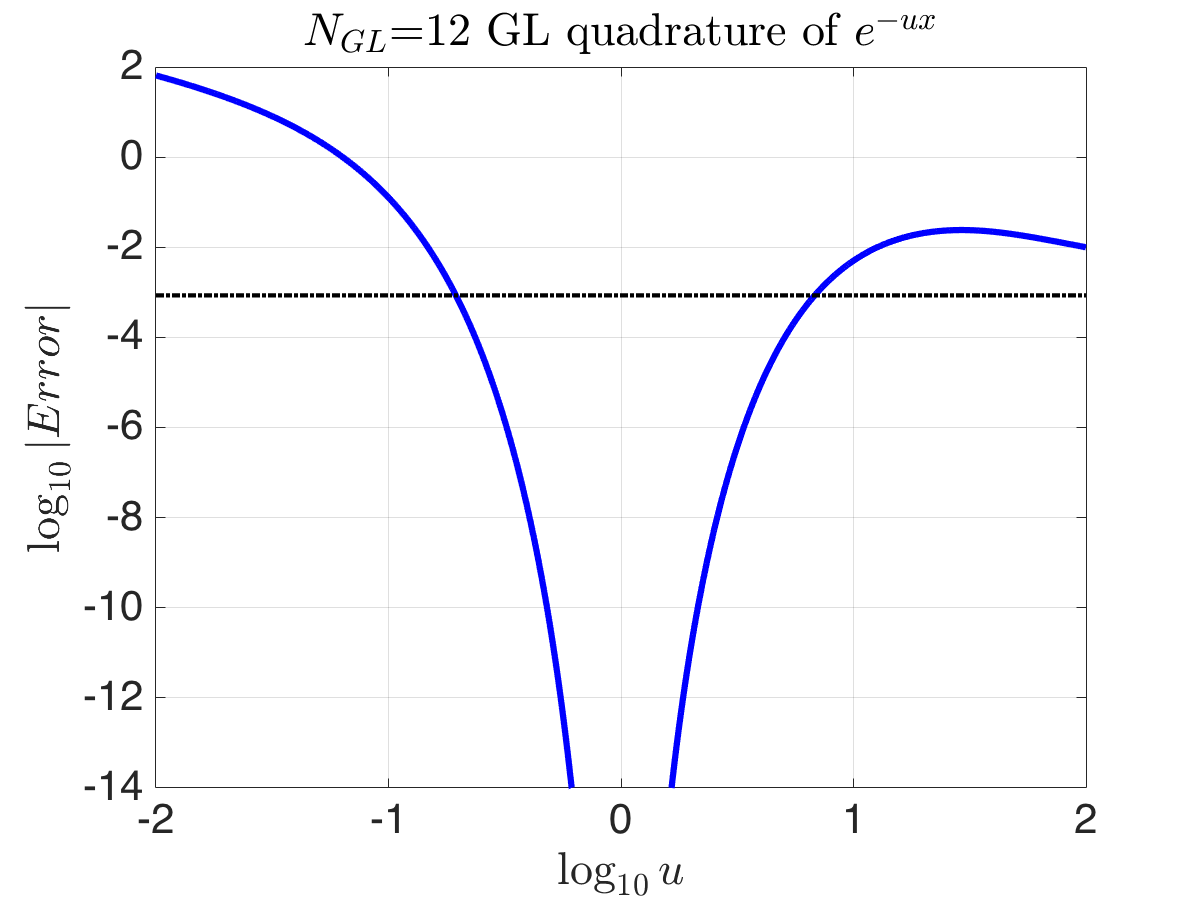}
\caption{Error of GL quadrature of $e^{-u\tau}$ with 12 quadrature points as a function of $u$.  The solid blue curve is the error versus $u$.  The dashed black horizontal line represents the choice of $\zeta$ giving equal errors at $u=\zeta E_{g}<1$ and $u=\zeta E_{BW}>1$ for the case $E_{BW}/E_{g}=23$.
}
\label{fig:choosea_errorofu}
\end{figure}
We first note that the error is exactly zero at $u=1$ since GL quadrature is exact when integrating $e^{-\tau}$. Figure~\ref{fig:choosea_errorofu} shows a plot of the error versus $u$.  The the error curve is symmetric around $\ln u=0$, especially when smaller error values are of interest which is the case herein.  Namely, our error function, to a good approximation, is even in $\ln u$ about $\ln u=0$.

Second, the interband energies $\Delta$ range from $E_{g}$ to $E_{BW}$.  Examining Fig.~\ref{fig:choosea_errorofu}, the lowest errors are sampled as $u$ ranges from its lowest value of $\zeta E_{g}$ to its highest value of $\zeta E_{BW}$. Therefore, it is reasonable to choose $\zeta$ such that $u=\zeta E_{g}<1$ and $u=\zeta E_{BW}>1$ straddle $u=1$ and have the same error.  For a symmetric error function about $\ln u=0$, this requires $-\ln(\zeta E_{g})=\ln(\zeta E_{BW})$ which yields the geometric mean, $\zeta^{-1}=\sqrt{E_{BW}E_{g}}$.  The geometric mean becomes exactly optimal as $N^{(\tau,GL)}$ is increased and the errors are also reduced when $E_{BW}/E_{g}$ is close to unity (the many windows limit).

\subsection{Number of GL quadarature points for fixed error}
\label{app:ngl}
\begin{figure}
\centering
\includegraphics[width=2.5in]{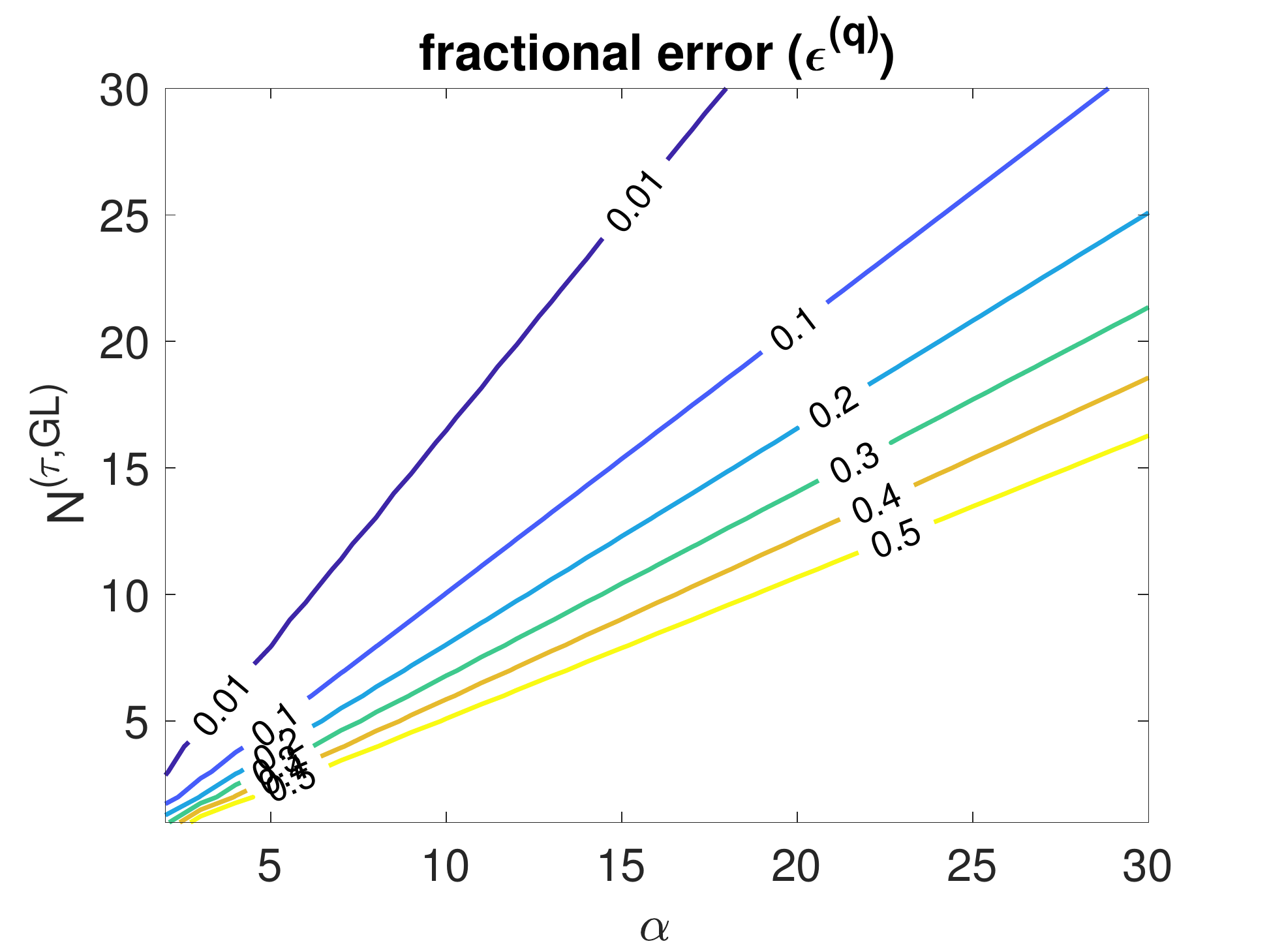}
\caption{Fractional error ($\epsilon^{(q)}$) of quadrature for $1/E_{BW}$ with respect to $\alpha$ and $N^{(\tau,GL)}$. $\alpha
$ is defined as $\sqrt{E_{BW}/E_g}$. Each contour is labeled by its $\epsilon$ value.  With a fixed fractional error, $N^{(\tau,GL)}$ is linear to $\alpha$.}
\label{fig:errorbw}
\end{figure}

Fixing $\zeta_{lm}^{-1}=\sqrt{E_{BW}^{lm}E_g^{lm}}$ as an optimal choice to balance the error in the discrete approximation to the time integrals in an energy window pair, 
the maximum fractional error of Eq.~(\ref{eq:erroru}), $\epsilon^{(q)}$, occurs at the largest energy transition (i.e., the error in computing the inverse energy $1/E_{BW}^{-1}$ via quadrature), $N^{(\tau,GL)}$ quadrature points as
\begin{equation}
\label{eq:errorbw}
\epsilon^{(q)}(\alpha) =  1- \alpha\sum_k^{N^{(\tau,GL)}} w_k \exp{\left[(1-\alpha)\tau_k\right]}.
\end{equation}
where $\alpha=\sqrt{E_{BW}/E_g}$.  The analogous equation for the fractional error to computing $E_g^{-1}$ has $\alpha$ replaced by $1/\alpha$, and is equal to the error of Eq.~(\ref{eq:errorbw}) due to optimal error-matching choice of $\zeta_{lm}$. Figure~\ref{fig:errorbw} displays a contour plot of the fractional quadrature error, $\epsilon^{(q)}$ of Eq.~(\ref{eq:errorbw}), which shows that $N^{(\tau,GL)}$ is essentially linear in $\alpha$ for any fixed choice of fractional error.  Numerical analysis of the contour plot shows that an accurate and compact explicit relation between the variables is
\begin{equation}
\label{eq:Ngl}
N^{(\tau,GL)}(\alpha;\epsilon^{(q)}) =\alpha(0.4-0.3\ln{\epsilon^{(q)}})\,.
\end{equation}
This equation fits the data well for $\epsilon^{(q)} \le 0.125$ but the range can be extended by adding 0.5 to the formual without loss of accuracy. Hence, the computational cost for computing the contributions from a set of interband transitions can be simply and accurately estimated given a maximum allowed {\it a priori} fractional error.

\section{Optimal sets of energy windows}
\label{app:winmin}
\begin{figure}
\includegraphics[width=2.5in]{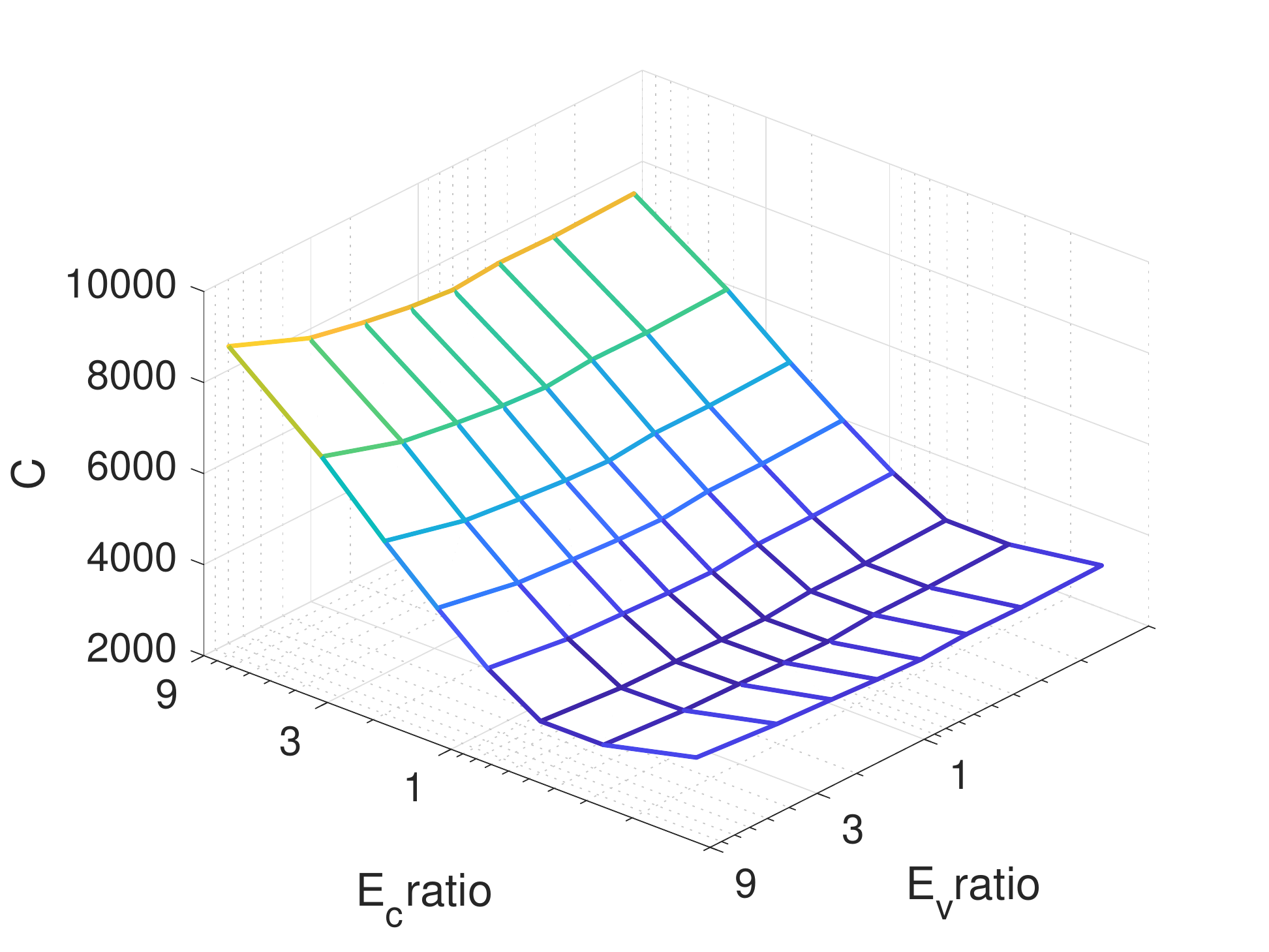}
\caption{Computational cost of $2\times2$ fixed windows for a 16-atom Si system with 399 states (32 occupied and 367 unoccupied states) and 1 k-point.}
\label{fig:Ccost}
\end{figure}

We describe our prescription to determine the optimal number and placement of energy windows in the range of $E_c$ and $E_v$. This is accomplished by minimizing the computational cost function $C^{(GL)}(\epsilon^{(q)})$ of Eq.~(\ref{eq:Ccost}). In this appendix, we omit the fractional error level $\epsilon^{(q)}$ as it does not affect the result of optimal set of energy windows. To motivate the discussion, consider a $2\times2$ window scheme where the two free parameters are the dividing energy values $E_v^*$ and $E_c^*$ in the valence and conduction bands, respectively, that determine the boundaries of the windows.  These are converted to dimensionless numbers via $E_c^{\text{(ratio)}} = \frac{E_c^*-E_{c}^{min}}{E_{c}^{max}-E_c^*}$ and $E_v^{\text{(ratio)}}=\frac{E_v^*-E_{v}^{min}}{E_{v}^{max}-E_v^*}$.  Figure~\ref{fig:Ccost} shows the dependence of the cost $C^{(GL)}$ on two ratios for the case of a flat densities of states. The function $C^{(GL)}$ is a smooth function of the window boundaries and we find that this smoothness is not confined to $2\times2$ windowing but carries over to large number of windows.

\begin{figure}
\includegraphics[width=2.5in]{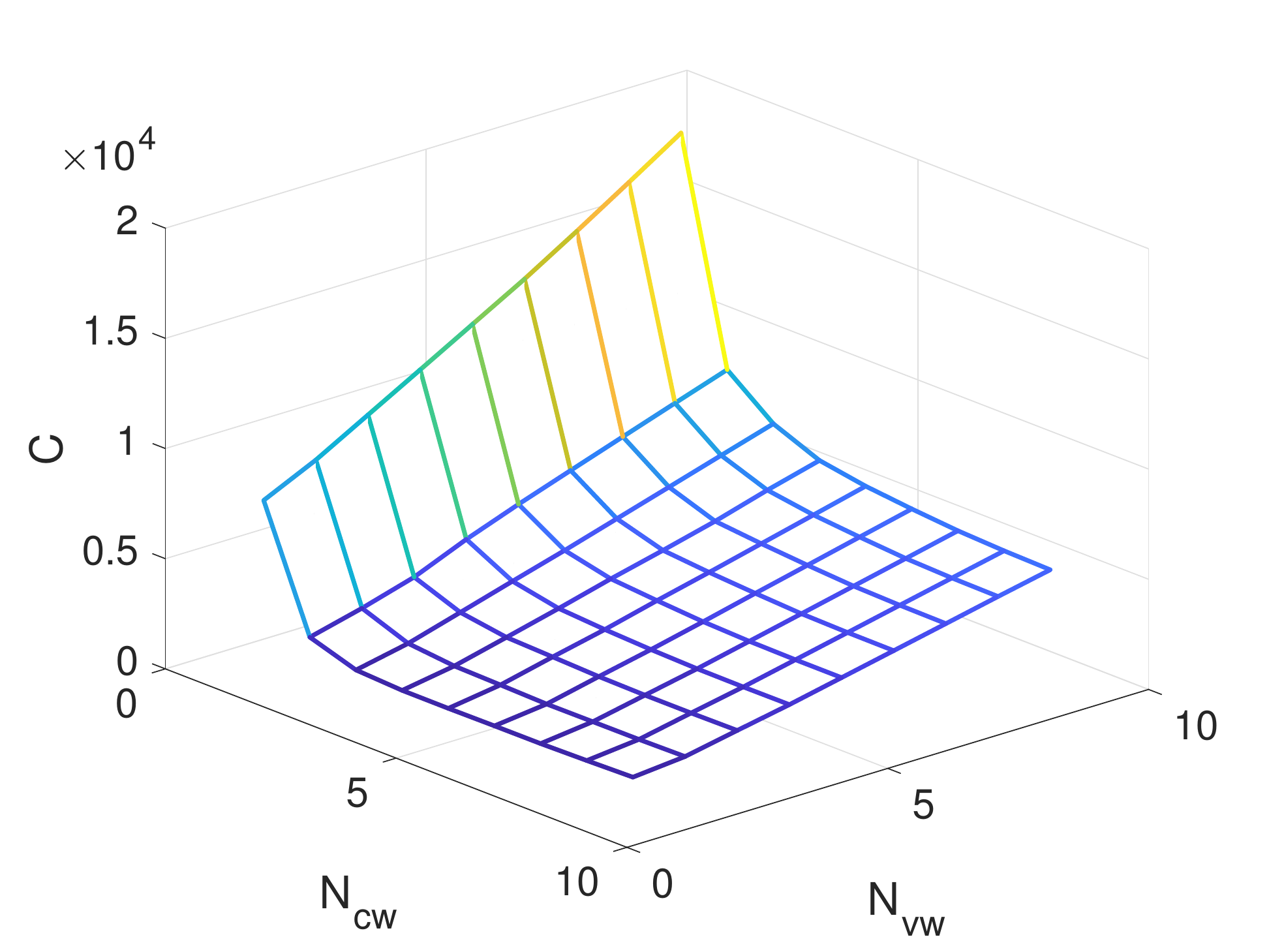}
\caption{Minimized computational cost function $C^{(GL)}$ for a 16-atom Si system for each set of number of windows $(N_{vw},N_{cw})$. The total number of bands is 399 (32 occupied and 367 unoccupied states) and 1 k-point was employed to generate the Kohn-Sham eigenenergies.}
\label{fig:costSivsnwindows}
\end{figure}

Since $C^{(GL)}$ is a smooth function of the sizes of the energy windows, for a given number of windows ($N_{vw}$,$N_{cw}$) and some initial starting window sizes (e.g., all equal), we use a simple gradient descent algorithm to minimize $C^{(GL)}$ as a function of the window size to find the minimum value for those window sizes $C^{(GL)}(N_{vw}$,$N_{cw})$.  Next, we vary the number of windows $N_{vw}$ and $N_{cw}$ over some reasonable range (in practice from 1 to 9 has been more than sufficient for all the systems we have considered), tabulate the minimized $C^{(GL)}(N_{vw}$,$N_{cw})$, and find the global minimum and optimal sizes of windows.  Figure~\ref{fig:costSivsnwindows} illustrates the minimal cost as function of the number of windows for bulk Si with a 16-atom cell comprising 32 valence and 367 conduction bands.  For this case, the minimum number of computation occurs at $N_{vw}=1$ and $N_{cw}=5$. For a more refined optimization in solid state calculations, when computing the polarizability $P^q$ for some momentum transfer $q$, we can apply the above cost minimization scheme to the energies involved in each $k$ and $k+q$ combination  (i.e., $E_{v,k}^l$ and $E_{c,k+q}^m$)  when computing the  $k$ sampling over  the first Brillouin zone (here $D(E)$ acquires band indices). 

\section{Weight function for window pairs with energy crossings}
\label{app:weight}
We develop a weight function and associated quadrature for the case when $F(x;\zeta)$ must be evaluated for energy differences $x$ that are both positive and negative within a window pair - i.e. energy crossings.  A standard choice in the GW literature is to employ a Lorentzian broadening parameter $\gamma>0$ to regularize the singularity
\begin{equation}
F(x) = Im\frac{\gamma}{1-i\gamma x} = \frac{x}{x^2+\gamma^2}\,.
\end{equation}
in the spirit of the additional scattering that typically ameliorates resonances in real materials.
This odd function in $x$ is continuous, approximates $1/x$ when $|x|\gg\gamma$, and has a separable form as a Fourier integral
\begin{equation}
F(x) = \gamma\,Im \int_0^\infty d\tau\, e^{-\tau}e^{i\tau x\gamma}\, .
\label{eq:integralFgamma}
\end{equation}
The exponential weight function implies that the most appropriate quadrature method for approximating the integral is the usual Gauss-Laguerre quadrature.  Hence, this $F(x)$ can be used to separate the sums over $n$ and $p$ when computing $\Sigma(\omega)$.  The  difficulties with this choice are practical.  First, the quadrature grids needed for reasonable errors can become large. Second, the function approaches $1/x$ only when $|x|\gg\gamma$ such that if $\gamma$ is not small compared to the width of the energy windows being employed, there will be sizable errors across window boundaries when we switch from $F(x)$ to $1/x$.  On the other hand, if we make $\gamma$ small to avoid this matching error, the steepness of $F(x)$ near the origin, which is directly related to the slow decay of $e^{-\gamma u}$ in the integral form of $F$ in Eq.~(\ref{eq:integralFgamma}), requires a large quadrature grid.

\begin{figure}
\includegraphics[width=3in]{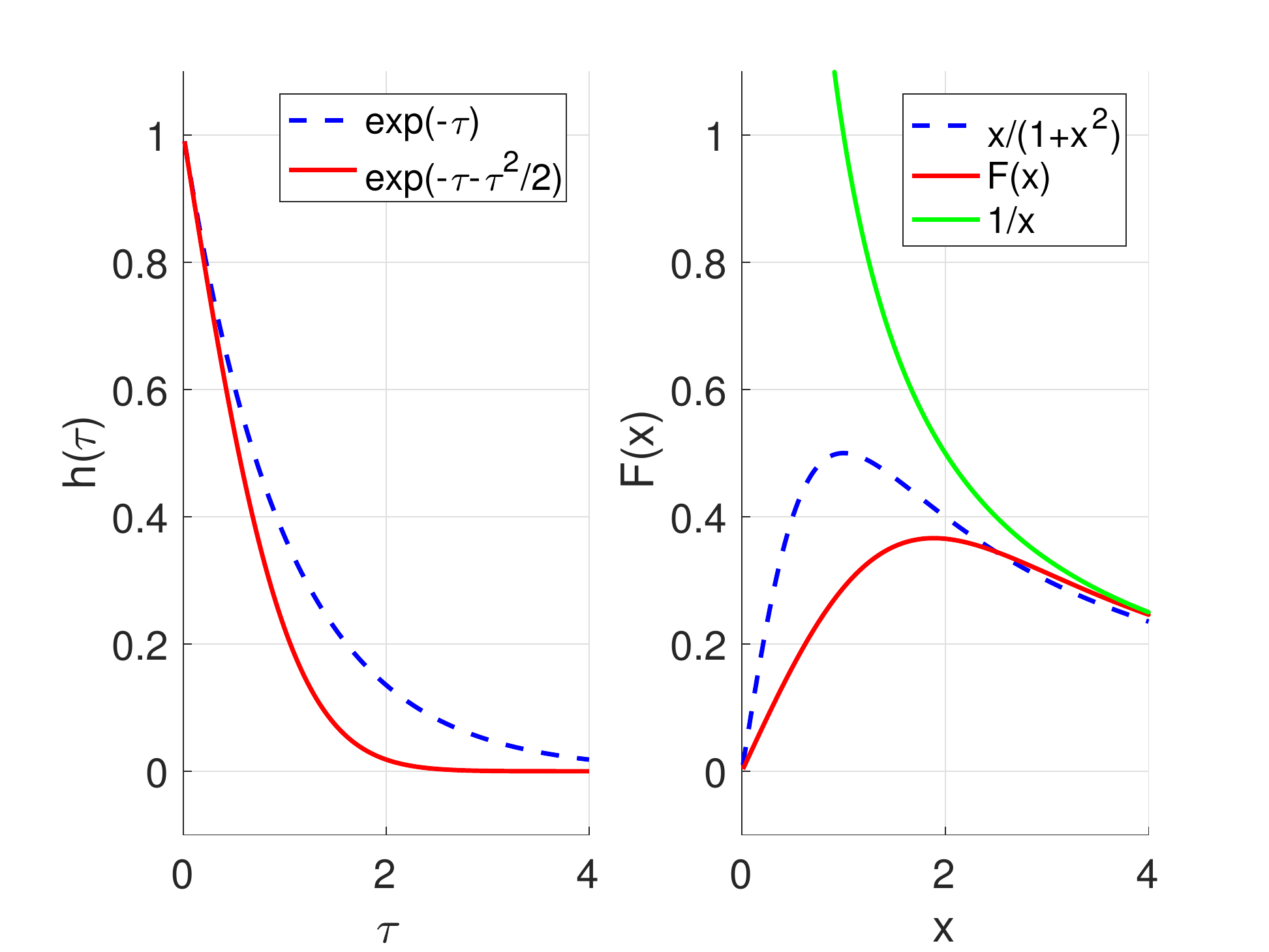}
\caption{Left: comparison of the two weight functions discussed in the text.  The blue dashed curve is the exponential weight $\exp(-\tau)$ associated with the Lorentzian broadening approach; the solid red curve is the new proposed weighing function.  Right: appropriate Fourier transforms of the weights.  The exponential weight $e^{-\tau}$ (dashed blue) corresponds to $x/(1+x^2)$ while the weight $h(\tau)=\exp(-\tau-\tau^2/2)$ corresponds to $F(x)$ (solid red).  For comparison, the target function $1/x$ is shown as well (short dashed green).  $F(x)$ is smoother for small $x$ and approaches $1/x$ more rapidly for large $x$ than $x/(1+x^2)$.}
\label{fig:Fwcomparison}
\end{figure}

We alleviate these difficulties by taking advantage of the freedom afforded in choosing the functional form of $F(x;\zeta)$ in Eq.~(\ref{eq:Fdef}).  Instead of using the weight function $h(\tau;\zeta)=|\zeta| e^{-\tau}$ with $\zeta=i\gamma$ that gives the Lorenzian function, we find that a minimal change of the weight function is sufficient for our purposes.  Our proposed weight
\[
h(\tau;\zeta) = |\zeta| \exp(- \tau -\tau^2/2)
\]
which falls off much faster for large $\tau$ and will thus generate a much smoother $F(x)$ for small $x$.  However, since its behavior for small $\tau$ is the same as the $e^{-\tau}$, the associated $F(x)$  also approaches $1/x$ asymptotically at large $x$.  In fact, the choice of having a ratio of exactly $1/2$ between the linear and quadratic parts of the exponential defining $h$ is not arbitrary: this precise ratio guarantees that $F(x;\zeta)=1/x+O(1/x^5)$ for large $x$ while any other choice has corrections of $O(1/x^3)$.  We also note the tranform $F(x)$ can be written, in closed form, in terms of the generalized error function
\begin{eqnarray}
F(x) &=& \zeta Im\left \{\sqrt{\frac{\pi}{2}}e^{-\frac{(x\zeta+i)^2}{2}}
\left [ 1 + i\mathrm{erfi}\left (\frac{x\zeta+i}{\sqrt{2}}\right)\right ]\right \}. \nonumber \\
\end{eqnarray}
Figure~\ref{fig:Fwcomparison} shows a comparison of the two weight functions and their computed Fourier transforms $F(x)$ 
for $\eta=1$.  

The weights and nodes for a Gaussian-type quadrature for weight function, $\exp(- \tau -\tau^2/2)$,
can generate using the standard procedures embodied in the matlab code provided in Appendix \ref{app:matlabcode}.  Table~\ref{tab:quaderrors} shows how the quadrature grid must be each specified error when using the Lorentzian generating weight $e^{-\tau}$ as well as the improved weight $ \exp(-\tau-\tau^2/2)$ for an energy window of unit width.  To generate this table, we specify a maximum percentage error and then find the broadening $\zeta^{-1}$ so that $F(x)$ differs from $1/x$ by less than the specified error when $x=1$.  We then find the size of a quadrature grid $N^{(\tau,HGL)}$ so that the difference between the quadrature approximation of Eq.~(\ref{eq:quadFx}) and the true $F(x)$ is below the same error level for all $x$ in the window (i.e., $0\le x\le 1$).  It is clear that the new weight and associated quadrature is at least an order of magnitude more efficient in generating its transfrom than the standard weight $e^{-\tau}$ in generating its transform, the Lorentzian.

\begin{table}[t!]
\centering
\begin{tabular}{c|c|c}
\%  error & $N^{(\tau,GL)}$ ($w=e^{-\tau}$) & $N^{(\tau,HGL)}$ ($w=e^{-\tau-\tau^2/2}$) \\\hline
5  & 6 &  1 \\
1   & 24  &  1  \\
0.1  & 124 &  5  \\
0.01  & 547 & 15 \\
0.001 &  2216 & 36 
\end{tabular}
\caption{Size of quadrature grid needed for a maximum specified percent error for the two weight functions discussed in this section.  The energy window has unit width ($x=1$).}
\label{tab:quaderrors}
\end{table}

\section{Hermite-Gauss-Laguerre quadrature grid size at fixed error}
\label{app:HGLgrid}
A necessary input to the cost function whose minimization determines optimal window placement, is the number of grid points required to generate error, $\epsilon^{(q)}$, in the time integrals.  Figure~\ref{fig:2x2w}(b) shows a 2$\times$2 windowing example that includes overlapping windows for the computation of $\sum_{a,b}1/(E_a-E_b)$ using the Hermite-Gauss-Laguerre (HGL) quadrature and weight function of the previous appendix. That is, the sign of the denominator changes for the window pairs $\{E_{a,1},E_{b,1}\}$ and $\{E_{a,2},E_{b,1}\}$, and we use our continuous odd function $F(E_a-E_b;\zeta)$ with weight function $h(\tau;\zeta)=|\zeta|\exp(-\tau-\tau^2/2)$ for these window pairs. Again, for window pairs without energy crossing, the time integrals are discretized using Gauss-Laguerre quadrature and these windows are not considered further.

\begin{figure}
    \centering
    \includegraphics[width=2.4in]{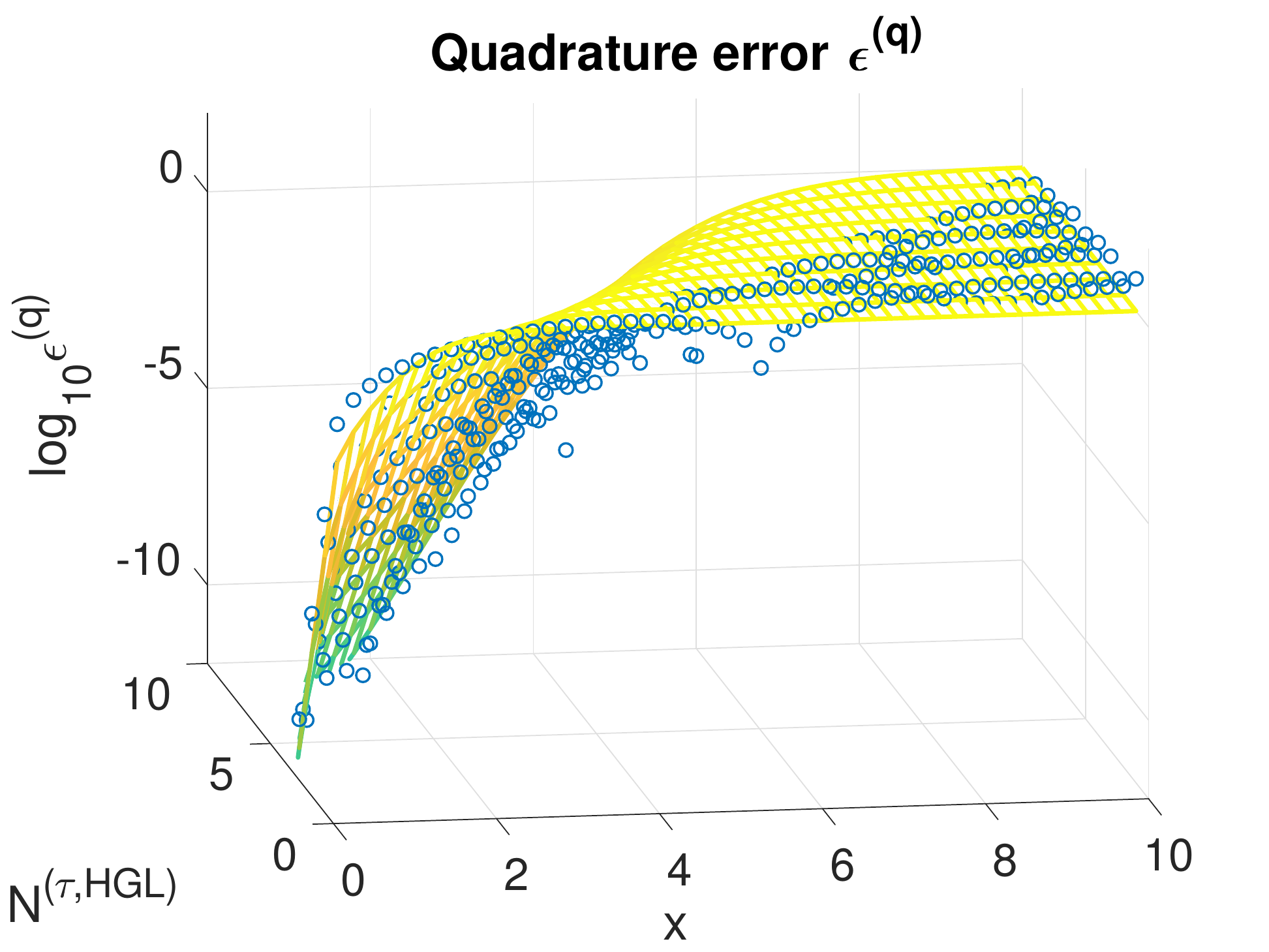}
    \includegraphics[width=2.4in]{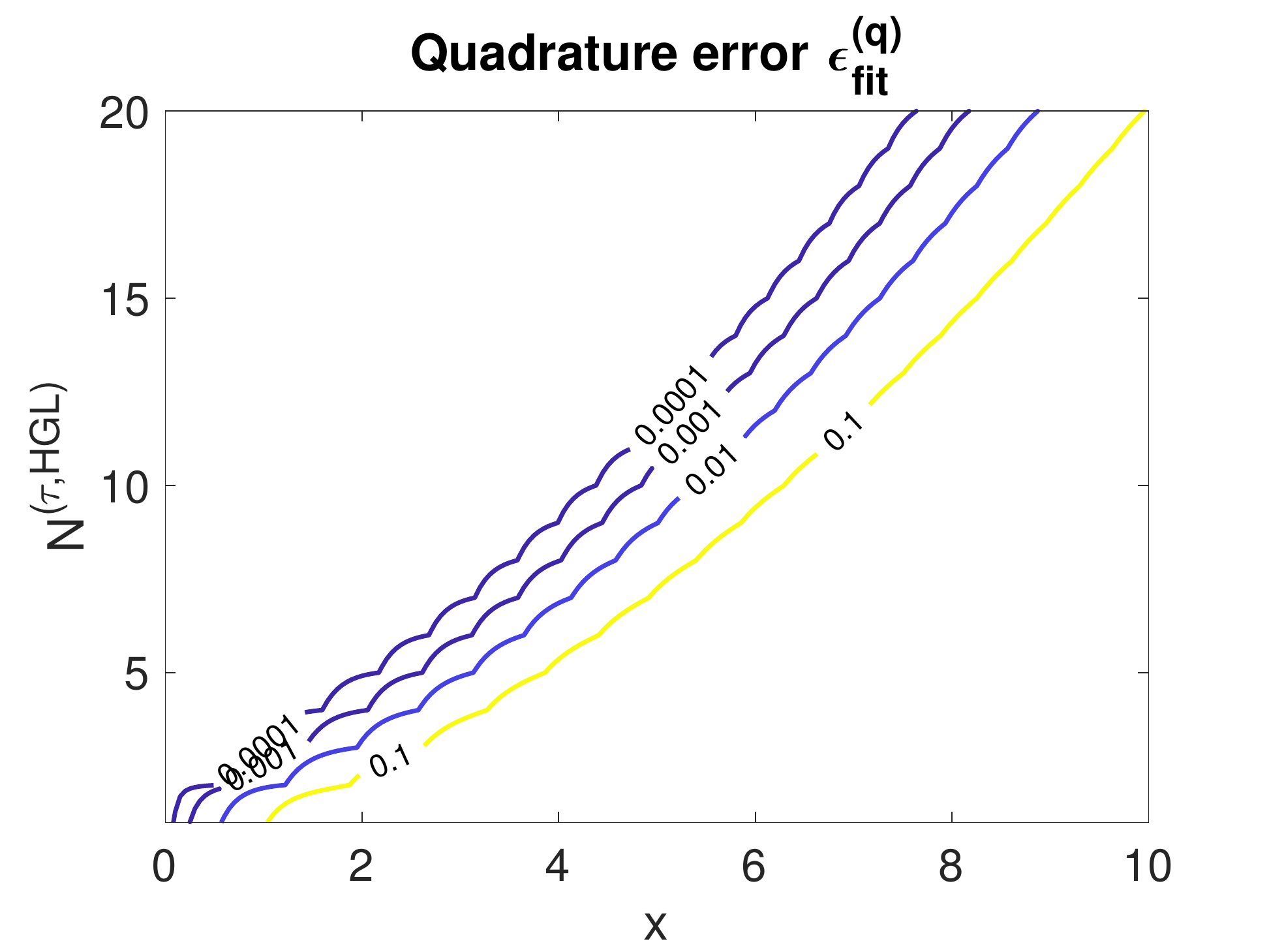}
    \caption{Quadrature error as function of $x$ and $N^{(\tau,HGL)}$. In the upper plot, exact fractional quadrature error ($\epsilon^{(q)}$) is indicated as blue dots along with the fitted function $\epsilon^{(q)}_{fit}$ (Eq.~\ref{eq:Qfit}). In the lower plot, the contour lines of $\epsilon^{(q)}_{fit}$ are shown. Each contour line is well represented with the quadratic function of $x$ (Eq.~\ref{eq:Nfit}).}
    \label{fig:Qnx}
\end{figure}

We first seek quantitative relationship between the number of quadrature points $N^{(\tau,HGL)}$, the energy difference $x=E_a-E_b$, and the fractional error of the quadrature for the case of overlapping energy windows. The fractional quadrature error is defined as
\begin{equation}
\epsilon^{(q)} = \frac{|F(x)-\sum_{j=1}^{N^{(\tau,HGL)}}w_j\sin{(\tau_j x )}|}{|F(x)|}
\label{eq:Q}    
\end{equation}
where, again, 
\[
F(x) = Im\int_0^{\infty}d\tau e^{\left (-\tau-\tau^2/2\right)}e^{i\tau x} 
\]
and have standardized the analysis by setting the energy scaling variable $|\zeta|=\gamma=1$.
Here, $F(x)$ is computed to very high accuracy via numerical integration of the above integral (but could be computed using open source routines for the complex error function). Figure~\ref{fig:Qnx} displays the function $\epsilon^{(q)}(x,N^{(\tau,HGL)})$: due to the presence of the sine function in $\epsilon^{(q)}$, the quadrature error $\epsilon^{(q)}$ is oscillatory as a function of $x$ and finding a simple relationship between $\epsilon^{(q)}$, $x$ and $N^{(\tau,HGL)}$ is challenging.

We propose the following analytical fit $\epsilon^{(q)}_{fit}$,
\begin{multline}
    \epsilon^{(q)}_{fit}=\tanh{(x^{2N^{(\tau,HGL)}})}\times\\
    \exp{\left[-(1+3.3N^{(\tau,HGL)})e^{-0.68x^2/N^{(\tau,HGL)}}\right]}
    \label{eq:Qfit}
\end{multline}
 which is also plotted in Fig.~\ref{fig:Qnx}.
 Direct analytical inversion of Eq.~(\ref{eq:Qfit}) to obtain $N^{(\tau,HGL)}$ as a function of $x$ and $\epsilon^{(q)}_{fit}$ is not feasible. However, we found a good estimate is 
 \begin{equation}
 N^{(\tau,HGL)} (x; \epsilon^{(q)})=
 c_2(\epsilon^{(q)})x^2+c_1(\epsilon^{(q)})x+c_0(\epsilon^{(q)})
 \label{eq:Nfit}
 \end{equation}
where
\begin{eqnarray}
c_2&=&-0.0036\ln\epsilon^{(q)}+0.11\ , \nonumber \\ 
c_1&=& -0.0043(\ln\epsilon^{(q)})^2-0.13\ln\epsilon^{(q)}+0.54\ , \nonumber \\
c_0&=&-0.204\ln\epsilon^{(q)}-0.29\ , \nonumber
\end{eqnarray}
and $x$ is the scaled bandwidth (i.e. scaled, here, by unit $\gamma$).

\section{Energy windowing with energy crossings}
\label{app:cost4overlap}
Minimization of the cost function to determine window placement for cases in which there are energy crossing,  for instance, in the computation of the self-energy $\Sigma(\omega)$, is given.  In perfect analogy with the cost for the static $P$ calculation, we take separate energy windows for occupied (valence $v$) case $\omega-E_v+\omega_p$ and the unoccupied (conduction $c$) case $\omega-E_c-\omega_p$. We allow the number of energy windows $N_{\omega_pw}$ and $N_{vw}$ or $N_{cw}$ to range from 1 to 9, and for each such choice $(N_{\omega_pw},N_{vw})$ or $(N_{\omega_pw},N_{cw})$, we minimize the computational cost via a simple gradient descent method.  When a window pairs have energy crossing, we employ Eq.~(\ref{eq:Qfit}) to estimate the quadrature size, while for all other window pairs we use GL quadrature and Eq.~(\ref{eq:Ngl}) to find the size of the quadrature grid. 

\begin{figure}
    \centering
    \includegraphics[width=2.4in]{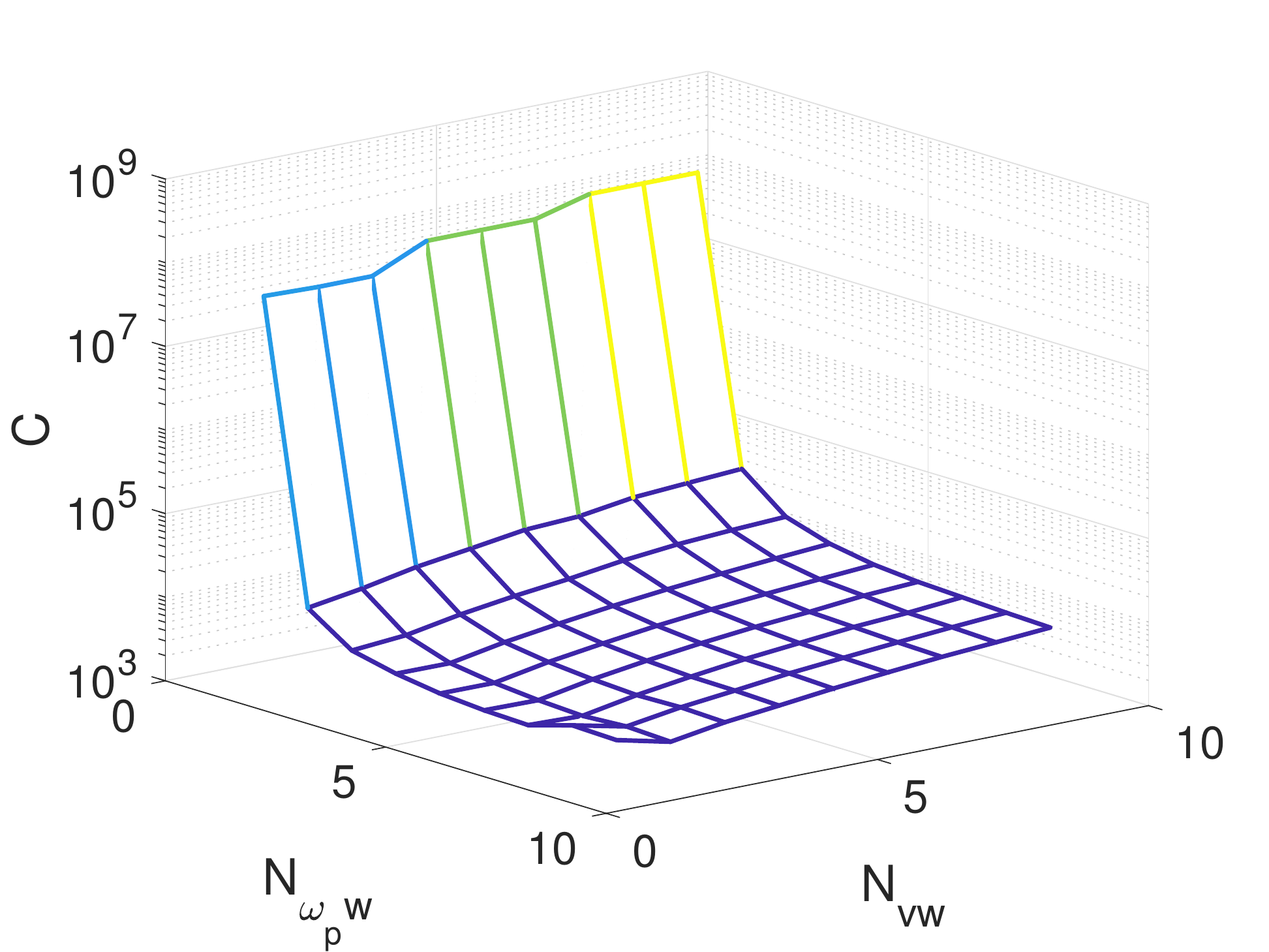}
    \includegraphics[width=2.4in]{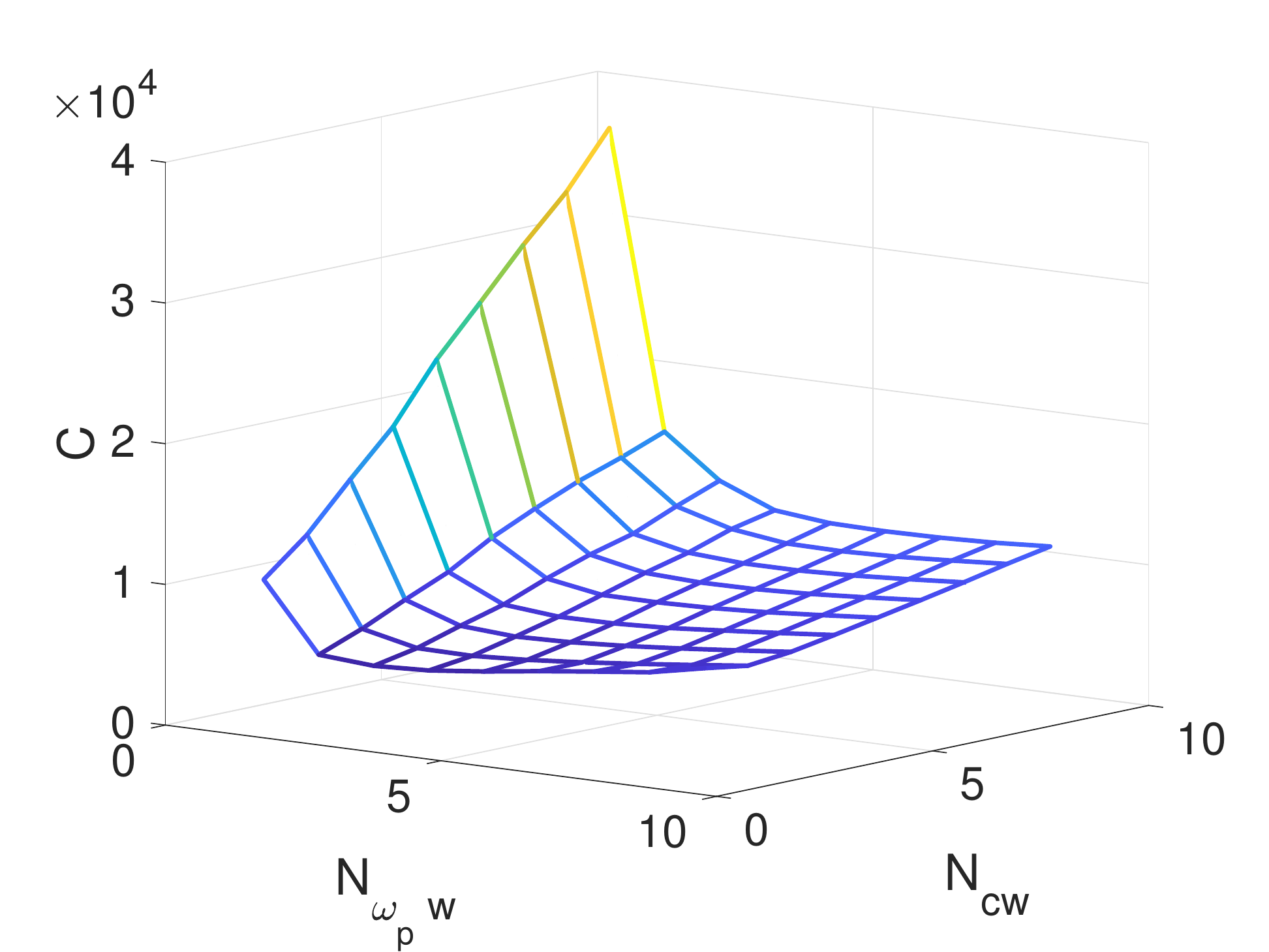}
    \caption{Finding the optimized energy windows as a function of the number of energy windows. In this figure, we choose $\omega=E_{v,min}$ (valence band minimum energy), and the upper and lower plots show the minimum cost for the valence contribution to $\Sigma$ and the conduction contribution,  respectively. The system under consideration is bulk Si.}
    \label{fig:sigmacost}
\end{figure}

Figure~\ref{fig:sigmacost} shows the result of the cost minimization of the model function 
\begin{equation}
G = \sum\frac{1}{\omega-E_n\pm\omega_p}
\end{equation}
using input frequencies from an 8-atom crystalline Si supercell cell and fixed fractional error ($\epsilon^{(q)}$) of 0.01. A total of 32 valence bands and 382 conduction bands are used, and the self-energy is evaluated at energy $\omega$ set to the valence band minimum energy.  The valence band ranges from -0.21 to 0.23 Ha, the conduction band from 0.25 to 2.29 Ha, and the 425 plasmon modes have energies ranging from 0.31 to 45.5 Ha.  We use the actual density of states when calculating the computational cost $C^{lm}$. As we select $\omega=-0.21$, only the valence band sum for $\Sigma$ has the sign change requiring HGL quadrature. For the conduction contribution for $\Sigma$, no sign change occurs, and we simply utilize GL quadrature for all \{$\omega_p,E_c$\} pair. For the valence band sum, the optimal windowing is $N_{vw}=2$ and $N_{\omega_pw}=7$, while for the conduction band $N_{cw}=1$ and $N_{\omega_pw}=3$.

\section{Interpolation method}
\label{app:interp}
\subsection{Theory}
In real space, the static random phase approximation (RPA) irreducible polarizability matrix is 
\begin{equation}
P_{r,r'} = -2\sum_{v}^{N_v} \sum_{c}^{N_c} \frac{\psi_{r,v}^*\psi_{r,c}\psi_{r',c}^*\psi_{r',v}}{E_c-E_v}
\label{eq:P}
\end{equation}
One advantage of working in a real-space basis is that the sum over products of wave functions is separable so one can come up with cubic scaling algorithms if one can make separable approximations to the energy denominator.  We begin by rewriting $P$ as
\[
P_{r,r'} = -2\sum_{v}\psi_{r,v}^*\,A(E_v)_{r,r'}\,\psi_{r',v}
\]
where the matrix $A$ is defined as 
\[
A(z)_{r,r'} = \sum_{c} \psi_{r,c}\psi_{r',c}^*/(E_c-z)\,.
\]
For a system with an energy gap $E_{g}$, the denominator $E_c-E_v$ is always positive with a minimum value of the gap $E_{g}$.  Furthermore, the matrix $A$ must be evaluated only for energies $z$ within the range of valence band energies $E_v$.  Hence, the calculation of $P$ uses $A(z)$ for values of $z$ where it is smooth in $z$.  This means we can use interpolation: we first tabulate $A(z)$ for a range of $z$ values ranging over the valence band energies.  This tabulation costs $N_zN_cN_r^2$ which is cubic since the valence band width is an intensive quantity and the number of points $N_z$ needed for a fixed accuracy is a fixed, intensive number.  Next, to compute $P$, we sum over $v$, and for each $E_v$ we interpolate $A$ to that energy by using the  tabulated $A$.  This calculation is also cubic and costs $N_iN_vN_r^2$ where $N_i\le N_z$ is the number of tabulated $z$ values needed for interpolation (e.g., $N_i=2$ for linear interpolation).

An efficient interpolation scheme should require a small number of $z$ points $N_z$ as well as a modest interpolation cost $N_i$.  In our case, the energy dependence requiring interpolation is given by $1/(E_c-z)$ which is most rapidly changing for the largest values of $z$ near the top of the valence $E_{v}^{max}$ band and when $E_c$ takes on its smallest value at the conduction band minimum $E_{c}^{min}$.  Hence, an efficient interpolation scheme will use a non-uniform $z$ grid that appropriately concentrates sampling points near $E_{v}^{max}$.

The next section below describes the approach we use to find optimal interpolation grids $z_j$ for the case of linear interpolation (i.e., two-point nearest neighbor interpolation with $N_i=2$) when sampling over the entire range of valence band energies.  We note higher order interpolation schemes with $N_i>2$ can be used as well that will reduce the number of grid points needed for a fixed error but require more work to perform the interpolation.  In our experience, the higher order interpolations do not in the end improve performance at the same level of error when compared to the simpler linear interpolation method.

Regardless of the precise interpolation scheme used, all such interpolation methods will have errors that decrease as a power of the number of grid points $n$.  As the data presented in the main text shows, the Fourier-Laplace transform based methods turn out to have superior error properties (their errors fall off exponentially in $n$).

\subsection{Energy grids for interpolation}

The function of $z$ that we wish to interpolate over $z$ is
\[
A(z)_{r,r'} = \sum_c^{N_c}\frac{\psi_{r,c}\psi_{r',c}^*}{E_c-z}\,.
\]
The function is steepest in $z$ close to the top of the valence band $E_{v}^{max}$ when the energy difference in the denominator is small.  In fact, we will consider the worse case scenario and focus on the stiffest and steepest term in the entire sum  which is for the case $E_c=E_{c}^{min}$, the conduction band minimum energy.  Hence the most difficult to interpolate term is given by the dimensionless function
\[
f(z) = \frac{E_{gap}}{E_{c}^{min}-z} \equiv \frac{1}{1+x},
\]
where $z=E_{v}^{max} - xE_{gap}$, and the scaled energy variable $x$ satisfies $0\le x\le (E_{v}^{max}-E_{v}^{min})/E_{gap}$.  

The question is how to pick a grid of $\{x_j\}$ values with $n$ points where $x_1=0$ and $x_n=(E_{v}^{max}-E_{v}^{min})/E_{gap}$.  For simplicity, we will be using linear interpolation, so that given some $x$ between two grid points $x_j\le x\le x_{j+1}$, the linear interpolation is $f^l(x)=[f(x_j)(x_{j+1}-x)+ f(x_{j+1})(x-x_j)]/\Delta x_j$ where $\Delta x_j = x_{j+1}-x_j$.  Calculus then provides an analytical expression for the maximum error $f^l(x)-f(x)$ in the interval $x_j\le x\le x_{j+1}$.  For large $n$ and thus small spacings $\Delta x_j$, the lowest order term for the error is
\[
(f^I-f)_{max} \approx \frac{(\Delta x_j)^2}{4(1+x_j
)^3}\,.
\]
We wish to bound this error by a fixed fractional error tolerance, $\epsilon^{(q)}$, for all  $j$,
\begin{equation}
\frac{(\Delta x_j)^2}{4(1+x_j
)^3}\le\epsilon^{(q)}\,.
\label{eq:errinterp}
\end{equation}
which then in principle determines the grid points $x_j$.  In practice, exact solution of this equation is very difficult, so we again appeal to the large $n$ limit where $x_j$ can be viewed as a function $x(j)$ of a continuous argument $j$ so we approximate $\Delta x_j\approx dx/dj$.  Then Eq.~(\ref{eq:errinterp}) turns into an ordinary differential equation with specified boundary conditions.  The solution is
\[
x(j) = \frac{1}{(1-(j-1)\sqrt{\epsilon^{(q)}})^2}-1\,.
\]
Since $x(n)=(E_{v}^{max}-E_{v}^{min})/E_{gap}$ is known, this determines $n$ for each $\epsilon^{(q)}$.  And finally we have $z_j = E_{v}^{max} - x_jE_{gap}$.

The above choice of grid bounds the error when evaluating the function once.  However, when using the interpolation to compute $P$ from $A$, we will be evaluating the interpolation over many values across the valence band which approximate an integral.  Hence, a more appropriate error control scheme will not only consider the error in interpolating $f(x)$ but also the fact that narrower intervals of $x$ will be sampled less often (assuming a smooth and roughly flat density of states).  Hence we should instead bound the error in the function times the size of the interval:
\[
\Delta x_j\times \frac{(\Delta x_j)^2}{4(1+x_j
)^3}\le\epsilon^{(q)}
\]
Repeating the above exercise, the grid appropriate to this error bound is given by
\begin{equation}
x(j) = \exp\left([4\epsilon^{(q)}]^{1/3}(j-1)\right)-1\,.
\label{eq:expgrid}
\end{equation}
As before, the fixed value of $x(n)$ then determines $n$ at fixed $\epsilon^{(q)}$, and we  use the $x_j$ to get the energy grid points $z_j$.  The results in the main text are based on use of this second (exponential) grid of Eq.~(\ref{eq:expgrid}).

\section{Details of KS-DFT and GW computations}
\label{app:details}
We perform DFT calculations to obtain the single particle wave functions and energies for the GW calculations. The plane wave pseudopotential supercell approach is used as implemented by the Quantum Espresso software package \cite{QE}. The band structure is input into the GW computations presented in the main text.

For Si, we use the local density approximation (LDA) for exchange and correlation as parameterized by Perdew and Zunger \cite{PZ}. The Si norm-conserving pseudopotential is generated with the valence configuration of 3$s^2$3$p^2$3$d^0$ with the cutoff radii of 1.75, 1.93, and 2.07 a.u. for $s$, $p$, and $d$ channels. The plane wave cutoff is 25 Ry, and the lattice parameter is fixed to the experimental value of 5.43 \AA. 

For MgO, GGA-PBE is used for the exchange-correlation functional \cite{PBE}. Both Mg and O are represented by norm-conserving pseudopotentials generated with the valence configuration of $\rm 3s^2$ and $\rm 2s^22p^43d^04f^0$ for Mg and O, respectively. The plane wave cutoff is 50 Ry, and the lattice parameter is fixed to 8.42\AA. 

For the data in the main text on $\epsilon_\infty$ and the COHSEX band gap, we sample the gamma point of a 16 atom unit cell for both materials. For the G$_0$W$_0$ band gaps for Si, a 4$\times$4$\times$4 sampling of the primitive cell equivalent to a 2$\times$2$\times$2 sampling of the 16 atom cell is used. The total number of bands was 399 for Si and 433 for MgO. For the CTSP-W method, $\{N_{vw}=1,N_{cw}=4\}$ and $\{N_{vw}=1,N_{cw}=4\}$ are employed to treat both MgO and Si.

To create the data on computational load versus the number of atoms (Fig.~4 in the main text), we chose Si with the following number of $k$-point and bands:  52 bands with 8 k points for the 2-atom cell, 104 bands with 4 k points for the 4-atom cell, 208 bands with 2 k points for the 8-atom cell, and 416 bands with 1 k points for 16-atom cell. For the CTSP-W method, $\{N_{vw}=1,N_{cw}=5\}$ are used.

For Al, we use LDA for exchange and correlation as parameterized by Perdew and Zunger~\cite{PZ}. The plane wave cutoff is 50 Ry, and the lattice parameter is fixed to 3.99 \AA. For the data in the main text on $P_{0,0}$ error, we use 8 Al atoms in a super cell, and sample 2 k points. Total 400 bands are used to create the data. We use Gaussian spreading method to calculate occupation number with $\beta^{-1}=0.03$Ry broadening. For the CTSP-W method, $\{N_{vw}=1,N_{cw}=7\}$ are employed.

\section{Code for the new quadrature}
\label{app:matlabcode}
The nodes and weights can be found, for example, by using the two matlab functions provided below.
\begin{verbatim}
function [x,w]=GLquad(n)
% function [x,w]=GLagIntP(n)
% Gauss-Laguerre integration: return nodes x 
% and weights w for a
% quadrature grid with n points

% This is basically the Golub-Welsch method
J=diag(1:2:2*n-1)+diag(1:n-1,1)+diag(1:n-1,-1);
[v,l]=eig(J);
[x,ix]=sort(diag(l));
w=v(1,ix)'.^2;
return

function [xmat,wmat] = myweightquad(n)
%function [xmat,wmat] = myweightquad(n)
% Return all nodes (xmat) and weights (wmat)
% for quadratures up to % n points for weight 
% w(x)=exp(-x-x^2/2).  These are organized in
% matrices. xmat are the nodes and wmat 
% are the weights.  Each column is for a 
% quadrature size going from
% 1 to n (left to right).  Thus the lower
% triangle is padded with zeros.

% Figure out number of grid points
% so that the biggest moment (2n)
% is well converged.  We do 
% Gauss-Laguerre quadrature to
% do these integrals over the weights!
Iold = 0;
for nx=round(10.^[1:.2:7])
    [xq,wq] = GLquad(nx);
    weight = exp(-xq.^2/2);
    I = sum(wq.*weight.*xq.^(2*n));
    if Iold>0
        err = (I-Iold)/I;
        if abs(err)<1e-14 
            break
        end
    else    
    end
    Iold = I;
end

% Build polynomials as we go
% and figure out the recursion
% relation coefficients as we go
p = zeros(length(xq),n+1);
p(:,1) = 1;
a = zeros(n,1);
b = zeros(n,1);
for j=1:n
    xpp = sum(wq.*xq.*weight.*p(:,j).^2);
    pp = sum(wq.*weight.*p(:,j).^2);
    a(j) = xpp/pp;
    if j>1
        ppm1 = sum(wq.*weight.*p(:,j-1).^2);    
        b(j) = pp/ppm1;
    end
    if j>1
        p(:,j+1) = ...
            (xq-a(j)).*p(:,j)-b(j)*p(:,j-1);
    else
        p(:,j+1) = (xq-a(j)).*p(:,j);
    end
end

% Prepare for Golub-Welsch
b = b(2:end);
b = sqrt(b);
mu0 = sum(wq.*weight);

% Build Golub-Welsch J matrix, 
% eigen decompose it, and get weights and
% nodes for each value of j=1,...,n  
% (i.e. all weights and nodes for
% quadratures up to size n)
J = diag(a) + diag(b,1) + diag(b,-1);
xmat = zeros(n,n);
wmat = zeros(n,n);
for j=1:n
    Jcut = J(1:j,1:j);
    [v,d] = eig(Jcut);
    d = diag(d);
    [~,is] = sort(d);
    d = d(is);
    v = v(:,is);
    x = d;
    w = v(1,:).^2*mu0;
    w = w';
    xmat(:,j) = [x' zeros(1,n-j)]';
    wmat(:,j) = [w' zeros(1,n-j)]';
end

return
\end{verbatim}

\end{document}